# Early Detection of Mental Stress Using Advanced Neuroimaging and Artificial Intelligence

Fares Al-Shargie [a,b]

[a]Biosciences and Bioengineering Research Institute, Department of Electrical Engineering, American University of Sharjah, United Arab Emirate

[b]Centre for Intelligent Signal and Imaging Research, Department of Electrical and Electronic Engineering, University, Teknologi PETRONAS, Malaysia

* **Correspondence**: fyahya@aus.edu

## ABSTRACT

People spend most of their time in the workplace, often with high workloads and time pressure, a practice that contributes to increased stress levels. An accurate stress assessment method may thus be of importance to clinical intervention and diseases prevention. While different neuroimaging modalities have been proposed to detect mental stress, each modality experiences certain limitations. In this book, we introduced the state of the art of mental stress and investigated whether fusion of electroencephalographic (EEG) and functional near infrared spectroscopy (fNIRS) signals could help improve the detection rate of mental stress. The study proposed novel approaches to fuse the EEG and the fNIRS signals in the feature-level using joint independent component analysis (jICA) and canonical correlation analysis method (CCA) and predated the level of stress using machine-learning approach. EEG and fNIRS signals were decomposed into a set of features in the temporal and spatial domain. The jICA and CCA were then developed to combine the features to detect mental stress. The jICA fusion scheme discovers relationships between modalities by utilizing ICA to identify sources from each modality that modulate in the same way across subjects. The CCA fuse information from two sets of features to discover the associations across modalities and to ultimately estimate the sources responsible for these associations. The study further explored the functional connectivity (FC) and evaluated the performance of the fusion methods based on their classification performance and compared it with the result obtained by each individual modality.

The jICA fusion technique significantly improved the classification accuracy, sensitivity and specificity on average by +3.46% compared to the EEG and +11.13% compared to the fNIRS. Similarly, CCA method improved the classification accuracy, sensitivity and specificity on average by +8.56% compared to the EEG and +13.03% compared to the fNIRS, respectively. The overall performance of the proposed fusion methods significantly improved the detection rate of mental stress, $p < 0.05$. The FC significantly reduced under stress and suggested EEG and fNIRS as a potential biomarker of stress.

# TABLE OF CONTENTS

# LIST OF FIGURES

# LIST OF TABLES

CHAPTER 1

INTRODUCTION TO MENTAL STRESS

## 1.1 Motivation

Stress constitutes an ever-growing problem in our society. It has become part of our daily life, and many people suffer from it. Stress is defined as the non-specific response of the body and mind to any demand of change [1]. People spend most of their time in the workplace, often burdened with high workloads and time pressure, which contributes to an increase of our stress levels. Stress is listed as the second most frequent work-related health problem in Asia and Europe. A majority of Malaysians are suffering from more stress-related illnesses due to the problems at work. A recent survey by Regus found that 70% of Malaysian employees were afflicted with diseases which stemmed from rising stress levels at work [2]. In 2002, stress in the workplace cost €20 billion to the enterprises of EU15 [3] and in 2005, 22% of working Europeans were reported as suffering from it [4]. According to a recent study, 51% of European workers have confessed that stress is common in their workplace, and it is estimated that 50–60% of all lost working days in European enterprises are due to work-related stress and psychosocial imbalances [3]. Similarly, in the United States and Australia, stress costs over $300 billion and more than $14.2 billion per a year accordingly [5].

Stress causes the activation of the hypothalamus-pituitary-adrenocortical axis (HPA axis) and the sympathetic nervous system (SNS) leading to an increase in the stress hormone (cortisol) in the adrenal cortex [6]. The continuous release of cortisol has a direct impact on our body, function and structure of the brain. It increases blood pressure, promotes the formation of artery-clogging deposits and causes brain changes that may contribute to anxiety, depression, and even addiction. Stress has been recognized as one of the major factors contributing to chronic disorders and productivity losses. Long-term exposure to stress has been linked to a variety of health problems such as heart disease, obesity, diabetes, stroke and depression [7, 8]. Besides, it increases the size and activity of the amygdala which is involved in storing memories associated with emotional events [9]. In order to avoid stress and achieve the highest level of performance and help diminishing the risks, it is necessary to detect stress in its early stages, i.e. when it is still limited to acute or episodic stress.

## 1.2 Background

Stress can be measured and evaluated based on psychological, physiological and behavioural responses. The psychological evaluation of stress can be carried out by means of self-report questionnaires or by being interviewed by a psychologist [10]. However, these questionnaires only offer information about the current stress levels of the patient and not about the stressors or the evolution of the stress levels. These tests can be taken from time to time, but may not be suitable for detecting the subtle changes which can indicate the presence of a major problem in its early stage. Furthermore, questionnaires are subjective and require the full attention of the user. ''People can suffer lapses in memory about the emotional tone of a day in as little as 24 h'', which means that we are not always aware of our real stress levels and that methods such as self-report questionnaires may indicate an incorrect stress level measurement [11, 12].

A more objective measure is the cortisol and alpha amylase level. The cortisol level can be estimated from urine, hair, sweat, blood and saliva [13]. Measurements from blood and saliva reflect real-time circulation of cortisol whereas others reflect cortisol production over time. Salivary cortisol has been established as a biomarker to evaluate stress in clinical and bio-behavioural studies [14]. Previous studies have shown that the cortisol level increases during stressful events and the occurrence of negative emotions [15]. Similar studies have shown a significant increase in the salivary alpha amylase level in response to stressful tasks such as, playing video games, pre and post examination, Trier Social Stress Test (TSST), speech and counting, mental arithmetic, negative emotion, driving under stress condition and parachute jump [13, 16, 17]. However, the response time of cortisol is slow (in minutes) and its level is affected by circadian rhythms [18]. The concentration level of cortisol in the early morning is reportedly higher than that in the afternoon. Similar to psychological questionnaires, these methods are neither suitable nor practical when carrying out a continuous monitoring of stress levels. One of the researchers [19] has suggested that the continuous sampling of such biomarkers is not realistic. Actually, considering the preceding method, these kinds of measurements are only done when the affected himself or the people around him realize or suspect the severity of the situation.

Stress can also be measured through bio-signals. The more commonly used bio-signals in evaluating stress are heart rate, blood pressure (BP) and skin conductivity (SC) [20, 21]. Heart rate and blood pressure are reported to increase under stress conditions. The heart rate variability (HRV) has also been established as an instantaneous quantitative measure of the Autonomous Nervous System (ANS) activity associated with stress [22]. Stress can cause a decrease and increase in the high and low-to-high frequency components of the heartbeat interval signals. Skin conductivity, on the other hand, varies with the changes in the skin moisture level revealing the changes occurring in the sympathetic nervous system. However, these methods are also sensitive to other biological changes, humidity cardiovascular and skin diseases. Any robust assessment on mental stress alone may therefore be challenging.

Stress also affects individual behaviour. Some of the induced changes are well-known, for example being much more irritated or angry, but these are not easily measurable. Other possible behavioural changes have been investigated, by for example analyzing individuals interacting with technological devices in order to verify their relation to stress and in order to create a reliable way to measure it [23]. Behavioural measurements for stress recognition are much less frequent than the physiological ones in the state of the art. They have probably not been sufficiently studied, and thus, stress detection results in general are not as accurate as those yielded by physiological methods [3]. These methods also require major human intervention, including manually recognizing and interpreting visual patterns of behaviour. As stress originates from the brain, it is highly desirable to measure and evaluate the stress using advanced non-invasive neuroimaging techniques.

## 1.3 Non-invasive measurement

Functional neuroimaging constitutes a powerful tool in assessing the prefrontal cortex (PFC) function in human subjects. Mental stress is known to be one of the risk factors for neuropsychiatric disorders such as bipolar disorders, schizophrenia, anxiety and depression [7, 24, 25]. Stress disrupts creativity, problem solving, decision making, working memory and other prefrontal cortex (PFC)-dependent activities [26-28]. Animal and human studies have demonstrated the detrimental effects of glucocorticoids (stress hormone) on PFC functioning [29] and identified it as the brain region most susceptive to mental stress. A variety of external stress treatments [30, 31] have shown the potential to remedy the PFC functioning in animals and human subjects.

Exposure to uncontrollable stress rapidly evokes chemical changes in the brain that impair the higher cognitive functions of the PFC while strengthening the primitive brain reactions [27]. It has been appreciated for decades that uncontrollable stress drives mental illness, including aforementioned cognitive disorders. New evidence furthermore suggests that it may also contribute to the cognitive deterioration of Alzheimer's disease. These disorders particularly afflict the most newly evolved pyramidal cell circuits in the association cortex, circuits that are uniquely regulated at the molecular level.

The PFC subserves our highest order cognitive abilities by generating the mental representations that are the foundation of abstract thought and the basis for flexible, goal-directed behavior. In primates, the PFC is topographically organized: the dorsolateral PFC (DLPFC) guides thoughts, attention and actions, while the orbital and ventromedial PFC (VMPFC) regulate emotion (Fig. 1.1a). The DLPFC has extensive connections with the association cortices and the dorsal aspects of the striatum for the regulation of thought and action [32]. In contrast, the most caudal and medial aspects of the PFC project to limbic structures such as the amygdala, ventral striatum, hypothalamus and brainstem for control of the autonomic nervous system (Fig. 1.1b). These PFC areas, along with the insular cortex, are taken to be critical for the mental suffering aspects of pain [31]. These areas receive projections from more rostral and lateral PFC and provide opportunities for the integration of cognitive and emotional processing. The PFC circuits are usually positioned to either facilitate or inhibit processing and thus allow for a flexible, top-down control. Human data suggest that the right hemisphere may be particularly important in inhibitory control.

The integrity of DLPFC function is often tested in working memory tasks, where information must be held in the mind and constantly updated to guide accurate and flexible responding. Under conditions when a subject feels alert, safe and interested, the phasic release of catecholamines strengthens the higher cognitive functioning of the PFC, thus allowing for the top-down regulation of thought, action and emotion. During stress exposure, high levels of catecholamines take the PFC 'off-line' while strengthening the functions of more primitive circuits, for example the conditioned emotional responses of the amygdala and the habitual actions of the basal ganglia. The amygdala activates brainstem stress systems that in turn activate the sympathetic nervous system.

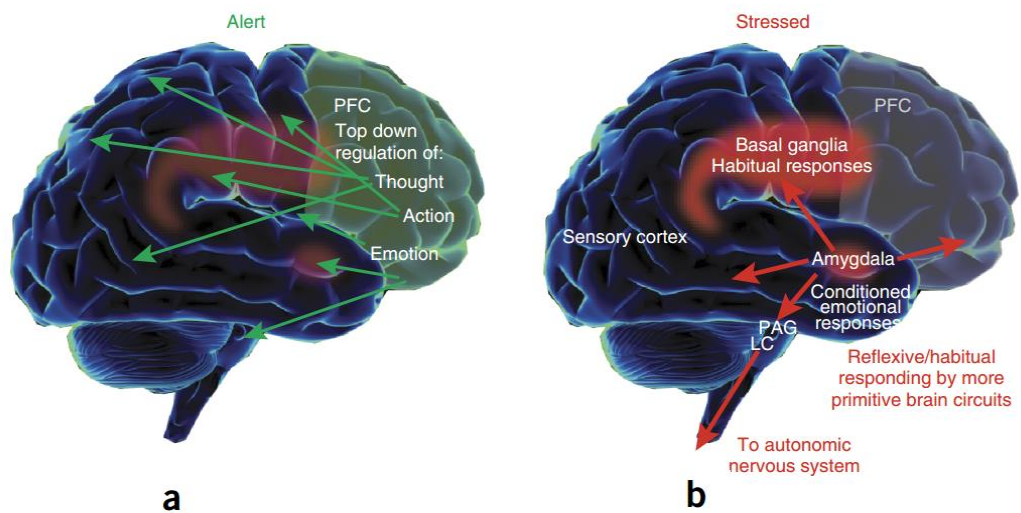

Figure 1.1: Changes in the brain system controlling behavior under conditions of alert safety versus uncontrollable stress [31].

In order to study the development of stress, continuous measurements are required to study the mental state between consecutive tasks over a long period of time [33, 34]. According to previous studies [29, 30], solving arithmetic tasks under time pressure induces mental stress on the PFC. These studies use neuroimaging technologies (e.g. positron emission topography [PET] and functional magnetic resonance imaging [fMRI]) which have good spatial resolution yet limitations in terms of temporal resolution and susceptibility to motion artifacts. Furthermore, these neuroimaging techniques constrain the test subject to a fixed position and a laboratory setting [35].

Electroencephalography (EEG) is a possible alternative neuroimaging technique that does not possess the same limitations. Unlike functional magnetic resonance imaging and positron emission tomography, modern EEG hardware is light-weight and portable enough to be used during unconstrained full-body motion. Unlike Functional Near-Infrared Spectroscopy (fNIRS), the EEG has temporal resolution in the order of a few milliseconds that makes it suitable for measuring cortical changes during workplace activities [36, 37]. EEG signals are categorized by the frequency bands Delta (1-4 Hz), Theta (4-8 Hz), Alpha (8-12.5 Hz) and Beta (12.5-30 Hz). Each frequency band can be used to describe the mental state of a person. Alpha and beta frequency power is linked to negative mood, stress and depression [38-42].

Only a few studies have used EEG to study mental stress. The brain region under study depends on the type of stimuli or tasks (visual, working memory or audio). Hill and Castro found high beta rhythm activity in the sensory motor area during stressful healing task [43]. Seo and Lee identified a similar high beta wave in the frontal and occipital lobe when negative images were presented to induce stress [44]. Choi et al. found a positive correlation between EEG beta power rhythms with stress in the temporal lobe [45]. Thompson and Alonso both proved an increase of beta waves associated with a decrease of alpha waves in the anterior cingulate and frontal anterior cortex [46, 47]. Gärtner et al. reported that the frontal theta decreased during stressful mental arithmetic tasks [48, 49]. Harmony et al., however, reported high delta waves while solving difficult mental arithmetic tasks [50]. Separately, Marshall and Lopez-Duran identified a negative correlation between the EEG alpha power rhythm and stressful events in the prefrontal cortex where the alpha rhythm decreased with stress [51, 52]. In order to detect mental stress, pattern recognition approaches are often adopted [53]. Studies have shown that EEG signals can be used to classify mental stress in the resting state [3, 5, 54-56]. However, EEG has traditionally been thought of as possessing poor spatial resolution and being highly prone to motion artifacts [57]. This can be overcome by combining the EEG with another neuroimaging modality that has a complementary nature.

Functional Near-Infrared Spectroscopy (fNIRS) constitutes a new neuroimaging technique that measures the cerebral hemodynamics associated with neural activity. The technique sends near-infrared light (in the wavelength range 695 - 850 nm) directly into the head [58]. Based on the absorptivity, the change in concentrations of oxygenated (*O₂Hb*) and deoxygenated (*HHb)* hemoglobin can be estimated using modified Beer-Lambert law [59]. Hence, fNIRS is a promising alternative, achieving some middle ground in spatial and temporal resolution as well as mobility between EEG and fMRI techniques [60-63]. Note that, the spatial resolution in fNIRS is based on how deep it penetrate the scalp not based on the number of optodes as the case of EEG system. In other words, the spatial resolution of fNIRS relies on the area of activation. The fNIRS has found its applications in cognitive and behavioural studies. Commonly used tasks to activate the prefrontal cortex (PFC) include mental arithmetic, word generation, colour-word matching, Stroop task, mental rotation, working memory task and inhibition [64-66]. More recently, fNIRS has been accepted as an assistive tool to differentiate depression, bipolar disorder and schizophrenia [67, 68]. Additionally, the measurements have been shown to be consistent with fMRI and EEG [69, 70].

## 1.4 Problem formulation and research question

Previous studies have evaluated stress by using questionnaires, physiological variables and recently based on advanced neuroimaging tools. However, questionnaires remain subjective and require the full attention of the user [11, 12, 71].

These questionnaires only offer information about current stress levels of the user, do not identify the stressors and do not explain the evolution of the stress levels. These tests can be taken from time to time, yet may not be suitable for detecting the subtle changes which can indicate a major health problem in its early stage. Physiological variables including cortisol level, heart rate, blood pressure and skin conductivity are also known to be affected by circadian rhythm, cardiovascular disease, and humidity [18, 72-74]. Furthermore, the concentration level of cortisol in the early morning is higher than that in the afternoon. Hormonal measurements are intrusive, costly and slow methods of analysis [75]. In addition to psychological questionnaires, these methods are not deemed suitable or practical for carrying out a continuous monitoring of stress levels. Furthermore, the stress assessment is often done after the effects have already become obvious to the subject or the people close to him or her. Changes can be detected by performing measurements every several months and by comparing the results, yet this may not be sufficient in order to detect any subtle changes related to early stages of stress [76]. Advanced neuroimaging tools such as functional magnetic resonance imaging and positron emission tomography have good spatial resolution, yet are limited in terms of temporal resolution and susceptibility to motion artifacts. Additionally, these tools require the test subjects to remain still during scanning [35]. Electroencephalography (EEG) constitutes a possible alternative neuroimaging technique that does not possess the same limitations. The EEG has temporal resolution in the order of a few milliseconds, which makes it suitable for measuring cortical changes during workplace activities [36, 37]. However, the EEG is generally considered as having poor spatial resolution (compare to excellent fMRI modality) and being highly prone to motion artifacts [57]. Functional near-infrared spectroscopy (fNIRS) has allowed human cortical activity to be measured during unconstrained movements, the temporal resolution is sub-second, and the spatial resolution is on order of 1 cm$^2$ at best. In order to overcome these constraints, it is advisable to combine the EEG with fNIRS modality to provide a complementary nature. Few studies have combined EEG and fNIRS signals in the feature level as well as in the decision-level fusion [77-79]. One major drawback of these studies is that, they did not discover the associations across the modalities as the fusion was performed by concatenation or by combining the output of the classifiers. Another problem with these approaches is that they failed to improve the overall classification performance as interlink/correlation between modalities was not investigated. Developing a fusion model to discover the association across modalities is very important to provide complementary nature and results in improving the overall diagnosis and classification performance.

## 1.5 Hypothesis

This study hypothesize that developing feature-fusion model to combine the EEG and the fNIRS features can assess mental stress more accurately than using either technique alone. The EEG and the fNIRS possess several advantages over other neuroimaging modalities (e.g. Magnetic Resonance Imaging MRI, Positron Emission Tomography PET) as they are non-invasive, portable, less expensive, safe for long-term monitoring, and reported to be a good complementary [80, 81]. The integration of these two modalities offers a means to partially overcome the limitations of each of the individual modalities by combining their complementary aspects in one single analysis. The proposed multimodal fusion is likely to lead to a more robust means of stress detection.

## 1.6 Objectives

This thesis aims at achieving the following objectives:

1- To detect the level of stress based on the cortical activity and cortical connectivity under normal-control and stress conditions while performing an arithmetic task;

2- To develop a fusion model of EEG and fNIRS features to discover the association across them and examine if the model can improve mental stress detection over individual modality.

**1.7 Thesis layout**

Chapter one introduces the motivation, background, hypotheses and objectives for this work. Chapter two reviews the physiology of stress and the psychological stress elucidation methods. A detailed account on the studies focusing on the development of biomarkers of stress is presented. The purpose of this review chapter is manifold. Firstly, to identify the research gaps that can help in formulating the problem statement and driving the objectives for this thesis. Secondly, the replication of the results based on the data presented in this work so that the proposed method can be compared with the previously applied methods and to review the literature involving psychological evaluation, physiological evaluation and behavioral evaluation. Psychological evaluations mainly focused on questionnaires and the way patients were interviewed by psychologists. Physiological evaluations include hormone levels, heart rate variability (HRV), galvanic skin response (GSR), Skin Temperature (ST), Electromyography (EMG), respiration, and blood volume pulse (BVP), EEG signals and hemodynamic responses are measured by functional magnetic resonance imaging (fMRI) and fNIRS modalities. In addition, physical signals for measuring stress include pupil diameter (PD) and thermal images. The behavioural evaluations focused on key-stroke and mouse dynamics, posture, facial expression and speech analysis. In short, the second chapter highlights the important benefits and drawbacks of each method. The third chapter focuses on the proposed fusion methods and elaborates the sub-process such as the features extraction and selection, building classification models and validation. In addition, the methods used for the localization of stress on the PFC subregions are elaborated in topographical mapping. Finally, functional connectivity is investigated to study the effects of stress on intra- and inter-hemispheric. Chapter four discusses in detail the results computed by the proposed individual modality as well as the fusion methods. Chapter five concludes the study with a summary of the results and implications for future work in mental stress assessment.

CHAPTER 2

STRESS ASSESSMENT METHODS: LITERATURE AND BACKGROUND

This chapter discusses about mental stress. First, the chapter explains the physiology of stress and stress stimulus followed by an extensive review about methods assessing mental stress. These include psychological assessment, physiological assessment and behavioural assessment methods. Subsequently the different types of advanced neuroimaging systems were explained, including EEG, fMRI and fNIRS together with their application to neurological assessment types including stress. Focusing on the EEG and fNIRS systems, the chapter explains basic operation of near infrared spectroscopy followed by optical properties of brain tissue and the Beer-Lambert law for calculation of hemodynamic signals. Lastly, most related studies are summarized and compared.

## 2.1 Physiology of stress

Stress is unpleasant, even when it is transient. A stressful situation, whether environmental such as a looming work deadline or psychological such as persistent worry about losing the job can trigger a cascade of stress hormones that produce well-orchestrated physiological changes. A stressful incident can make the heart pound, breathing quicken and muscles tense. This combination of reactions to stress is also known as the "fight-or-flight" response because it evolved as a survival mechanism that enables people and other mammals to react quickly to life-threatening situations. The carefully orchestrated yet near-instantaneous sequence of hormonal changes and physiological responses helps the distressed individual to fight the threat off or flee to safety. Unfortunately, the body can also overreact to stressors that are not life-threatening, such as traffic jams, work pressure, and family difficulties.

Over the years, researchers have learned not only how and why these reactions occur, but have also gained insight into the long-term effects stress has on physical and psychological health. Over time, repeated activation of the stress response takes a toll on the body. Research suggests that prolonged stress contributes to high blood pressure, promotes the formation of artery-clogging deposits, and causes brain changes that may contribute to anxiety, depression, and addiction. More preliminary research suggests that chronic stress may also contribute to obesity, both through direct mechanisms (i.e. causing people to eat more) or indirectly (decreasing sleep and exercise). The stress response begins in the brain (see Fig 2.1). When someone confronts an oncoming car or any other danger, the eyes and ears send the information to the amygdala, an area of the brain that contributes to emotional processing. The amygdala interprets the images and sounds. When it perceives danger, it instantly sends a distress signal to the hypothalamus. When someone experiences a stressful event, the amygdala, an area of the brain that contributes to emotional processing, sends a distress signal to the hypothalamus. This area of the brain functions like a command center that communicates with the rest of the body through the nervous system so that the confronted person has the energy to fight or flee.

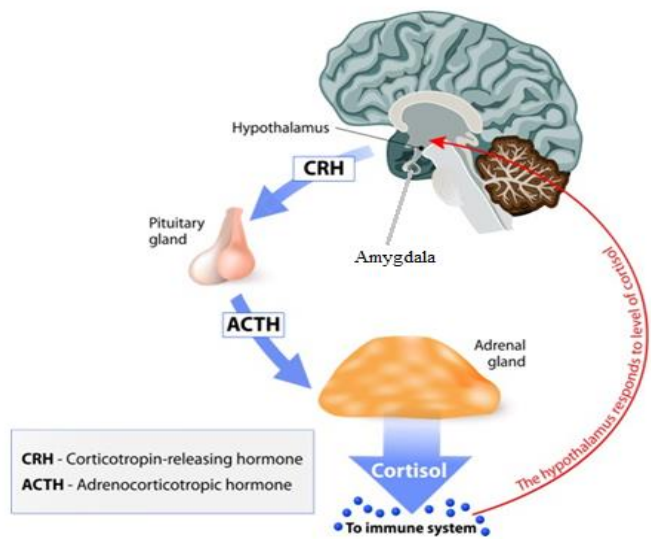

Figure 2.1: Stress response [32].

The hypothalamus functions very similar to a command center. This area of the brain communicates with the rest of the body through the autonomic nervous system that controls such involuntary body functions as breathing, blood pressure, heartbeat, and the dilation or constriction of key blood vessels and small airways in the lungs called bronchioles. The autonomic nervous system has two components, the sympathetic nervous system and the parasympathetic nervous system. The sympathetic nervous system functions like a gas pedal in a car. It triggers the fight-or-flight response and provides the body with a burst of energy so that it can respond to immediate danger. The parasympathetic nervous system acts like a brake and promotes the "rest and digest" response that calms the body down after the danger has passed.

Once the amygdala sends a distress signal, the hypothalamus activates the sympathetic nervous system by sending signals through the autonomic nerves to the adrenal glands. These glands respond by pumping the hormone epinephrine (also known as adrenaline) into the bloodstream. As epinephrine circulates through the body, it induces a number of physiological changes. The heart beats faster than normal and pushes blood to the muscles, heart, and other vital organs, and the pulse rate and blood pressure go up. The person undergoing these changes also starts to breathe more rapidly. Small airways in the lungs open wide. In this way, the lungs can take in as much oxygen as possible with each breath. Extra oxygen is sent to the brain, thus heightening alertness. Sight, hearing, and other senses become sharper. Meanwhile, epinephrine triggers the release of blood sugar (glucose) and fats from temporary storage sites in the body. These nutrients flood into the bloodstream, supplying energy to all parts of the body.

All of these changes happen so quickly that the distressed individual is not aware of them when they happen. In fact, the wiring is so efficient that the amygdala and hypothalamus start this cascade even before the brain's visual centers have had a chance to fully process what is happening. That is the reason why people are able to jump out of the path of an oncoming car even before they think about what they are doing. As the initial surge of epinephrine subsides, the hypothalamus activates the second component of the stress response system known as the HPA axis. This network consists of the hypothalamus, the pituitary gland, and the adrenal glands.

The HPA axis relies on a series of hormonal signals to keep the sympathetic nervous system (the "gas pedal") pressed down. If the brain continues to perceive something as dangerous, the hypothalamus releases corticotropin-releasing hormone (CRH) that travels to the pituitary gland and triggers the release of adrenocorticotropic hormone (ACTH), as shown in Fig 2.1. This hormone travels to the adrenal glands and prompts them to release cortisol (stress hormone). The body thus stays revved up and on high alert. Once the threat has passed, the cortisol level falls, and the parasympathetic nervous system (the "brake") then dampens the stress response [82].

## 2.2 Psychological stress elicitation

In order to carry out stress-related studies and research, it is necessary to provoke the stress response on the desired subject at the required moment. For this purpose, several stress elicitation methods have been developed and validated. The Stroop Color-Word Interference Test and mental arithmetic tasks are the most commonly tests used in previous studies. Both methods have been frequently studied [83]. Similarly, car driving [20] and watching visuals with stressful content and playing computer games are also considered as stress eliciting tasks [32, 84-88]. In a recent study [89], the speed and difficulty of a game have been varied in order to provoke stress and calm reactions alternately on subjects. Public speaking tasks [90] and the Cold pressor test [91] have also been used to induce stress. Finally, Dedovic et al. and others [33, 34, 92] have tested and validated arithmetic task difficulty with fMRI as a stress eliciting method. The aforementioned methods altered physiological as well as biological changes in the body that include increasing the heartbeat, blood pressure, skin conductance and the level of cortisol or stress hormone. Figure 2.2 summarizes the overall physiological changes occurring after exposure to the stress stimulus.

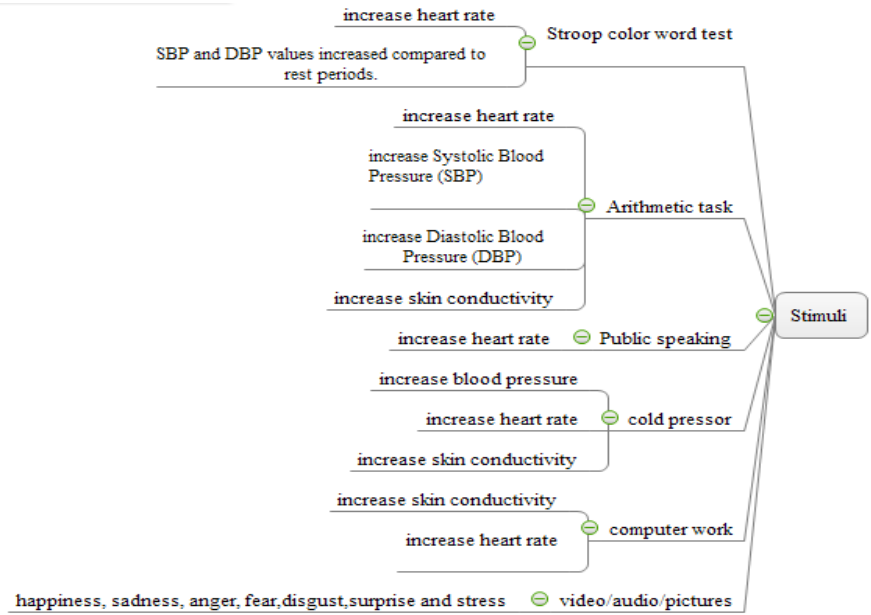

Figure 2.2: Stress inducement methods.

## 2.3 Measuring stress levels

In human subjects, the Sympathetic Nervous System (SNS) provokes the stress response carrying psychological, physiological and behavioural symptoms [21]. Psychological response is understood as ''of or relating to the mind or mental activity'' [23] and does not involve the execution of an action. Physiological responses are part of the normal functioning of a living organism or bodily part [24], therefore, they are non-voluntary actions or responses and very hard or impossible to notice by external observation. On the other hand, behavior is defined as ''the manner of conducting oneself'' [25]. Unlike the physiological response, it involves an action that can be controlled or changed relatively easily in a voluntary way and externally observed.

Psychological responses comprise the increase of strong negative emotions, such as anger, anxiety, irritation or depression [93] and can also make our emotional responses more intense. People tend to feel more worried, frustrated, and hostile with the consequent effects on our relationships [94]. From a physiological point of view, the increase of SNS activity changes the hormonal levels of the body and provokes reactions like sweat production, heart rate increase and muscle activation [95].

Respiration becomes faster and the blood pressure increases [96]. As a consequence of changes in the muscles controlling the respiratory system and the vocal tract, speech characteristics also change. The skin temperature decreases together with that of the hands and feet [26] and the Heart Rate Variability (HRV) decreases [27]. The pupil diameter may vary.

Finally, behavioural reactions include eye gaze and blink rate variations, in addition to changes in facial expressions or head movement [97]. When working in an office environment, interaction patterns with the computer can be affected, together with the General Somatic Activity or the body's agitation level. Performance related to the accuracy and cognitive response, such as the logical thinking [94], attention and working memory can also be affected and lead to a decreased productivity and a tendency to make mistakes. Some people may also abuse tobacco, alcohol and drugs [98]. While the analysis of physiological changes has been the objective of many stress studies, other areas such as behavioural changes have yet to be studied in more depth.

### 2.3.1 Psychological evaluation

The psychological aspects of stress are usually evaluated by means of self-report questionnaires or through interviews by psychologists. Self-report questionnaires constitute the most commonly used method to measure stress levels in humans based on specific scales (e.g. Stress Self Rating Scale (SSRS) [99, 100], Medley Hostility Scale, and Stress Response Inventory (SRI) [101, 102]. However, these questionnaires only offer information about the current stress level, yet do not help in identifying the stressors or explain the evolution of the stress levels. These tests are usually taken at certain intervals and are not suited to detect the subtle changes indicating a major health problem at its early stage. Additionally, these assessments require major human intervention, including manually recognizing and interpreting visual patterns of behavior in observational studies. In fact, these types of assessment are taken only when the affected himself or the people around him or her realize or suspect a severe problem at which time it is already too late for constructive intervention in the vast majority of the cases. Besides, questionnaires are known to be highly subjective.

People can suffer lapses in memory about the emotional tone of a day in as little as 24 hours [71], which means that they are not always aware of our real stress levels and that methods such as self-report questionnaires may result in an incorrect stress level measurement.

### 2.3.2 Physiological evaluation

The Autonomic Nervous System (ANS) responsible for involuntary actions is made up of the Sympathetic and the Parasympathetic Nervous System. The exposure to stressful events causes dynamic changes in the ANS where the activity rate in the Sympathetic Nervous System (SNS) increases and the activity rate in the Parasympathetic Nervous System (PNS) activity decreases. Alternatively, activities in the PNS dominate during resting state activities. SNS and PNS regulate the galvanic skin response, heart rate variability, blood pressure and brain waves, which are the main measures for stress. Symptoms of stress appear as time progresses, which makes continuous recordings of physiological signals significant to monitor variations and detect stress reliably. The most common physiological signals for detecting stress include hormone levels, heart rate variability (HRV), galvanic skin response (GSR), Skin Temperature (ST), Electromyography (EMG), respiration, and blood volume pulse (BVP). In addition, physical signals for measuring stress include the pupil diameter (PD), thermal images, eye gaze, voice characteristics, and face movement. The overall physiological and physical measures being investigated in the existing body of research literature are illustrated in Fig. 2.3.

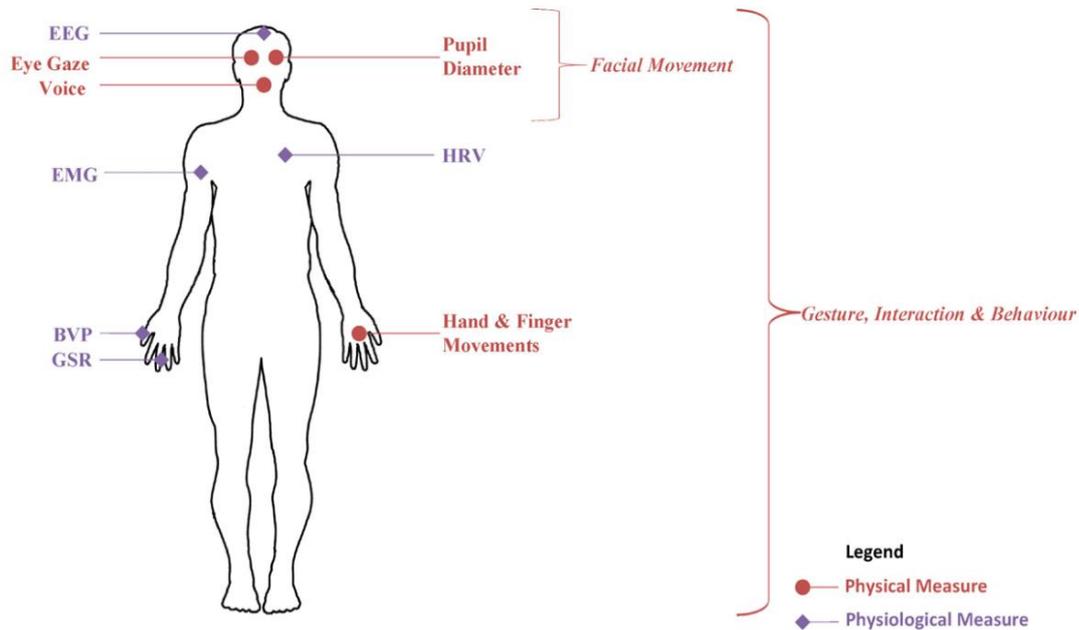

Figure 2.3: Common physiological and physical measures used to detect stress.

### 2.3.2.1 Hormone levels

The stress response changes the endocrine and immune systems by releasing adrenaline and cortisol hormones [54] from the adrenal cortex and the adrenal medulla respectively. Cortisol levels follow a daily cycle in healthy people characterized by peak values in the early morning, decreasing values during the whole day and lowest values at night. Under stress, the ability to regulate cortisol levels decreases keeping them high even at night and changing the typical patterns [103]. Cortisol levels are considered a reliable biomarker of psychological stress [14] and can be estimated from urine, hair, sweat, blood and saliva, the latter being preferred in research due to its non-invasive nature [13]. Among these markers, saliva has the largest variability in concentration to the occurrence of stress. Saliva samples are also very easy to obtain and can be collected at any time. Therefore, saliva is perfectly suited for research in natural environments such as the home and the workplace. Measurements from saliva and blood reflect real-time circulation of cortisol, whereas other measures reflect cortisol production over time. Although cortisol levels measured in blood can offer better inter-individual differences, they are difficult to acquire simultaneously.

Clinical and bio-behavioral studies have shown that cortisol levels increase during stressful events and the occurrence of negative emotions [15, 104, 105]. Additionally, cortisol plays an important role in the regulation of various physiological processes such as blood pressure, glucose levels, and carbohydrate metabolism. The main drawback of cortisol is that its response time is slow (in minutes) and its level can be affected by the subject's sex, age, circadian rhythms, meals, drugs, and salivary flow rate [18]. It has been reported that the level of concentration is higher in the early part of the day than in the afternoon.

Salivary alpha amylase (sAA) constitutes another alternative stress measure. It has been established as a biomarker for the psychological stress response within the SNS [106, 107]. Several studies have shown a significant increase in salivary alpha amylase level in response to stressful tasks such as playing a video game, before and after an examination, Trier Social Stress Test (TSST), speech and counting task, mental arithmetic task [16], negative emotion, driving under stress conditions and parachute jumping [108, 109]. The changes in α-amylase levels are more remarkable than those in salivary cortisol after the same task of stress. Compared to salivary cortisol, alpha amylase is more stable, easy to use, and highly sensitive to physiological stressors and psychological stress.

However, similar to the psychological questionnaires, these methods are not suitable or practical for carrying out a continuous monitoring of stress levels. Some researchers [19] have suggested that the continuous sampling of such biomarkers is not realistic. In fact, these kinds of measurement are only done when the affected himself or herself or the people around such an individual realize or suspect the severity of the case. Changes can be detected by performing measurements at an interval of every several months and through comparing results, yet this may not be enough to detect subtle changes related to the early stages of stress [76]. As Sharma and Gedeon [54] state, these methods ''require major human intervention, including manually recognizing and interpreting visual patterns of behavior''. Unfortunately, individuals may not recognize these signs and allow them to go untreated [96]. Vizer et al. [76] also suggests that ''current tests are usually administered in a physician's office or a rehabilitation facility, causing inconvenience for the patient, using valuable healthcare resources, and making frequent monitoring unrealistic". Hormonal measurements are intrusive, costly, and slow methods of analysis [75]. Consequently, the realization of a person being over-stressed often comes too late and at a time when obvious health problems have already manifested themselves.

### 2.3.2.2 Electrocardiogram (ECG)

The electrocardiogram (ECG) constitutes a popular non-invasive measure to detect cardiovascular conditions by measuring the electrical activity generated by the heart. ECG signals are one of the most used signals in stress detection research as they reflect activity of the heart directly, which is clearly affected by ANS changes [110]. It can be easily measured by placing some electrodes on specific places of the body and measuring the potential difference. The number of electrodes and their positions varies. The simplest and most effective way is to place three electrodes one on the right arm, one on the left arm and the last one on the left leg respectively.

The most typical and useful features computed with an ECG are those related to the Heart Rate Variability (HRV). Stress studies have used ECG signals successfully, such as in Cinaz et al. [111] who considered a threefold classification problem to separate office workers' mental workload into low, medium and high groups using only an ECG signal and nine HRV features (eight time domain features and the LF/HF ratio), achieving correct predictions for six out of seven subjects using Linear Discriminant Analysis (LDA). Similarly, Wijsman et al. [95] measured ECG in combination with skin conductance, respiration and EMG of the Trapezius muscles. Using five features from those signals including the heart rate they achieved an accuracy of 80% and 69.1% in the non-stress and stress detection respectively.

In a study completed by Palanisamy et al. [83], HRV, ECG, EMG, EDA and ST were measured, and a total of 148 features were extracted. The classification accuracy of each of the signals was analyzed individually. The results showed that the ECG and the HRV performed better in stress detection compared to the other signals. Precisely, a maximum classification rate of 93.75% was achieved with the HRV followed by the ECG with 76.25% and a minimum classification rate with EDA signals of 70.83%.

Melillo et al. [112] investigated the effect of stress on the HRV parameters under real-life conditions, unlike most of the state of the art works which have analyzed them in laboratory settings. They selected two critical moments to measure the ECGs of students, while they were taking an oral exam and again after their holidays. They used 13 non-linear HRV features and LDA to classify stressed and relaxed situations and achieved an accuracy of 90%, thus affirming the potential of these signals for real life stress detection. All these results suggest that ECG and HRV features allow distinguishing between different mental workloads and stress levels.

Furthermore, it has been proven that the ECG can be monitored continuously. Okada et al. [19] developed a continuous stress monitoring system for office workers based on ECG recordings aided by an accelerometer used for activity recognition and motion artifact removal. The RRI (R-R Interval), HRV spectrum and Tone and Entropy information were extracted from the ECG. After a three-day experiment, the availability of the system was validated and the feasibility of a continuous monitoring system approved. However, the developed system needed an offline analysis to be carried out by an expert. The main drawback of this method is that it is sensitive to other biological changes as well as cardiovascular diseases [72]. The robust assessment on mental stress alone can therefore be considered as challenging.

### 2.3.2.3 *Electrodermal Activity (EDA)*

EDA is one of the best real-time correlates of stress [20]. It is linearly related to arousal and it has been widely used in stress and emotion detection studies [20]. Some researches consider EDA measures as the basis for analysing the performance of other signals [75]. A highly relevant stress related research was carried out by Healey et al. [20], where a real-live driving task was analysed with hand and foot EDA, together with three other physiological measurements, namely ECG, Trapezius muscle EMG and respiration. Three levels of stress were induced on the test subjects by making them drive through a highway (medium stress level), through a city (high stress level) and have rest periods (no stress). A total of 22 features were extracted, and a LDA classifier was used achieving a recognition rate of 95%, 94.7% and 97.4% for low, medium and high stress levels respectively. From the viewpoint of a continuous monitoring of stress levels, the EDA and HRV were found to serve as the best correlates of real-time stress. De Santos Sierra et al. [113] created individual stress templates for 80 individuals using EDA and HR signals and a fuzzy logic algorithm. An accuracy of 93.5% was achieved for a two-way classification problem suggesting that both signals indeed possessed the potential for detecting stress levels precisely and in real-time. Other researches are not consistent with these results. Seoane et al. [114] concluded that cardiac and respiratory activity was a better stress indicator than EDA, ST and speech, and Palanisamy et al. [83] affirmed that EDA offered a lower classification accuracy than ECG, HRV, EMG and ST signals. Together with the heart rate, this type of measurement is affected by humidity and skin disease.

### 2.3.2.4 Blood Pressure (BP)

Blood pressure is defined as the pressure of the blood against the inner walls of the blood vessels that can be measured using a stethoscope and a sphyngomanometer [115]. It is proven that stress increases the blood pressure depending on the experienced stress levels. Nevertheless, Hjortskov et al. [73] state that blood pressure is not as good an indicator as HRV to detect stress situations. They discovered an increased blood pressure during the whole experiment, both in stress situations and control situations, and observed no differences in terms of the decrease in control situations when there was no exposure to stressors. They stated that this could be explained by the fact that unlike HRV which is regulated by the "central command", blood pressure is regulated peripherally and is influenced by local conditions in working muscles which can mask the changes of mental workloads. Thus, BP may not be as suitable as other physiological measurements for detecting subtle stress responses in real-time.

### 2.3.2.5 Skin Temperature (ST)

Skin temperature (ST) at constant room temperature may vary for different reasons such as fever, malnutrition, physical exertion and physiological changes [74]. If the other variables are controlled, the effect of any physiological changes can be appreciated. Physiological variations in the ST mainly come from localized changes in blood flow caused by the vascular resistance or arterial blood pressure that are in turn influenced by ANS activity [116] suggesting that stress level changes ST. It can be easily measured by placing a temperature sensor in contact with the skin. ST has been measured in many stress and emotion detection researches [114]. However, not all of them agree on the effects that stress and emotions have on this parameter. Some of them affirm that the finger temperature rises with stress. In the experiment carried out by Palanisamy et al. [83], the ST measured under the armpit increased in most of the subjects. Notwithstanding, other studies suggested that the finger temperature decreases under stress. Skin temperature of facial features, such as the nose and forehead can be an effective indicator in objectively evaluating human sensations such as stress and fatigue. Nakayama et al. [35] found out in a research carried out with monkeys that a decrease in nasal ST is suffered when negative emotions arise. However, recent researches on facial thermal imaging suggest that the facial temperature in some parts rises when a person feels stressed [117]. Some other researchers simply state that the ST does not provide much information on the subjects' emotions [74]. This may be true when considering universal temperature patterns, yet the reason for these disagreements may probably lie in the fact that the temperature response hardly depends on each individual. Further research is needed to clarify this ambiguity.

### 2.3.2.6 Electromyogram (EMG)

An Electromyogram (EMG) measures the electrical activity of the muscles by placing electrodes over the muscle of interest. As it is known that stress elevates muscle tone, many researches have been done to analyse the potential of the EMG for measuring stress. Stress has been found to provoke involuntary reactions on facial and Trapezius muscles [83]. In the study carried out by Wijsman et al. [118], it was verified that a significant increase in Trapezius muscle activity is suffered during mental stress. This increase in Trapezius muscle activity is translated into the increase of the EMG amplitude and a decrease in the amount of gaps, i.e. short periods of relaxation. They also observed that low frequency contents increase significantly under stress situations, thus demonstrating that the EMG signals provide useful information in detecting mental stress.

Wei [89] affirmed that the EMG is more effective than respiration signals for detecting stress levels. In this study, the EMG and respiration measurements which resulted in the extraction of a total of 37 features from these signals. The LDA classifier was used for a two-way classification problem, and the results showed that the EMG signals provided a more relevant information than respiration signals achieving 92.8% of accuracy discriminating relax and stress states with the EMG and 86.7% with respiration. As in regard to most of the physiological measurements, obtaining an EMG can be obtrusive for certain situations. In order to make them more practical and realistic, Taelman et al. [119] developed a biofeedback EMG recording shirt for daily use. However, this type of measurement can only be only used for evaluating physical stress.

### 2.3.2.7 Respiration

In 1973 researchers from the department of psychology of the Peking University discovered that when the stress level changes, the speed and depth of respiration system also change [89]. Due to this finding, respiration has been measured in many stress related researches [20, 95, 120] together with other physiological signals. Respiration can be measured with a pneumotachometer (or pneumotachograph). Nevertheless, a device of this nature may be very intrusive and consequently the possibility of estimating respiration rates from an ECG signal has already been analysed [121] with satisfying results. Unfortunately, the literature suggests that respiration monitoring is not as useful as other physiological signals. Healey et al. [50] found out that the contribution of respiration signals to stress detection was far from being as evident as the EDA or HRV contribution. Wei [89] also qualified respiration signals as less effective in stress classification than the EMG signals.

*2.3.2.8  Blood Volume Pulse (BVP)*

Blood Volume Pulse is the measure of the volume of blood that passes over a photoplethysmographic (PPG) sensor with each pulse. Photoplethysmography consists of measuring the blood volume of the skin capillary beds in the finger, relying on the capability of blood for absorbing light. The BVP has not been used as frequently as other signals in stress detection researches. Zhai and Barreto [122] measured it together with other three physiological signals, and a competent prediction method was developed. The biggest contribution of this signal in the literature is probably that it allows to measure information on the HRV non-intrusively. Chigira et al. [123] took advantage of this property and described a photoplethysmographic mouse to measure heart activity in office workers in a completely transparent way. More specifically, the blood volume of the fingers is measured using a near-IR light and a photo-detector by which enough IBI (Inter Beat Interval) precision is achieved to compute HRV features.

*2.3.2.9 Pupil Diameter (PD), eye gaze and blinking*

Pupil Diameter, eye gaze and blink rates can be measured with infrared eye tracking systems or with image processing techniques applied to visual spectrum images of the eyes. Pupil dilations and constrictions are governed by the ANS. Thus, the PD exhibits changes under stress situations [124], and the existing literature suggests that it can positively contribute to the problem posed herein. Liao et al. [125] affirmed that pupils are dilated more often under stress situations. Similar study by Barreto [65] used PD as stress inferring information, together with BVP, EDA and ST. Their results indicated the validity of the chosen signals and features for a two-way classification problem, i.e. to distinguish between stressed and not stressed people, as an accuracy of 90.10% was achieved. Barreto et al.[126] also carried out significant researches related to the PD activity under stress stimuli. They verified that the PD measured before and after the stress stimulus show different statistics and that the mean of the PD signal is significantly more relevant than the mean ST, the mean BVP and the mean of BVP period for the identification of affective states [127]. Sakamoto et al. [128] measured PD variability in the same frequency bands as HRV. They concluded that the LF/HF ratio of the PD variability can effectively replace the LF/HF ratio of HRV, thus validating its use in stress recognition.

Recently, a work of Ren et al.[129] affirmed the high ability of the PD features to discriminate between stress and relaxed situations. In fact, their results showed that the PD outperforms the EDA features. In their study, 42 individuals were subjected to an experiment where stress was induced by a Stroop test while their EDA and PD signals were being measured. A self-assessment test with two questions was answered by all the subjects in order to verify the stress eliciting method, and only those subjects who reported a higher stress response than a certain threshold were selected. T-test based labels were also computed to select relevant sections of the data. A total of three features were extracted from each one of the signals. Five different classifiers were used to create the stress models in which four out of five gave the best results using the questionnaire-based labels and the PD features. Only one out of five used the combination of both the EDA and PD features, and the worst in all the cases was achieved by using only the EDA based features. When t-test-based labels were used instead of the questionnaires, similar results were achieved. The PD features outperformed all the others in three out of five cases. In the other two cases, the combination of both yielded the highest accuracy while there was no case in where the EDA outperformed the PD. The highest accuracy of 88.71% was achieved with the Naive Bayes algorithm using only PD features and self-reported stress levels as ground truth.

### 2.3.2.10  Thermal Imaging (TI)

Several existing studies state that stress can be measured from thermal images due to the temperature changes suffered from stressed individuals [35, 84, 130]. Facial temperature can be easily measured using an Infrared camera, which is a completely unobtrusive method, thus making it interesting for office-place applications. In the past few years, this technique has been included in the set of stress measuring methods. In 2009, Levine et al. [35] have used the TI to analyze the activation of the corrugator muscle placed on the supraorbital area which may indicate mental stress. They concluded that the progressive and sustained corrugator muscle warming was experienced by all the subjects under stress conditions. They also affirmed the possibility of detecting subtle changes using this method due to the lack of adipose tissue above the corrugator muscle, thus minimizing the thermal inertia needed for provoking changes on the surface. Norzali et al. [117] confirmed the previous information verifying that supraorbital temperature changes under stress situations. They further found out that the blood flow under stress situations also increased in periorbital and maxillary areas. A stress detection system using a combination of both thermal and visual spectrum (VS) facial data has also been tested by Sharma et al. [84]. Facial expressions were analyzed in visual images while temperature changes have been detected in thermal images. Spatio-temporal features were extracted from recordings where the subjects' faces were registered while watching stressed and not-stressed films. A classification accuracy of 85% was achieved using LBP-TOP (Local Binary Pattern – Three Orthogonal Planes) features for the VS and LBP-TOP and HDTP (Histogram Dynamic Thermal Patterns) features for the TI. The promising results obtained with the TI have led other researches to analyse the facial blood flow under stress situations with even more sophisticated methods. Recently, Chen et al. [131] developed a stress detection system based on hyperspectral imaging, improving the TI technique for situations where big temperature changes or changes in subjects' sweating due to reasons other than stress may arise. The aim of using this technique was to detect the tissue oxygen saturation on facial tissues as the increase in facial blood flow under stressful situations suggests that oxygen saturation may also vary. Results prompt the use of this method for stress detection as increased StO2 levels have been detected around the eye sockets and forehead areas, but further research is needed for verifying its viability in real-time and real-life situations.

### 2.3.3 Behavioural evaluation

Behaviour regards expectations of how a person or a group of people behave in a given situation based on established protocols, rules of conduct or accepted social practices [132]. Stress affects individual behaviour. Some of the induced changes are well-known, for example being much more irritated or angry, yet these are not easily measurable. Other possible behavioural changes have been investigated, for example by analyzing people's interactions with technological devices in order to verify their relationship with stress and to create a reliable way to measure it. The advantage of measuring behavioural responses is that unlike physiological measurements, they can normally be executed in a totally unobtrusive way and in some cases, without the need of expensive extra equipment.

### 2.3.3.1 Keystroke and mouse dynamics

Under keystroke dynamics is understood the study of the unique characteristics that are present in an individual's typing rhythm when using a keyboard or keypad [133]. In the same way, mouse dynamics are affected by the subject's characteristics when moving it or clicking on its buttons. Keystroke and mouse dynamics have been widely analysed in the security area for authentication of people and for emotion recognition as summarized by Kolakowska et al. in their recently published review [134]. Stress detection has also been the objective of some studies based on keystroke and mouse dynamics. In 2003, Zimmermann et al. [135] first mentioned the possibility of using mouse and keyboard dynamics information to measure the affective state of the user. Thenceforth, many other researchers have tried to implement a method based on different features extracted from these devices. One of the biggest advantages of using a keyboard and a mouse for this purpose is that the developed technique is not intrusive and there is no need of any special hardware. Vizer et al. [76] highlighted other advantages like allowing to monitor information continuously leading to the possibility of an early detection and permitting to easily extract baseline data. Moreover, the article states that these kinds of systems can be introduced in the users' everyday life without the need to change their habits.

Individual writing patterns are considered to be stable enough for security applications, yet small variations attributed to stress and other situational factors on these patterns have been detected. Some researchers, as for example Hernandez et al. [75] affirm that relevant information about the affective and cognitive state of the user can be provided by keyboard dynamics. In their study, a pressure sensitive keyboard and a capacitive sensing mouse was used in order to detect stress levels in users. Self-reports and physiological signals were used as reliable assessment techniques. The results showed that 79% of the participants increased significantly the typing pressure and that 75% had more contact with the mouse under stress. This may be inconsistent with other studies [125] that affirmed that the mouse button is clicked harder when stress is decreasing. Surprisingly, no significant differences were found in terms of the amount of characters introduced, task duration or typing speed between the stressed and relaxed conditions.

Other authors disagree that keyboard dynamics provide relevant affective and cognitive information. Alhothali [136] stated that typing speed, key latency and key duration are only weakly correlated to emotional changes. Nevertheless, others [137] have already gone a step further accepting the reliability of behavioural biometrics based on keyboard and mouse usage patterns to assess stress levels by using it to extract personality traits.

### 2.3.3.2 Posture

It has been proven that posture is a good indicator about the feelings of the worker towards the tasks they are carrying out [138]. Thus, individual postural behaviour may also provide important information about stress levels. Anrich et al.[139] have tried to verify this hypothesis by analyzing the changes in the posture of office workers using a pressure distribution measuring system installed in their chairs. The spectra of the norm of the pressure center (CoP) were used as postural feature. A total of 28 men were stressed using the MIST [92], and it has been verified that the amount of fast movement increases during stress tests compared to control tests, and that the spectra of the CoP obtained in the two tests also differed. Using spectral information, 73.75% of accuracy was achieved when separating stress situations from cognitive load, suggesting that postural behaviour contains information related to stress levels. Others [71] have analyzed posture using visual techniques. Specifically, a Kinect has been used for detecting the interest levels of the office workers. Using techniques such as depth information and skeletal tracking, the inclination of the person and consequently an indicator of the workers' motivation was deduced.

### 2.3.3.3  Facial expressions

Facial features can provide insight to feelings and mental states for individuals including stress. When conversing with an individual, a person can get feedback from facial features which they can act accordingly, e.g. the person might cut a long story short when they observe and realize that the individual is showing signs of frustration, agitation or preoccupation by fewer nods, reduced facial muscle movements or frequent eye movements to other objects in the surroundings

Stress classification models have been developed from facial feature data and results show that facial expressions can be used to measure stress [133]. When responding to stressors, facial expressions indicate biological responses reliably that are commonly used to assess stress. Online analysis of facial expressions can be used to predict behaviour and events (e.g. car accidents) in real-time [140]. Facial muscle movements have been used to determine stress. Increased head and mouth movements indicate increase in stress [113]. In a recent work [141], unlike in most stress detection studies, a mathematical stress model was created rather than considering the main objective as a classification problem. Instantaneous facial expressions were analyzed from images by creating an emotional percentage mixture model and relating it to stress levels. Moreover, the seven basic emotions, without any image for reference, were also related to stress. The equation for evaluating stress quantitatively from facial expressions was estimated, yet no further information about its performance was provided.

### 2.3.3.4 Speech analysis

Most researchers agreed to the fact that stress changes human vocal production [142]. More precisely, under stress situations, changes in pitch and in the speaking rate are usual, together with variations in features related to the energy and spectral characteristics of the glottal pulse [143]. Speech analysis has primarily caused interest as it can be easily measured in a completely unobtrusive way. Nevertheless, voice-based stress analysis can be ineffective in both quiet and noisy spaces [144], due to the lack of speech recordings and the excessive noise. Most of the research conducted in voice stress recognition has been carried out in laboratories or in quiet environments. However, there are exceptions, such as the research carried out by Lu et al. [143] where stress detection in indoor and outdoor acoustic environments was executed by using mobile phones. An accuracy of 82.9% indoors and of 77.9% outdoors was achieved. Demenko et al. [145] analyzed call-centre recordings, including stress and no-stress speech which achieved 84% accuracy distinguishing between the two classes, using LDA classifier and nine features extracted from amplitude and pitch information. In the laboratory experiment carried out by Kurniawan et al. [146], speech and EDA were measured and used to create a universal stress model and an inter-individual stress model, both with independent information given by each signal and combining the two of them. Three two-way classification problems were considered, namely recovery vs. workload, recovery vs. heavy workload and light workload vs. heavy workload. The generated results showed that the selected speech features (Mel Frequency Cepstral Coefficients and pitch) were more efficient in stress detection than the selected EDA features. Furthermore, the combination of both types of signals did not show any improvement, and the inter-individual model outperformed the universal model. Thus, the best result for the most difficult case was achieved using the inter-individual model with the SVM classifier and speech features which measured 92.6% accuracy. This result suggests that stress detection can be done by means of speech features whereby the subject is placed in an environment with good acoustic conditions.

## 2.4 Measuring brain activities

Activity in the brain is defined as the activity of neurons in a localized area as the result of specific cognitive function. Brain activity results from neurons firing (synapsing) and via the coupling with the vascular system (blood vessels) that increases metabolic demand, cerebral blood flow and consequential oxygen consumption [147]. The fluctuations of oxygen in the blood stream (from accompanying oxygenated hemoglobin ($O_2$Hb) and deoxygenated hemoglobin (HHb) changes in the red blood cells) as a result of the oxygen demand can be measured with optical techniques and the corresponding changes in intensity. Research tools can provide a relationship between the activity of specific regions of the brain and cognitive functions. The development of non-invasive research and clinical tools are accelerating the fields of brain and cognitive study.

Brain activity as a result of cognitive function can be measured non-invasively using advanced neuroimaging techniques either indirectly or directly. The techniques are briefly compared by Aslin [148]. Neuronal activity can be measured directly by electrical and magnetic field changes. Well-researched direct methods electroencephography (EEG) and magnetoencephalography (MEG) rely on the electrical and magnetic field respectively emanating directly from neurons in the brain. These techniques are characterized by high temporal resolution which is advantageous for cognitive processing. However, low spatial resolution makes it difficult to determine the source of activation and the corresponding activity pathway.

Brain activity can also be measured indirectly by changes in blood volume and oxygenation usually referred to as hemodynamics. Increased neural activity requires increased metabolic demand at the cellular level. Increased metabolic demand in turn requires oxygen delivered to the cell by the hemoglobin molecule present in the red blood cells. There exists a correlation between the dynamics of hemoglobin changes in the surrounding blood vessels and the neural activity detected in the localized area as researched and established by Buxton et al. [149].

Indirect methods of monitoring brain activity include positron emission tomography (PET) [150] and functional magnetic resonance imaging (fMRI) [151] which are both non-optical techniques, and functional near-infrared spectroscopy (fNIRS) [152] an optical technique. These techniques depend on these hemodynamic changes in the brain and are regarded as indirect since the key point the brain activation depends upon changes the blood volume and oxygen metabolism.

The PET technique is regarded as safe and does require the injection of a radioactive isotope. It relies on the detection of photons emitted as a result of the decay process. Typically, PET is used in clinical populations, for example in tumor detection, where the benefits outweigh the risks of the procedure. Characterized by higher spatial resolution than other techniques, PET measures the total blood volume rather than the blood oxygenation changes.

The fMRI technique relies on the detection of magnetic fluctuations in a resonant magnetic field. As deoxyhemoglobin is a paramagnetic molecule, neural activity can be detected as a result of its magnetic fluctuations. The technique is characterized by low temporal resolution and high spatial resolution (less than magnitudes of a millimeter). The techniques of fMRI, MEG and PET are expensive not only in terms of their manufacture cost but also the cost of the physical infrastructure to house the equipment. Additionally, a disadvantage especially when working with infants is that the techniques do require the subjects to be still and are restrictive not only with head movement but require the subjects to be lying down.

### 2.4.1 Functional magnetic resonant imaging (fMRI)

Functional magnetic resonant imaging or fMRI constitutes a technique for measuring brain activity. When the neural activity of a brain area increases, it consumes more oxygen. In order to meet this increased demand, the blood-flow increases to the active area. The fMRI detects these changes in blood oxygenation and flow. Thus, it can be used to produce activation maps showing which parts of the brain are involved in a particular mental process. Hayashi et al. [99] used fMRI technology in order to verify whether stress responses were evident in some brain regions, including the ones related to emotional and cognitive processing by stimulating stressed and not stressed people with audio-visual contents. The results did not show differences in the brain regions related to emotional processing, yet did show less activity in cognitive processing brain regions on stressed people. Also observed was that the superior and inferior parietal gyrus was significantly more activated by pleasant and unpleasant stimuli in people not suffering from stress than others, thus suggesting that attention deficits may take place even in the early stages of stress. Other studies described significantly reduced brain activations over the prefrontal cortex (PFC) [29, 32]. The PFC has been identified as the most sensitive to the detrimental effects of stress exposure and displayed behavioral and somatic responses to stress. The downside is that unlike the EEG, the fMRI does not offer good temporal resolution. Furthermore, this method is restrictive by nature and it does not allow monitoring in the workplace [35].

## 2.4.2 Electroencephalogram (EEG)

The Electroencephalogram (EEG) is considered as the most common source of information to study brain functions and conditions. It is a very complex signal and can be recorded non-invasively using surface electrodes attached to the scalp. It measures the electrical activity of the brain by connecting an array of electrodes onto the subject's scalp so that the electrical fluctuations are simultaneously recorded [37]. The placement of electrodes is controlled by the international 10-20 system as shown in Fig 2.4. The number of electrodes depends on the application and experiment. The EEG is the most studied non-invasive brain imaging device due to its fine temporal resolution, ease of use, and low set-up cost. Additionally, the EEG benefits from high temporal resolution that enables it to measure changes in the subject's cognitive activity within millisecond scale.

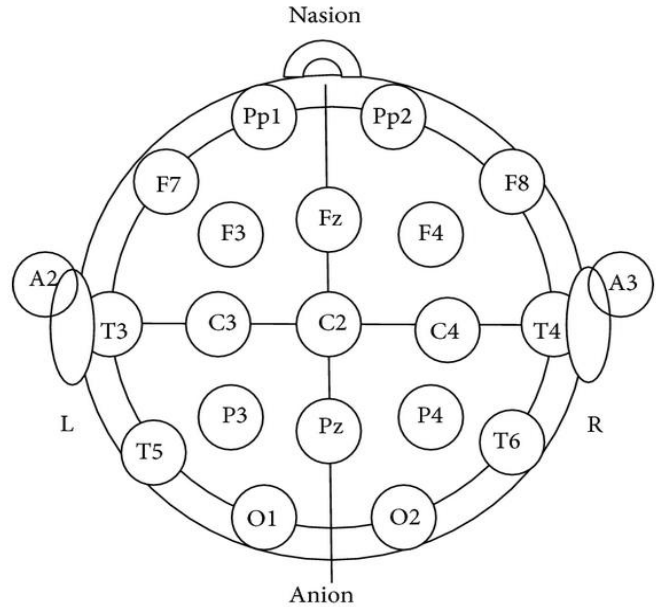

Figure 2.4: EEG electrode placement according to 10-20 international system [36].

EEG signals are usually studied by characterizing them into frequency bands. Alpha (8–13 Hz), Beta (13–30 Hz), Delta (0.1–4 Hz) and Theta (4–8 Hz) are the most common frequency bands used in previous studies. Each frequency band can be used to describe the mental state of a person. Alpha and beta frequency power is linked to negative mood, stress and depression [38-42]. Stress has also been related to changes in right frontal activity provoking frontal asymmetry. Whether stress can be reliably evaluated from the EEG remains unclear, yet there are some researches that suggest its validity. Few studies have used it to study mental stress. The brain region under study depends on the type of stimuli or tasks (visual, working memory or audio). Hill and Castro found high beta rhythm activity in the sensory motor area during stressful healing tasks [43]. Thompson and Alonso separately found an increase of beta waves associated with a decrease of alpha waves in the anterior cingulate and frontal anterior cortex [46, 47]. Another set of studies identified a positive correlation between the EEG beta power rhythms with stress in the temporal lobe to odor irritation and traffic noise [45, 153, 154].

Seo et al. [44] analyzed the relationship between the EEG, the ECG and cortisol levels when measuring chronic stress by using the stress response inventory and the self-assessment manikin tests as ground truth. A significant positive correlation was found between the cortisol levels and the high beta activity at the anterior temporal sites when people kept their eyes closed which affirmed the aforementioned relationship between the beta band and stress. Furthermore, the mean high beta power at the anterior temporal sites of the stress group was found to be significantly higher than that of the non-stress group. Rahnuma et al. [155] recorded the EEG data from frontal, central and parietal lobes of the scalp in order to create an emotion recognition system based on Russel's model of affect [156]. This model describes all the emotions based on arousal and valance by mapping stress to a negative valence and a very negative arousal coordinate. An accuracy of 91.4% was achieved which suggested that emotions and stress detection is possible by combining valance and arousal information obtained from the EEG.

Zhang et al. [157] monitored the EDA, HRV and EEG signals of 16 subjects under cognitive load and in relaxed states. Subjective stress ratings based on the State-Trait Anxiety Inventory (STAI) were used as labels. Discriminative spatial-spectral EEG components were computed by means of a spatial-spectral filtering in the theta, alpha, low beta, mid beta and high beta frequency bands. The mean EDA and 10 time-domain features of the HRV were also computed. A large margin unbiased regression approach was developed in order to overcome the inter-subject variability that showed these signals. The results suggested that the EEG features extracted using the proposed filtering technique outperformed both the EDA and the HRV features in discriminating both situations, with 87.5%, 75% and 62.5% accuracy respectively. Alpha waves reflect a calm, open and balanced psychological state. A significant decrease in alpha activity in response to stress situations over the anterior sites was found at [47]. Separately, Marshall and Lopez-Duran reported a negative correlation between the EEG alpha power rhythm and stressful events in the prefrontal cortex where the alpha rhythm is reduced with stress [51, 52]. On the other hand, theta power has been linked to task difficulty (theta rhythm decreases with increasing task difficulty) [49]. Gärtner [48] found that the frontal theta decreased with stressful mental arithmetic tasks. Similarly, Harmony et al. reported high delta waves, while solving difficult mental arithmetic tasks [50]. In order to detect mental stress, pattern recognition approaches are often adopted [53]. Table 1 summarizes some of the most commonly used expert systems in classifying the EEG signals individually or in combination with other physiological signals in stress related studies. However, low spatial resolution makes it difficult to determine the source of activation and the corresponding activity pathway.  To overcome, combining EEG with another neuroimaging technique that has the complementary nature is required.

### 2.4.2.1 EEG features

Several things are particularly important to consider when selecting a feature from EEG signals to be used for classification. First, the EEG signals contain information that are both spatially as well as temporally in nature. A feature that neglects patterns that occur either across electrodes or through time may discard important patterns that are present in the signal. The EEG features can be taken from the time-domain or frequency domain. From the time-domain, mean amplitude, mean amplitudes of Event Related Potential (ERP) components, variance, skewness, kurtosis and peak values are usually calculated. Since the EEG signals often contain patterns that are oscillatory in nature, those features that utilize frequency spectra are commonly used as the case in BCI systems.

Table 2.1: Previous studies related to EEG arousal and physiological signals classification.

| Author /Year | Physiological signals Used | Stressor | Number of Subjects | Expert system employed | Classification Accuracy | Remarks |
|---|---|---|---|---|---|---|
| Ishino (2003) [158] | EEG | Video and puzzle games | 1 | NN | 54.5%, 67.7%, 59% and 62.9% for happy, calm, sad and relax. | The occurrence of stress was not validated. The assessment was based on physiological signals which could be due to workload or negative emotion. |
| Ryu (2005) [159] | EEG and ECG | Arithmetic task | 10 | Multiple regression analysis | N/A only to study brain response. | The study used simple arithmetic task to induce stress but the task was simple and may not induce stress. It is believable that, the author induce workload not stress. |
| Chanel (2006) [39] | EEG, ST, BP and respiration | Video and image | 4 | Bayes and FDA | 55% for low and high arousal. | The sample size used in this study is not significant and there is no strong correlation between blood pressure and EEG fluctuations. |
| Chanel (2007) [160] | ST, BP, respiration and EEG | Recall event | 1 | LDA, SVM | 76% and 73% using EEG and peripheral signals. | The sample size used in this study is not significant and method of stress was not validated. |
| Lin (2008) [161] | EEG | Driving Simulator | 6 | K-NN and NBC | 71% to 77% between stress and rest. | The method used to induce stress was not validated. There was a distraction during the driving simulation and the sample size used in this study is not significant. |
| Chanel (2009) [56] | EEG, ST, BP, HRV and respiration | Recall memory | 11 | LDA, SVM, and RVM | 63% using time-frequency EEG and 70% using fusion features for negatively excited, positively excited, and calm-neutral states. | The recall process was not a significant method to induce stress. The author used a combination of physiological signals but limited accuracy was achieved. |

| Hosseini (2010) [162] | ST, HRV and EEG | Pictures induction calm-neutral and negative-excited | 15 | SVM and Elman network | 84.1% for two categories, calm and stress using psychological signals and 82.7% using EEG signals. | This type of study did not validate the procedures of inducing stress. This is an emotional study and the classification was about mood state not a stress study. |
|---|---|---|---|---|---|---|
| Saidatul (2011) [163] | EEG | Mental arithmetic task | 5 | NN | 91.17% using Burg Method, 88.36% using Welch Method and 85.55% using Yule Walker for stress and relax. | Simple arithmetic task was used and limited sample size. There was no validation of stress inducement procedures. The author induced workload not stress. |
| Rahnuma (2011) [155] | EEG | Negative videos and images | 4 | MLP | 71.69 %, 60.74%, 71.84% and 65.94% for happy, calm, sad and relax. | The study used small sample size and very complex features which results in poor flexibility and computation time. |
| Khosrowabadi (2011) [164] | EEG | Before and after examination | 26 | K-NN, SVM | 90% for stress and relax. | The study used many type of features which make the system complex and the method of inducing stress was not validated. |
| Sharma (2013) [165] | EEG, ECG, ST, BP, eye gaze and pupil diameter signals | Video: stress and non-stressed film | 25 | GA+SVM, GA+ANN | 95% using all physiological signals and 91% using EEG signals alone. | The author built a very complicated system to detect stress and was not able to eliminate the noise. Additionally, the method used to induce stress was not validated. |

EEG Electroencephalography, ECG *electrocardiogram,* ST *skin temperature,* BP *blood pressure,* HRV *hear rate variability,* NN *neural network,* LDA *linear discriminate analysis,* RVM *relevance vector machine,* K-NN *k-Nearest Neighbor,* MLP *multilayer perceptron,* NBC *naive Bayes classification,* GA *genetic algorithm.* FDA *Fischer's linear discriminant analysis*

*2.4.2.2 Frequency domain*

Fast Fourier Transform (FFT) is considered as a common method in representing the EEG features. This method employs mathematical means for EEG data analysis. Characteristics of the acquired EEG signal to be analyzed are computed by power spectral density (PSD) estimation in order to selectively represent the EEG sample signal. However, four frequency bands contain the major characteristic waveforms of the EEG spectrum [166]. The PSD is calculated by Fourier transforming the estimated autocorrelation sequence found through nonparametric methods. One of these methods is Welch's method whereby the data sequence is applied to data windowing producing modified periodograms. The information sequence $x_i(n)$ is expressed as:

$$x_i(n) = x(n+iD), n = 0, 1, 2, ..., M-1, i = 0, 1, 2, ..., L-1; \qquad (2.1)$$

Taking *iD* to be the point of start of the *i*th sequence, L is the length of the sequence and *2M* represents the data segments which are formed. The resulting output periodograms give

$$P_{xx}(f) = \frac{1}{MU}\left[\sum_{n=0}^{M-1} x_i(n)w(n)e^{-j2\pi fn}\right]^2 \qquad (2.2)$$

Here in the window function, U gives the normalization factor of the power and is chosen such that

$$U = \frac{1}{M}\sum_{n=0}^{M-1} w^2(n) \qquad (2.3)$$

where *w(n)* is the window function. The average of these modified periodograms gives Welch's power spectrum as follows:

$$P_{xx}^W = \frac{1}{L}\sum_{n=0}^{L-1} P_{xx}(f) \qquad (2.4)$$

*2.4.2.3 Wavelet Transform (WT) Method*

The wavelet transform method plays an important role in the recognition and diagnostic field as it compresses the time-varying biomedical signal that is comprised of many data points, into a small few parameters representing the signal [167]. As the EEG signal is non-stationary, the most suitable way for feature extraction from the raw data is the use of the time-frequency domain methods like wavelet transform (WT), a spectral estimation technique in which any general function can be expressed as an infinite series of wavelets [168]. Since the WT allows the use of variable sized windows, it gives a more flexible way of time-frequency representation of a signal. In order to get a finer low-frequency resolution, the WT long time windows are used. In contrast, in order to obtain high-frequency information, short time windows are used. Furthermore, the WT only involves multi-scale and no single-scale structures. In the WT method, the original EEG signal is represented by secured and simple building blocks known as wavelets. The mother wavelet gives rise to these wavelets as part of derived functions through translation and dilation, that is by shifting and compression and stretching operations along the time axis [169]. There are two categories for the WT; the first one is continuous while the other one is discrete [167].

### 2.4.2.4 Continuous Wavelet Transform (CWT) Method

The continuous wavelet transform method can be expressed as follows:

$$CWT(a,b) = \int_{-\infty}^{+\infty} x(t)\psi_{a,b}^*(t)dt \qquad (2.5)$$

in which $x(t)$ stands for the unprocessed EEG, where $a$ stands for dilation and $b$ represents the translation factor. The $\psi_{a,b}^*(t)$ denotes the complex conjugate and can be calculated by

$$\psi_{a,b}(t) = \frac{1}{\sqrt{|a|}}\psi\left(\frac{t-b}{a}\right) \qquad (2.6)$$

where $\psi(t)$ means wavelet. However, its major weakness is that the scaling parameter $a$ and the translation parameter $b$ of CWT changes continuously. Thus, the coefficients of the wavelet for all available scales after calculation consumes a lot of effort and yields plenty of unused information [167].

*2.4.2.5 Discrete Wavelet Transform (DWT)*

In order to address the weakness of the CWT, discrete wavelet transform (DWT) has been defined on the base of multi-scale feature representation. Every scale under consideration represents a unique thickness of the EEG signal [170]. An example of the multi-resolution decomposition of the raw EEG data *x(n)* is shown in Fig 2.5.

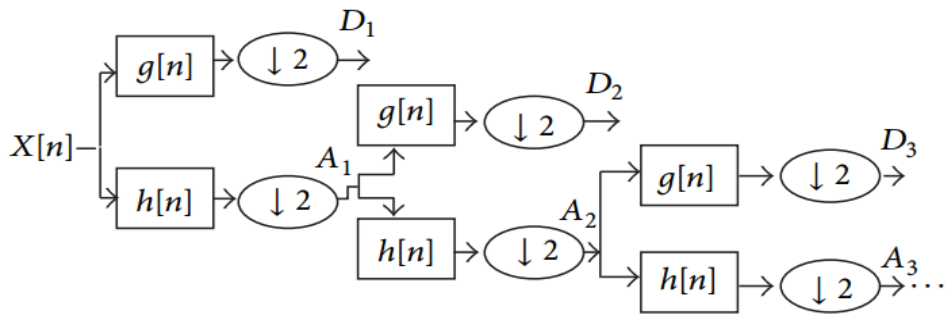

Figure 2.5: Decomposition of DWT.

Each step contains two digital filters, *g(n)* and *h(n)* and two down-sampled by 2. The discrete mother wavelet *g(n)* is a high pass in nature, while its mirror image *h*(n) is a low-pass in nature. As shown in Fig 2.5, each stage output provides a detail of the signal D and an approximation of the signal A, where the latest becomes an input for the next step. The number of levels to which the wavelet decomposes is chosen based on the component of the EEG data with dominant frequency [167].The relationship between WTs and filter *h*, that is, low pass, can be represented as follows:

$$H(z)H(z^{-1}) + H(-z)H(-z^{-1}) = 1 \qquad (2.7)$$

where, $H(z)$ represents filter's *h* z-transform. The high-pass filter's complementary z-transform is expressed as:

$$G(z) = zH(-z^{-1}) \qquad (2.8)$$

WT decomposed the EEG signals into a set of functions to obtain their approximation and the corresponding coefficients at different levels. Features can then be extracted from them. The wavelet family of Dubechies-8 (db8) was used in order to decompose the EEG signals into five frequency bands (delta, theta, alpha, beta and gamma). Table 2.2 gives a summary of the wavelet decomposition levels and their corresponding frequency bands.

Table 2.2: EEG frequency bands and wavelet decomposition levels.

| Decomposition level | Frequency bandwidth | Frequency band |
|---|---|---|
| DL1 | 64 Hz -128 Hz | Noisy signal |
| DL2 | 32 Hz -64 Hz | Noisy Gamma |
| DL3 | 16 Hz -32 Hz | Beta |
| DL4 | 8 Hz -16 Hz | Alpha |
| DL5 | 4 Hz -8 Hz | Theta |
| AL5 | 0-4 Hz | Delta |

### 2.4.3 Functional near infrared spectroscopy (fNIRS)

Functional near Infrared Spectroscopy (fNIRS) is a non-invasive brain imaging technology based on hemodynamic responses to cortical activation. The fNIRS measures blood flow through hemoglobin concentrations and tissue oxygenation in the brain [2]. The technique uses a light source to send near-infrared light (in the wavelength range 650-850 nm) into the head. By measuring the light sent at two wavelengths, the oxygenated and deoxygenated hemoglobin concentrations can be calculated using modified Beer-Lambert law [171]. Since 1990s, the fNIRS has been developed as a tool for neuroimaging studies focused on the assessment of cortical activities to cognitive and motor tasks [152]. The application of the fNIRS to functional neuroimaging is extensively studied by such groups as those headed by Britton Chance (University of Pennsylvania), David Boas (Massachusetts General Hospital), Theodore Huppert (University of Pittsburgh), Scott Bunce (Drexel University), and Enrique Gratton (University of Illinois at Urbana-Champaign). With further research, including the present study, the fNIRS may evolve into a synergistic and valuable complement to EEG and other psychophysiological measures for monitoring and predicting cognitive states.

The fNIRS interrogates the outer surface of the cortex only, and quantifies the changes in the oxygenated and deoxygenated hemoglobin concentration ($[O_2Hb]$ and $[HHb]$, respectively). It is portable, relatively low-cost, non-confining, non-invasive, and safe for long-term monitoring. The temporal resolution is sub-second, while spatial resolution is on the order of $1cm^2$ at best [172]. Measurements were shown to be consistent with the fMRI [69] and the EEG measurements [70]. Optical absorption increases as the hemoglobin concentrations increase, which in turn depends on the local neural activity [173]. This is analogous to the generation of BOLD signals in the fMRI via changes in deoxygenated hemoglobin. Both techniques quantify the neural activity indirectly via neurovascular coupling. The fNIRS and the fMRI sense the hemodynamic response [63], as opposed to the Event Related Optical Signal (EROS) and Electroencephalogram (EEG) measurements, which sense the activity of the neurons themselves (through cellular changes which affect optical scattering, or through electrical activity, respectively). A critical comparison between most common neuroimaging modalities is as shown in Fig 2.6.

Figure 2.7 illustrates how a near infrared imaging system works. When there is a stimulus (e.g. finger tapping) there is also neuronal activity in related areas of the brain (motor cortex) leading to changes in local cerebral blood flow and volume. This results in concentration changes of the blood-borne cells and molecules in the capillaries surrounding the neurons as well as changes of the blood oxygenation level (hemodynamic response), which can be detected by the fNIRS.

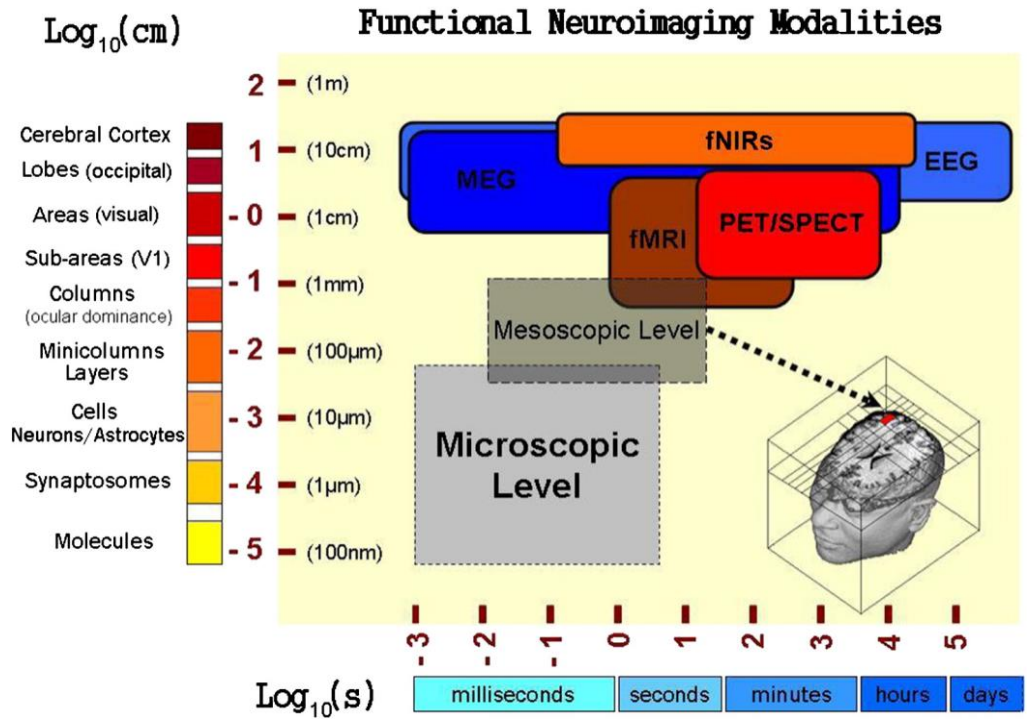

Figure 2.6: Spatial and temporal comparison of the functional neuroimaging modalities [63].

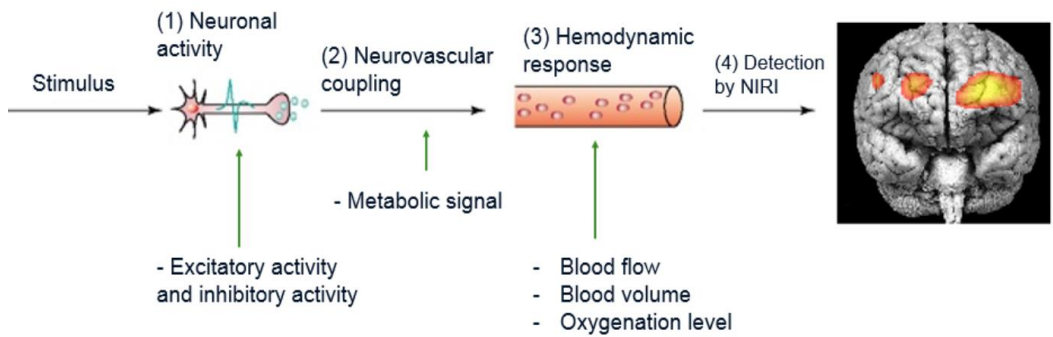

Figure 2.7: Neuronal activity leads to changes in the blood flow, volume, and oxygenation level [172].

### 2.4.3.1 Optical properties of brain tissue

As photons traverse through the tissue, they are either being absorbed, transmitted through or being scattered to a random direction (Fig. 2.8). If the photons are scattered, all of the three mentioned events can occur again. This happens due to the high scattering properties of brain tissue which makes it possible for emitted photons to be scattered back to the surface of the head and be detected by photodetectors [172]. The absorption coefficient and the scattering coefficient are two fundamental optical properties of brain tissue. The former provides information about the concentration of the composing chromophores (e.g., oxygenated hemoglobin and deoxygenated hemoglobin), the latter providing the form, size, and concentration of the scattering components.

The optimal wavelength range for optical imaging is 700 to 900 nm. In this range the light absorption is dominated by hemoglobin, and also the absorption by water is low enough that light can penetrate deep into the brain to be affected by the brain functional activity. For wavelengths smaller than 700 nm, light is absorbed by hemoglobin, while for wavelengths larger than 900 nm, light is absorbed by water so it cannot pass through tissue with an adequate penetration depth. The wavelength at which oxygenated and deoxygenated hemoglobin have the same absorption is called the isobestic point (808 nm). Below the isobestic point, the deoxygenated hemoglobin dominates light absorption, while above this point oxygenated hemoglobin dominates. More details on the optical wavelengths are as illustrated in Fig 2.9.

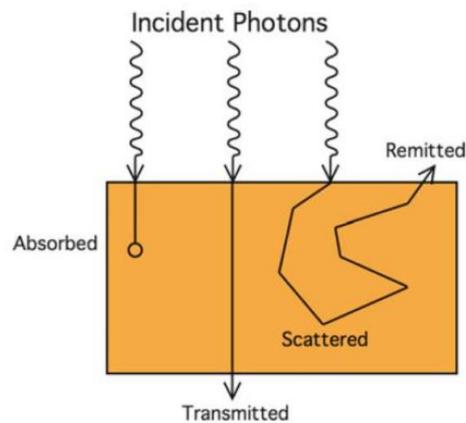

Figure 2.8: Photon interaction with tissue [172].

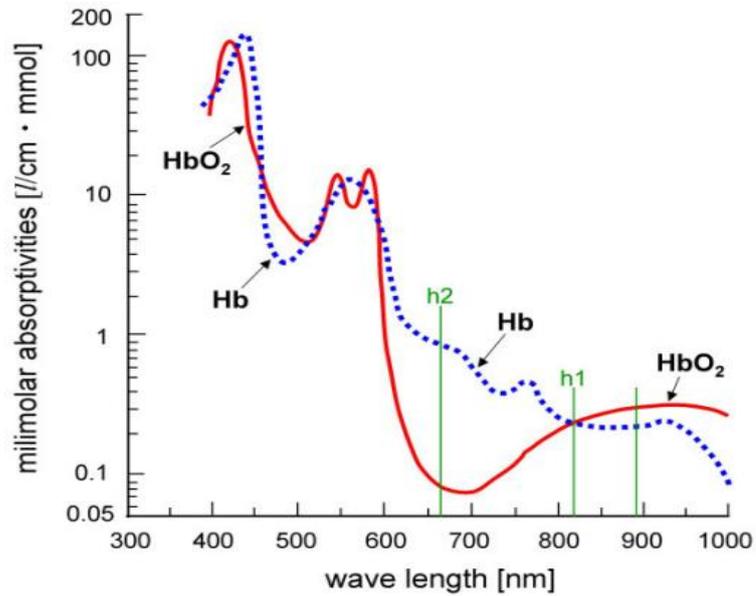

Figure 2.9: Optical wavelength range [172].

In near-infrared spectroscopy, the light source and detector are both placed on the surface of head. They have limited penetration depth and receive information only from the superficial structures of the brain. The penetration depth can be controlled to some extent by changing the distance between the source and the detector such that for greater source-detector distances the penetration of photon is more. However, as the distance increases, the intensity of the detected light decreases rapidly. Thus, there is a tradeoff between the penetration depth and intensity. In order to make functional activity measurements from the surface of the head with continuous wave instruments, optical delivery and detection is required at 3 to 4 cm apart. Thus, a probe is made up of a source or emitter and a detector, which straddles its measurement location on the head surface. The measurement location is typically the point through which a line normal to the head passes on its way to the deep brain region of interest. A clear illustration about the path flow is shown in Fig 2.10.

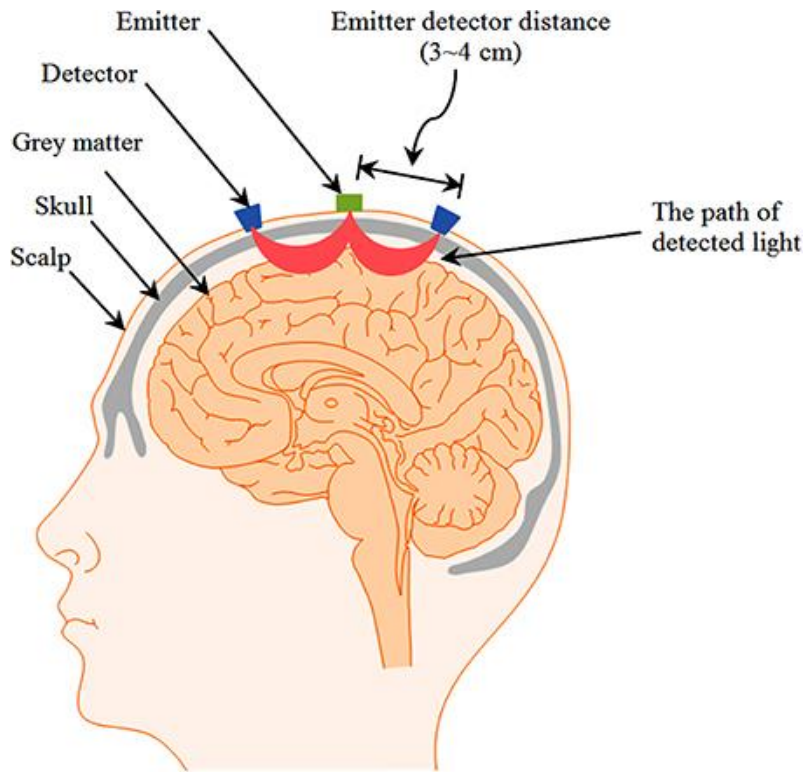

Figure 2.10: Path flow of infrared light intensity [174].

Measured optical intensity changes are due to the changes in the absorption of the light as it migrates from the source to the detector. This change in absorption contains information about the change in hemoglobin concentration ([Hb]) depending on the length of the optical path through the tissue undergoing hemodynamic changes [175]. Oxygenated and deoxygenated hemoglobin are the predominant absorbers in biological tissue in the red and near-infrared wavelength region used for the fNIRS.

Relative changes in the concentrations of both these species of hemoglobin over time are calculated using the Modified Beer Lambert Law (MBLL) from intensity measurements for each of two wavelengths per location. The modification refers to two changes to the well-known Beer Lambert Law, which describes the dependence of light attenuation on absorption, path length and chromophore concentration. One is the subtraction of a background scattering term, which is a significantly greater source of light attenuation than that due to absorption, yet assumed to be constant. The other is the inclusion of a differential path length factor (DPF) to account for the increase in path length due to the multiple scattering effects as the photons migrate through tissue.

Activations are quantified for a region of interest by comparing the hemoglobin concentration changes between species. A hemodynamic response is indicated by an increase in [$O_2Hb$] or a decrease in [HHb], over time periods on the order of 10-12 seconds (or longer if sustained) as regional blood flow increases to overcompensate for local metabolism [176]. Regional or time averages are taken of the activation quantity, and comparisons are made between the measurements obtained during different conditions such as task and rest. Care must be taken when directly comparing activation magnitudes as the sensitivity is not uniform across the probe locations including across participants and across wavelengths and depends on the make-up of the capillary bed probed by the photon migration path.

*2.4.3.2 Sensor Location and Sensitivity Challenges*

The sensitivity of the measurement technique relies on the degree to which the tissue sensed is affected by the hemodynamic changes in response to the experimental stimuli [175]. The systemic physiology which affects the hemodynamic changes within the vasculature sensed anywhere between the source and the detector confound the measured signals. In order to aid the understanding of the many factors affecting the sensitivity of the optical probes, this study introduce some important terms here. The distance the sourced photons travel overall through the tissue is termed the optical path length (depicted in Figure 2.11 in pink).

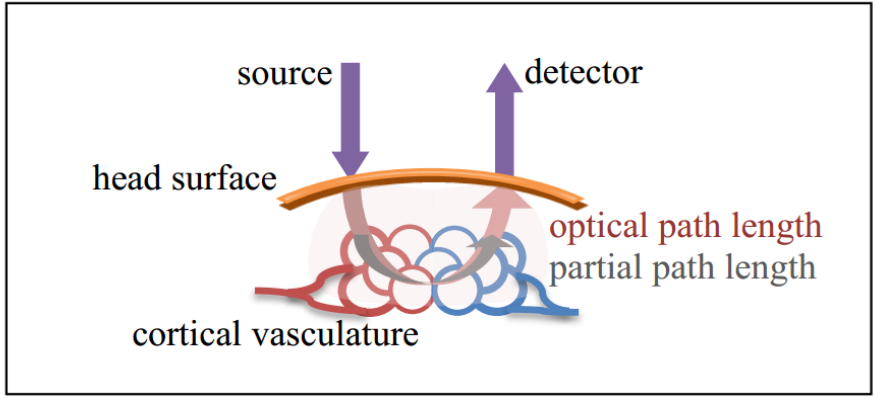

Figure 2.11: Schematic of the optical and partial path lengths relevant to the fNIRS [177]

This distance is greater than the chord distance between the source and the detector by the multiplicative differential path length factor (DPF). Brain activations do not cause changes in all of the tissue through which the light scatters but only a portion of it. The distance the light travels through the parts that do change with activation is termed as the partial path length (depicted in Figure 2.11 in gray). The light from each probe location encounters different superficial tissue, and the light of each wavelength takes a different path through that tissue, even if injected at the exact same location. The partial path length is not known. Therefore, because the Modified Beer Lambert Law calculation of hemoglobin concentration assumes that the absorption has occurred along the entire optical path length, the absorption by the hemoglobin and thus the calculated change in [Hb] is underestimated. In reality, a greater absorption has occurred over the shorter partial path length [173], thus contributing to the inherent error in the measurements.

The spatial extent of the sensitivity profile determines the spatial resolution. The attempt is made at each probe location to maximize the overlap of the sensed volume with the volume of tissue undergoing the hemodynamic activations of interest. This is typically done without detailed information on the underlying vasculature and anatomy. The reduction of the source-detector distance does not necessarily improve spatial resolution, as many factors affect the shape of the sensed volume relative to the location of the volume of interest. Errors are introduced due to the vascular and the superficial tissue non-uniformity, both across the head and between individuals.

With few probes, the extent, position and intensity of measured hemoglobin changes strongly depend on the probe arrangement [178].

However, if a high spatial density of regularly-spaced sources and detectors is used with image reconstruction methods, the measured activation can be a more accurate reproduction of the actual activation in both location and extent, especially for localized activations [178]. Thus, with a greater number of detectors, sensitivity and accuracy can be improved by using overlapping and neighboring probes placed symmetrically about the active region of interest. It is thus advantageous to develop probes with a reduced spatial footprint so that more sources and detectors can fit as unobtrusively as possible on the head.

### 2.4.3.3 Practical considerations for placing optodes

In order to best approximate this without the fMRI or the 3D tracking data, the optodes have been placed according to the International 10-20 system used for electroencephalography (as in this study) [179] rather than by trial and error which is not recommended. Maximizing the detected intensity is one way to avoid large vascular obstacles, yet this decreases the likelihood of accurately measuring the extent of the hemoglobin concentration changes ($\Delta[Hb]$) [175]. Maximum intensity at a reasonable detector gain setting can be caused by a lack of tissue absorption. When properly positioned such that the tissue volume sensed by the optode overlaps with hemodynamic absorbers, the optical intensity is reduced during activation. During activation, the optical intensity increases as [HHb] decreases while the offset by intensity decreases as [$O_2$Hb] increases.

In the other extreme, one may find a vessel that is too large and providing too much absorption and not enough signal intensity. What should be maximized instead is the modulation between the baseline (intensity higher, total [Hb] lower) and the stimulated activation (intensity lower, total [Hb] higher). The optical intensity at both these extremes should be within the dynamic range of the optical detectors. This is not readily determined without a controlled experiment or localizer task. Also, a large optical signal can be due to static or dynamic ambient light exposure if direct optode-to-skin contact is not maintained. Thus, the maintenance of baffling, detector sensitivity settings and sensor placement throughout the experimental trial are very important.

### 2.4.3.4 The importance of consistent and stable optode locations

Once selected, obtaining consistent signal at all locations across time and thus across stimulus conditions is important because the quantification of the neuronal activation with the fNIRS relies on the relative measurements. A consistent signal measurement is also important since the loss of signal at one optode can reduce the accuracy of the state classifier by removing an important predictor variable from the classification algorithm input. This is heavily influenced by the stability of the head probes. Many errors are introduced at the optical-tissue interface [180], and the signal quality is highly susceptible to mechanical instability. The existing methods for the fNIRS probe attachment in adults can be time-consuming and difficult to employ, are susceptible to motion artifact, and can be quite obtrusive and uncomfortable [181]. Indeed, many recent fNIRS studies have been restricted to frontal probe placement [182], thus avoiding absorptive and mechanical offset interference from the hair.

### 2.4.3.5 Previous fNIRS related studies

The fNIRS as a technique to measure stress on the PFC has not been yet investigated, yet has been applied in cognitive and behavioural studies. The commonly used tasks to activate the prefrontal cortex (PFC) include mental arithmetic, word generation, colour-word matching, Stroop task, mental rotation, working memory task and inhibition [64-66]. The PFC is related to the working memory which enables us to hold in mind and mentally manipulate information over a short period of time. These aforementioned tasks are used in the brain-computer interface (BCI) and attention studies [183-187]. More recently, the fNIRS has been accepted as an assistive tool to differentiate depression, bipolar disorder and schizophrenia [67, 68]. Additionally, the fNIRS has been successfully used in the classifying tasks from the rest state including the use of arithmetic. Table 2.3 summarizes the main features, classifiers and accuracy achieved by the previous arithmetic fNIRS studies. Although, none of these studies have addressed the problem of mental stress. The significant improvements in the classification accuracy confirmed that, it is possible to use fNIRS in studying mental stress. Besides, these studies have used arithmetic tasks to induce high workload which could serves as a baseline to this thesis and have shown a good classification accuracy. One of the main limitations of these studies is that, none of them have taken in consideration the classification sensitivity, specificity and the area under the ROC which could serve as a good indicator of mental workload/stress. Perhaps the most serious disadvantages of these studies are the limitation of sample size and no effective model to eliminate the noise was used. This encourage researchers to start in a new direction which is very important to the field of fNIRS as well as to the cognitive neuroscience studies. This thesis will work on studying the properties of temporal resolution of fNIRS and improve its limitation with the use of EEG signals.

## 2.5 Functional connectivity measure

Functional connectivity is defined as the temporal correlation of the activities of different neural assemblies [188]. In the existing literature, many neurophysiologic signals have been assessed through functional connectivity methods including the signals taken from a single unit and local field potential (LFP) recordings, EEG, Magnetoencephalography (MEG), PET, fMRI and fNIRS [189]. In the case of the EEG signals, human brain connectivity began to be measured using cross-correlation of pairs of EEG signals/ electrode locations in the 1960s. Higher correlations indicate stronger functional relationships between the related brain regions. Brain connectivity can be measured in the frequency domain as well as the time domain. The connectivity in the frequency domain is measured by the Magnitude Squared Coherence (MSC) or coherence. It allows the spatial correlations between the signals to be measured in different frequency bands as the case of EEG (delta, theta, alpha and beta band) [190]. Correlation, on the other hand is another measure of brain connectivity and can be calculated over a single epoch or over several epochs and is sensitive to both phase and polarity, independent of the amplitude. However, under normal physiological conditions, no strong and abrupt power asymmetries are expected to occur. Thus, the influence of power on coherence should be negligible and should produce results similar to those produced through correlation. Recently, multivariate Granger-causality-based (GC) measures provide a useful framework for establishing causal relations between neural populations. They have been successfully applied in determining the interactions at the subcortical and cortical levels. GC measures have been used extensively for intracranial signals while partial directed coherence (PDC) has been applied, for example, to find directed connectivity in the intracranial epileptic signals, see review [191]. According to the existing fMRI studies, stress reduces the PFC connectivity [192] in animals and human subjects. This study therefore propose the PFC connectivity to serve as an index to measure stress based on the magnitude squared coherence.

Table 2.3: Previous fNIRS related studies.

| Author/year | Region | Task | Features | Classifier | No of subjects | Accuracy % | Remarks |
|---|---|---|---|---|---|---|---|
| Naito et.Al 2007[193] | PFC | Arithmetic task | Amplitude of $O_2Hb$ | QDA | 17 | 80 | This type of study only classified workload induced by arithmetic task from rest state. |
| Power. 2010 [194] | PFC | Arithmetic task | Means of $O_2Hb$ | HMM | 10 | 77.2 | The author used simple arithmetic task and used machine learning to classify it from rest state. The validation of the study was based on small sample size. |
| Power. 2011[195] | PFC | Mental arithmetic | Slope $O_2Hb$ | LDA | 8 | 71.2 | The study succeed in classifying workload from rest state. However, robust analysis technique is needed to separate the hemodynamic from systemic noise. |
| Bauernfeind et al. 2011[196] | PFC | Mental arithmetic | Antagonistic $\Delta O_2Hb$ | LDA | 10 | 79.7 | The study achieved poor accuracy and the proposed task was very simple. |
| Abibullaev et al. 2011[197] | PFC | Mental arithmetic | Mean, Power, standard deviation of $O_2Hb$ | ANN | 4 | 80 | The author used many features which introduce computational cost. |
| Holper, 2011[198] | Motor cortex | Mental arithmetic | Mean, variance, skewness, kurtosis of $\Delta O_2Hb$ | FLDA | 12 | 80 | The study classified motor task from rest. This type of study was not specific to working memory. |
| Power. 2012b[199] | PFC | Mental arithmetic | Slope of $O_2Hb$ | LDA | 10 | 72.6 | The study did not improve the classification accuracy even they used different features. |
| Power et. 2013[200] | PFC | Mental arithmetic | Slope of $\Delta O_2Hb$, and $\Delta HHb$ | LDA | 1 | 71.1 | The study combined different features of oxy and deoxy hemoglobin but failed to improve the classification accuracy. |

| Stangle et.al 2013[177] | PFC | Mental arithmetic | Amplitude of $\Delta$ $O_2Hb$ | LDA | 12 | 65 | The study did not achieve good classification accuracy. |
|---|---|---|---|---|---|---|---|
| Schudlo et. 2014[201] | PFC | Mental arithmetic | Slope of $\Delta$ $O_2Hb$ $\Delta$HHb and $\Delta$Ht | LDA | 10 | 77.4 | This study combined all hemodynamic responses but failed to improve the overall classification accuracy. |
| Nasser et.al 2014[174] | PFC | Mental arithmetic | Mean value of $\Delta$ $O_2Hb$, $\Delta$HHb | LDA and SVM | 14 | 74.2 (LDA) and 82.1 (SVM) | The study reported different classification accuracy with different classifiers. This indicated that the features used in this study were not consistent. |
| Khan et al. 2014[202] | PFC | Mental arithmetic | Mean value of $\Delta$ $O_2Hb$, $\Delta$HHb | LDA | 12 | 80 | The study showed the potential of fNIRS in classifying cognitive from rest state but fails to obtain high accuracy using oxygenated hemoglobin alone. |
| Hwang et al. 2014[186] | PFC | Mental arithmetic | Mean value of $O_2Hb$, HHb and Ht | LDA | 7 | 70 | The study used combination of features but failed to improve the overall classification performance. |
| Hong et.al 2015[203] | Motor cortex and PFC | Mental arithmetic | Mean and slope of $O_2Hb$ | LDA | 10 | ›75 | The study showed the ability of fNIRS in classifying mental tasks from rest state but failed to classify motor from cognitive tasks. |
| Nasser et.al 2016[204] | PFC | Mental arithmetic | Mean, variance, skewness, kurtosis of $\Delta$ $O_2Hb$ | LDA, QDA, *k*NN, SVM, ANN | 7 | 71.6, 90.0, 69.7, 89.8, 89.5 | The study reported SVM as the best classifier for fNIRS signals but failed to justify the reason of obtaining different accuracy with different classifiers. |
| Shin, 2016[205] | PFC | Mental arithmetic | Mean, average slope of the $\Delta$ $O_2Hb$,HHb | LDA | 11 | EC ($75.6 \pm 7.3$); EO ($77.0 \pm 9.2$) | This type of study was a bout BCI and did not improve the classification accuracy compared to the state-of-the art. |

### 2.5.1 Magnitude Squared Coherence (MSC)

Cross-correlation and MSC are the most commonly used linear synchronization methods and are defined as follows. Consider two simultaneously measured discrete time series $X_n$ and $Y_n$, n=1,…N. Then the cross-correlation function $C_{xy}$ is defined as:

$$C_{xy}(\tau) = \frac{1}{N-\tau} \sum_{n=1}^{N-\tau} ((X_n - \overline{X}) / \sigma_x)((Y_{n+\tau} - \overline{Y}) / \sigma_y) \qquad (2.9)$$

where $\overline{X}$ and $\sigma_x$ denote mean and variance, respectively, while $\tau$ is the time lag. MSC or simply coherence is the cross spectral density function $S_{xy}$, which is derived via the FFT of Eq. (2.9) and normalized by their individual auto-spectral density functions. However, due to the finite size of the neural data, the true spectrum known as periodogram needs to be estimated using smoothing techniques (e.g. Welch's method [206]). Thus, the MSC is calculated as

$$C_{xy}(f) = \frac{\left| \langle S_{xy}(f) \rangle \right|^2}{\left| \langle S_{xx}(f) \rangle \right| \left| \langle S_{yy}(f) \rangle \right|} \qquad (2.10)$$

where $\langle . \rangle$ indicates window averaging. The estimated MSC for a given frequency f then ranges between 0 and 1.

### 2.5.2 Functional connectivity measure using fNIRS

With the development of multichannel fNIRS systems, it has become possible to measure interactions between brain regions other than traditional brain activation to derive fNIRS-based connectivity [207]. fNIRS-based connectivity is a novel analysis tool for fNIRS data from the perspective of functional interactions which could be complementary to fNIRS activation analysis [189]. It is widely assumed that functional connections reflect neuroanatomical substrates [208]. Strong temporal correlations of spontaneous fluctuations of distinct regions in the brain are found in frequency range (<0.1 Hz) during resting state [209]. Several fMRI studies with frequency component analyses have shown that functional connectivity is predominantly subtended by low frequency components of the data (<0.3 Hz) [210].

To date, fNIRS has been increasingly used not only to localize focal brain activation during cognitive engagement [211], but also to map the functional connectivity of spontaneous brain activity during resting and task states in normal people and patient with psychiatric disorders [212-223]. The technique has shown a comparable pattern to fMRI in resting state condition [213, 224]. On the other hand, fNIRS has a higher time resolution (sampling rate: >10 Hz) than fMRI (sampling rate: ~1 Hz), which prevents aliasing of higher frequency activity such as respiratory (~0.15-0.3 Hz) and cardiovascular activity (~0.6-2 Hz) into low frequency signal fluctuations [225] indicates a more reliable estimation of rest state functional connectivity. Studies using fNIRS demonstrated that spontaneous oscillations of cerebral hemodynamics include two distinguishable frequency components at low frequency (~0.1 Hz) and at very low frequency (~0.04 Hz) [212, 226, 227]. Although the mechanism underlying these signal fluctuations remains unknown, simultaneous recordings of cerebral hemoglobin oxygenation, heart rate, and mean arterial blood pressure showed that the systemic signal contribution to the hemodynamic changes in the frequency range (0.04–0.15 Hz) was 35% for $O_2Hb$ and 7% for HHb [228], suggesting that low-frequency fluctuations largely reflect hemodynamic responses to regional neural activities.

Functional connectivity at resting state has been demonstrated to exhibit distinct frequency-specific features. Wu et al. demonstrated that the correlations among cortical networks are concentrated within ultralow frequencies (0.01–0.06 Hz) [229]. Sasai et al. reported that functional connectivity between the homologous cortical regions of the contralateral hemisphere showed high coherence in the frequency range of (0.009–0.1 Hz) [212]. Medvedev et al. demonstrated hemispheric asymmetry within the functional architecture of the brain at low frequency range of (0.01–0.1 Hz) over the inferior frontal gyrus and the middle frontal gyrus [227]. Specifically, the study found a leading role of the right hemisphere in regard to left hemisphere as measured by granger causality. Xu et al. on the other hand, found reduced in coherence at specific frequencies, (0.06–2 Hz) and (0.052–0.145 Hz) in PFC, and (0.021-0.052 Hz) in motor cortex in healthy people at the end of driving task [222]. Another study demonstrated significant reduced connectivity in various frequency intervals; (0.0095–0.021 Hz) in homologous, (0.021–0.052 Hz) in front-posterior and (0.052-0.145 Hz) in the motor-contralateral in subjects with cerebral infarction compared to that of healthy group [230]. Based on the aforementioned studies, part of this study aims to investigate the frequency-specific characteristics of fNIRS on inter and intra-hemispheric PFC during active control and stress conditions.

The PFC subserves our highest order cognitive abilities, generating the mental representations that are the foundation of abstract thought and the basis for flexible, goal-directed behavior. In primates, the PFC is topographically organized: the dorsolateral PFC (DLPFC) guides thoughts, attention and actions, while the orbital and ventromedial PFC (VMPFC) regulate emotion.

### 2.5.2.1 Effect of fNIRS Scanning Duration on FC and Network Properties Stability

The effects of the fNIRS signal collection duration on the functional connectivity (FC) and network metrics, such as the nodal efficiency, nodal betweenness, network local efficiency, global efficiency, and clustering coefficient has been studied previously. It was reported that, with increasing scanning durations, the FC maps did not exhibit relatively large pattern variations across both Pearson-correlation as demonstrated by Fig 2.12A and cross-correlation in Fig 2.12B networks [231]. When contrasted with the relatively longer 10-min data acquisition duration, these results also revealed significant ($p < 0.001$) and strong correlation across each fNIRS signal time bin (the mean correlation coefficients $r = 0.98 \pm 0.03$ for Pearson-correlation and $r = 0.97 \pm 0.04$ for cross-correlation) as demonstrated in Fig 2.12C and Fig 2.12D respectively. This suggests that the short-time fNIRS signal acquisition duration, e.g., at 1 min, can also bring about highly similar FC maps as those calculated from 10-min scanning durations.

Additionally, for nodal efficiency, plots of these efficiency values showed approximately horizontal lines, with small difference between the magnitudes of nodal efficiency across the scanning duration as demonstrated in Fig 2.13A, Fig 2.13B left respectively. For nodal betweenness, plots of these betweenness centrality values showed relatively bigger changes, compared with nodal efficiency plots, between the magnitudes of betweenness values across the scanning duration  as demonstrated in Fig 2.13A, and Fig 2.13B right respectively. However, the map-map correlation analysis as shown in Fig 2.13C, and Fig 2.13D revealed significant (p < 0.001) and strong correlations between short and long scanning duration data for both nodal efficiency and nodal betweenness, indicating almost immediate stability for nodal efficiency and nodal betweenness.

For local efficiency, global efficiency, and clustering coefficient metrics, these plots of the network values also showed similar magnitudes with the increase in scanning duration for the both the Pearson correlation-based network and the cross-correlation-based network  as demonstrated in Fig 2.14A, and Fig 2.14B. The statistical analysis using paired t-tests further revealed no significant difference existed in local or global efficiency metrics between fNIRS signal acquisition duration (p > 0.05). This result also demonstrated that the network efficiency and clustering coefficient computed by using the 1.0-min fNIRS signal acquisition duration were no different compared to these measures calculated by using the 10-min fNIRS scanning time [231].

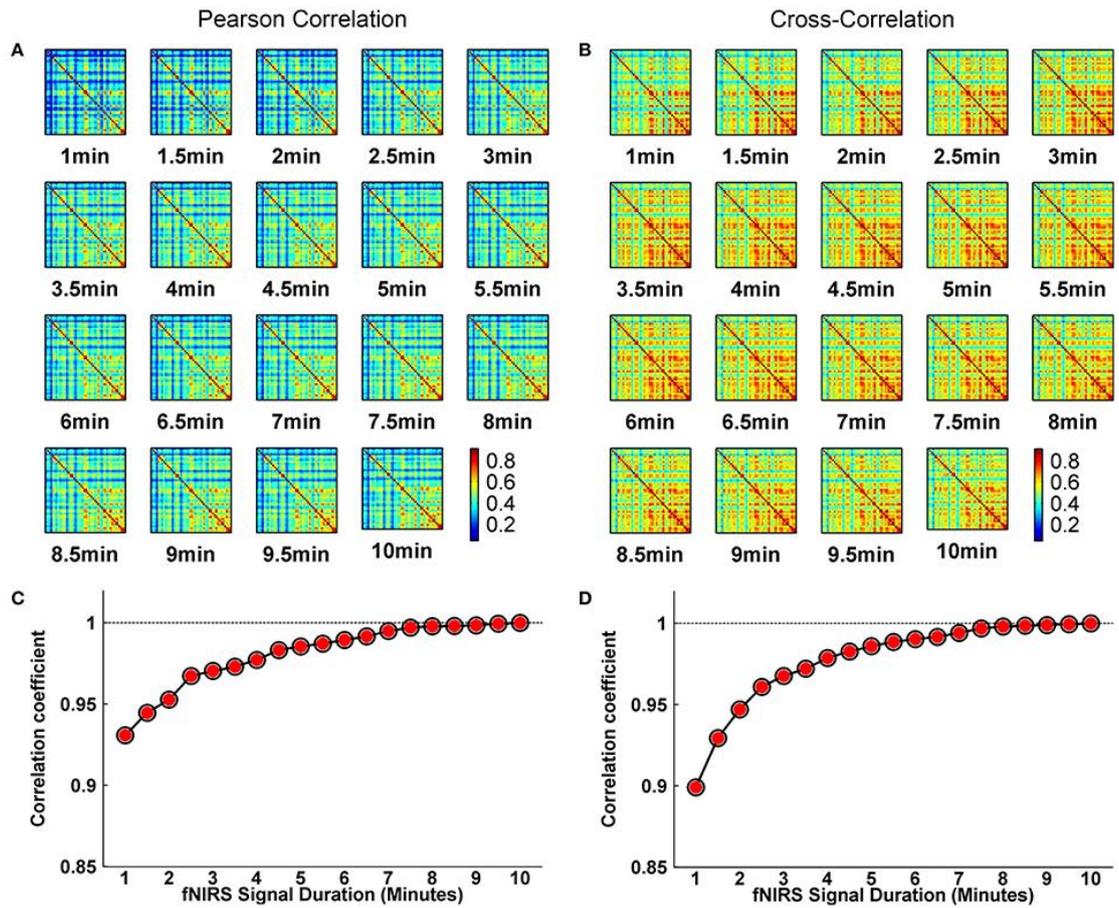

Figure 2.12: Effect of fNIRS signal acquisition duration on the stability of the spatial
FC map [231].

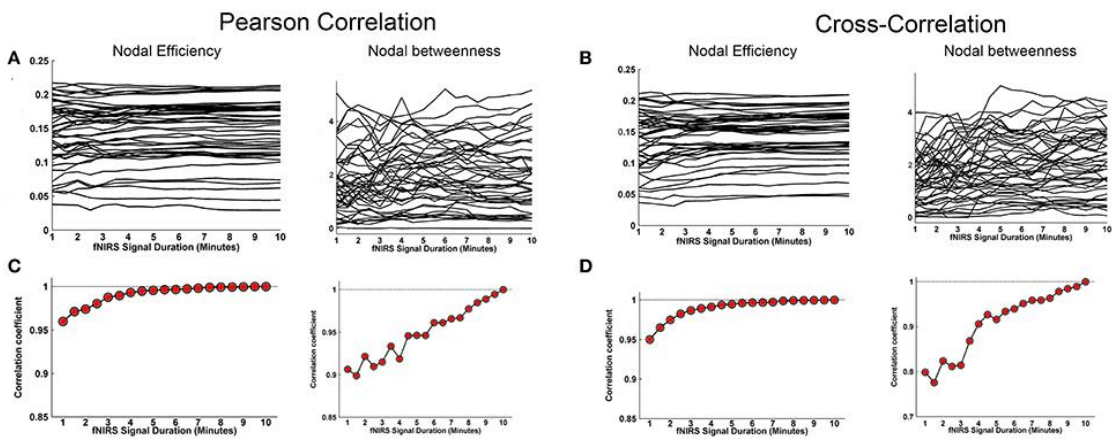

Figure 2.13: Effect of fNIRS signal acquisition duration on the stability of nodal
efficiency and nodal betweenness [231].

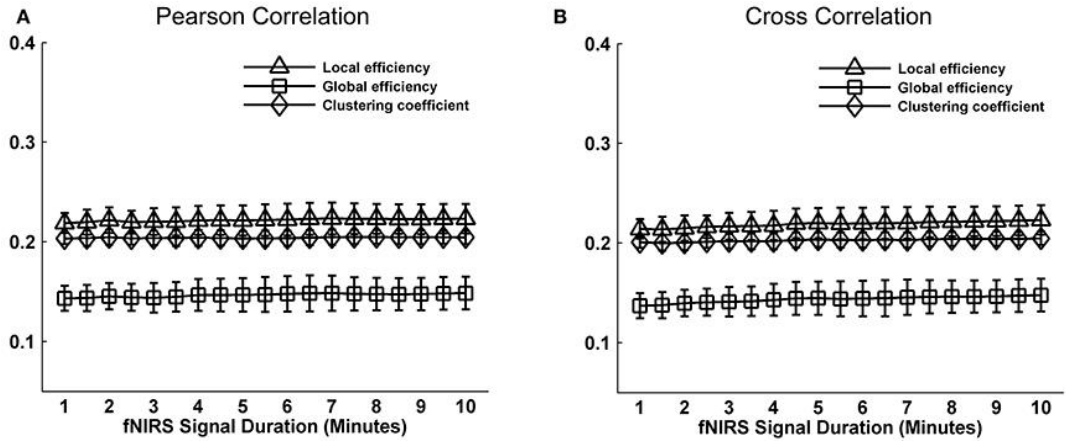

Figure 2.14: Effect of fNIRS signal acquisition duration on the stability of local efficiency, global efficiency, and clustering coefficient [231].

### 2.5.2.2 *Evaluation of Between-Run Reproducibility and Reliability*

Previous studies also further evaluated the effect of resting-state fNIRS signal acquisition duration on the reproducibility and reliability of FC and these network metrics mentioned above. For the Pearson-correlation network, the FC maps showed high similarity between the two runs at each fNIRS signal time bin as demonstrated in Fig 2.15A, and the ICC values of the FC also demonstrated approximately or equally excellent reliability as the scanning duration ranging from 1 to 10 min as demonstrated in Fig 2.15C [231]. Similarly, the nodal efficiency and nodal betweenness also showed good reproducibility between two scanning runs as demonstrated by Fig 2.16A and high reliability as demonstrated in Fig 2.16C with the increase in the scanning duration. This given results suggested that fNIRS data as short as a 1-min resting-state fNIRS signal can yield reproducible and reliable FC maps, nodal efficiencies and nodal betweenness.

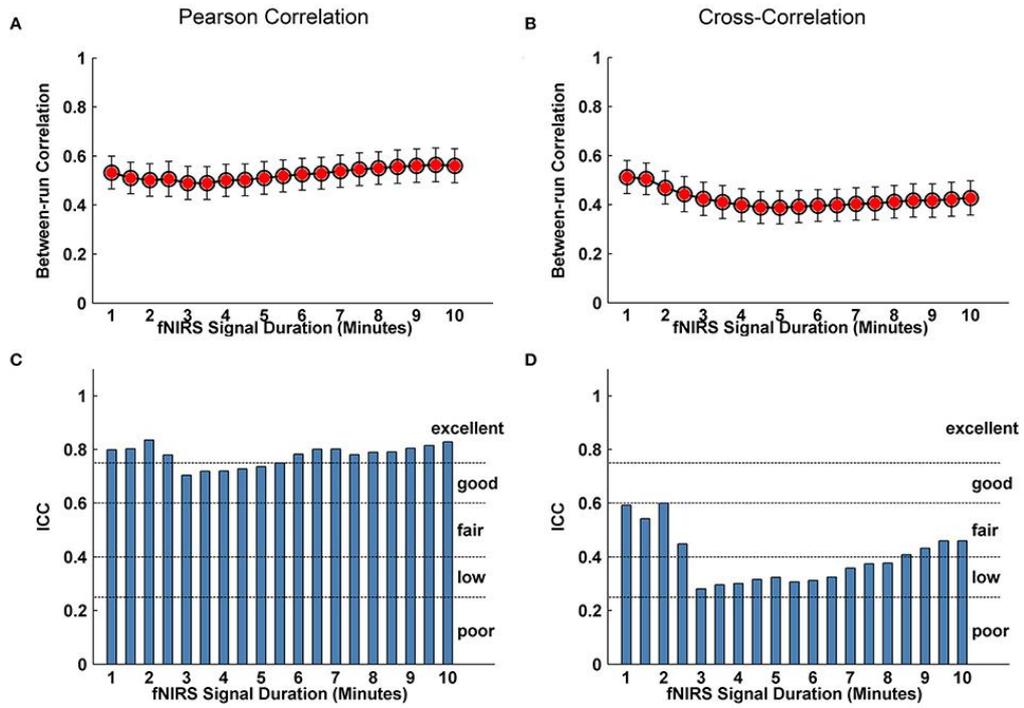

Figure 2.15: Evaluation of the effect of fNIRS signal acquisition duration on the reproducibility of spatial FC patterns [231].

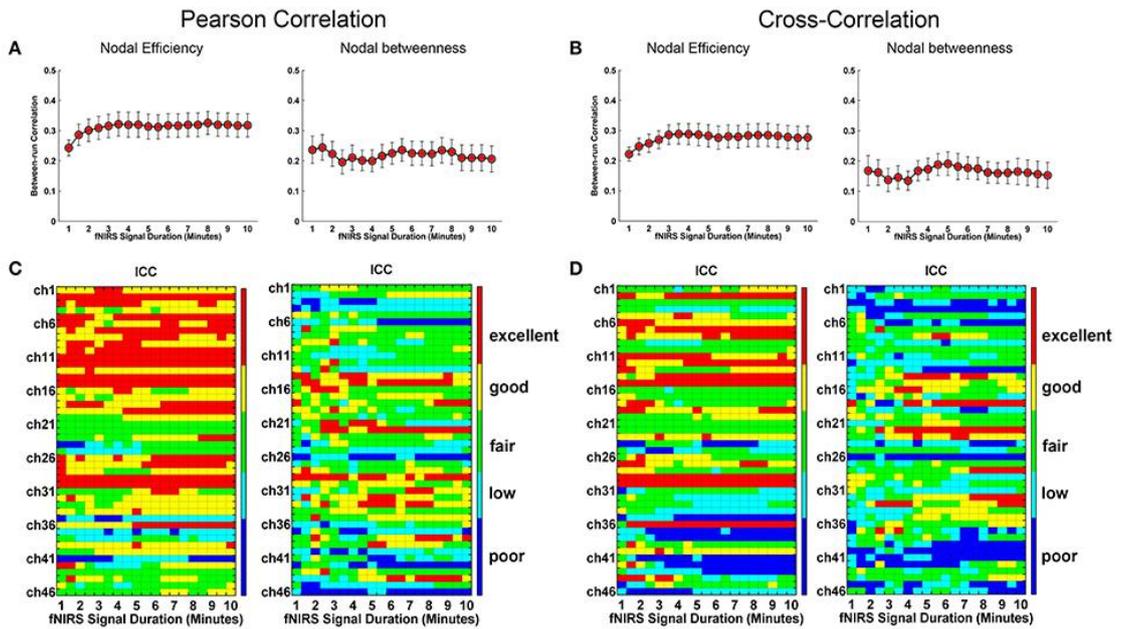

Figure 2.16: Evaluation of the effect of fNIRS signal acquisition duration on the reproducibility of nodal efficiency and nodal betweenness [231].

## 2.6 Simultaneous measurement of EEG+fNIRS

Simultaneous measurement of the EEG+fNIRS has been introduced recently. Up to this date, only few studies have looked into the possibilities of combining hemodynamic responses with their electrophysiological counterparts in a hybrid method aiming at improving the overall systems performance and accuracies. Fazli et al. proposed a hybrid sensory motor rhythm based on the brain computer interface (BCI) paradigm combining the fNIRS and the EEG signals [79]. A meta-classifier was constructed to combine the output probability of the individual classifier-modality. The results showed that the simultaneous measurement of the EEG+fNIRS can significantly improve the classification accuracy of the motor imagery by an average of +5%. Similarly, Khan et al. decoded four movement directions (left, right, forward, and backward) using the combined features of the fNIRS and the EEG modalities [202]. The researchers used the EEG features to classify left/right, and the fNIRS features to classify forward/backward. The study showed that the four different control signals can be accurately estimated using the hybrid fNIRS+EEG technology.

Additionally, Koo et al. proposed a novel hybrid BCI system using the fNIRS-EEG systems together to achieve online self-paced motor imagery based on the BCI [232]. The research team detected the occurrence of motor imagery using the fNIRS system and classified its type using the EEG system. Yin et al. combined the fNIRS and the EEG features in a motor imagery task and reported an improvement in the classification accuracy by +1% to +5% compared to the sole fNIRS and EEG feature set [233]. Putze et al. used the bimodal EEG+fNIRS for auditory and visual tasks in a BCI and demonstrated the efficiency of concatenated features in improving the versatility of the BCI system [77]. Blokland et al. recently examined the principle of combining these modalities in patients with tetraplegia and reported an improved accuracy in the brain switch control for some subjects [78]. The combination of the EEG+fNIRS technology is also applicable to language studies, cortical current estimation, tetraplegia and fatigue [78, 234-236]. Based on these studies, this thesis proposed integrating the EEG and the fNIRS modality which can significantly improve the detection rate of mental stress compared to the individual modality.

## 2.7 Previous Fusion methods

Recently, collecting multiple types of brain data from the same individual using various non-invasive imaging techniques (MRI, DTI, EEG, MEG, etc.) has become common practice. Each imaging technique provides a different view of brain function or structure. For example, functional magnetic resonance imaging (fMRI) measures the hemodynamic response related to neural activity in the brain dynamically; structural MRI (sMRI) provides information about the tissue type of the brain [gray matter (GM), white matter (WM), cerebrospinal fluid (CSF)]. Diffusion tensor imaging (DTI) can additionally provide information on structural connectivity among brain networks. Another useful measure of brain function is electroencephalography (EEG), which measures brain electrical activity with higher temporal resolution than fMRI (and lower spatial resolution). Typically these data are analyzed separately; however separate analyses do not enable the examination of the joint information between the modalities.

By contrast, combining modalities may uncover previously hidden relationships that can unify disparate findings in brain imaging [237]. For example, the spatial precision of fMRI could be complemented with the temporal precision of EEG to provide unprecedented spatiotemporal accuracy [238]. The combined analysis of fMRI and magnetoencephalography (MEG) measurements can lead to improvement in the description of the dynamic and spatial properties of brain activity [239]. In another case, using combined genetic and fMRI data achieved better classification accuracy than using either alone, indicating that genetic and brain function representing different, but partially complementary aspects [240]. Finally, a lower and different function-structure linkage is often found in patients with brain disorder such as schizophrenia [241], suggesting that combination of two brain modalities provides more comprehensive descriptions of altered brain connectivity. Therefore, a key motivation for jointly analyzing multimodal data is to take maximal advantage of the cross-information of the existing data, and thus may discover the potentially important variations which are only partially detected by each modality. Approaches for combining or fusing data in brain imaging can be conceptualized as having a place on an analytic spectrum with meta-analysis to examine convergent evidence at one end and large-scale computational modeling at the other end [242]. In between are methods that attempt to perform direct data fusion [243]. Joint-ICA is one of the most commonly used technique for data fusion. It has been successfully used for the fusion analyses of fMRI, EEG, and sMRI data [244]. The jICA examines intersubject covariances of sources to discover associations between feature datasets from different modalities. The jICA fusion scheme discovers relationships between modalities by utilizing ICA to identify sources from each modality that modulate in the same way across subjects. ICA is a popular data-driven blind source separation technique and has been applied to a number of biomedical applications such as for fMRI data in [19]. However, this technique poses a more constrained approach to the data fusion problem.

Recently, there has been increased interest in the use of CCA for feature fusion in various pattern recognition applications [245]. The CCA is used to fuse features for handwritten character recognition in which different kinds of features are extracted from the same handwriting data samples and CCA is performed on them to obtain two sets of canonical features which are combined to form a new feature set. This method not only finds a discriminative set of features but also somewhat eliminates redundant information within the features. In [14], the same idea was used in a block-based approach where the sample images were divided into two blocks and canonical features were extracted which were then combined linearly to obtain better discriminating vectors for recognition. In [16], CCA was used to fuse feature vectors extracted from face and body cues to form a joint feature and then utilize the multimodal information to discriminate between affective emotional states. In [17], CCA was used to fuse features from speech and lip texture/movement to form audiovisual feature synchronization which aids in speaker identification. Thus, CCA has been used to fuse two sets of features to obtain more a discriminative set of features for recognition problems. In this study, different approach of jICA and CCA are proposed in which the techniques are used to fuse information from two sets of features to discover the associations across two modalities in this case, the EEG and fNIRS and to ultimately estimate the sources responsible for these associations.

## 2.8 Statistical analysis and feature selection

In order to better understand the brain responses to different tasks under different conditions, statistical analysis is an option. Researchers usually evaluate the differences in brain responses using t-test or multi-factor analysis of variance [246].The principle behind t-test is that it compare the mean at two different conditions with certain of probability, mostly with p<0.05. Assuming the two distributions have the same variance, the t-test is calculated using the following equation:

$$t = \frac{\bar{X}_1 - \bar{X}_2}{s_{X_1 X_2} \cdot \sqrt{\frac{1}{n}}} \qquad (2.11)$$

where, $s_{X_1 X_2} = \sqrt{(s_{X_1}^2 + s_{X_2}^2)}$ . Here $s_{X_1 X_2}$ is the grand standard deviation, the order of $X_1 X_2$ represents the group for example in this study; the control and the stress condition respectively, $s_{X_1}^2 + s_{X_2}^2$ are the unbiased estimators of the variance. Note that the denominator of t denotes the standard error of the differences between the mean of the control and the mean of the stress group. The degree of freedom was set to $2n - 2$ , $n$ being the number of participants/features in each subject/group. For comparing two means (for example, at control condition $u_1$ and at stress condition $u_2$ ), the basic null hybothesis is that the measures are equal,

$$H_0 : u_1 = u_2$$

With three common alternative hypotheses,

$$H_a : u_1 \neq u_2,$$

$$H_a : u_1 < u_2, or$$

$$H_a : u_1 > u_2,$$

Each is chosen according to the nature of the experiment or study. In this study, the MATLAB build in function of *ttest2*(control, stress, 'tail') used to calculate the corresponding t and p values respectively. The tail here stands for the alternative hybothesis in which the population mean of data features at stress condition decreased as compared to the mean of control condition. In other words, if the population mean of control condition is higher (right tail) or lower (left tail) than that the population mean at stress condition. The t and p values were calculated for every single subject/channel in which the population mean of extracted feature points at the control condition were compared to the population mean of the extracted feature points in the stress condition in accordance with equation 2.11 and by the use of MATLAB build in function of *ttest2*. The differences in all the aforementioned responses (control versus stress) were considered statistically significant if *p value < 0.01*. The use of T-value here is due to its reputability as a good indicator to local activation/deactivation [246]. This study defined empirically a threshold value of t > 3 to be considered as the significant change in the $O_2Hb$ for FNIRS to better localize the effects of mental stress within the PFC subregions. The selection of t>3 is to obtain good focality point of the oxygenated hemoglobin to the stress on the PFC subregions. The given figures at; Fig.2.17 to Fig.2.19 show an example of the raw EEG data and the normality distribution of the EEG and the fNIRS data features.

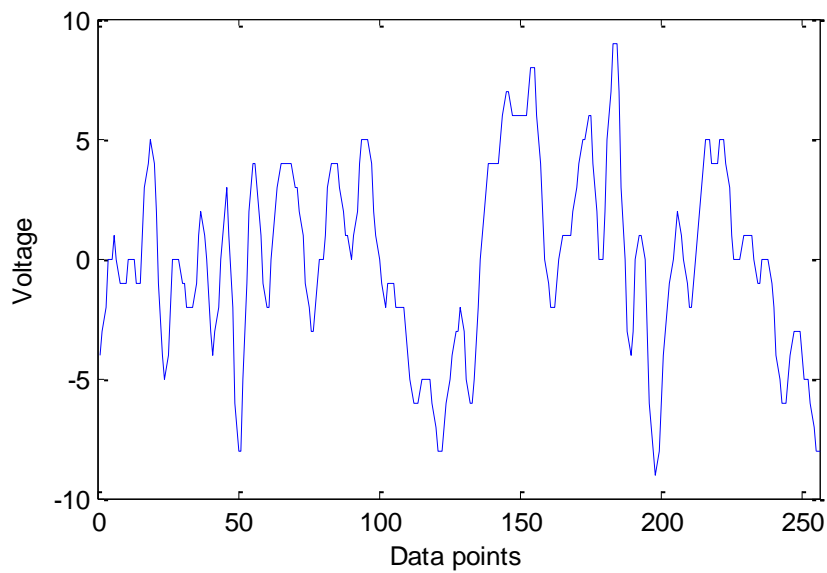

Figure 2.17: An example of raw EEG waveform at one second recording time.

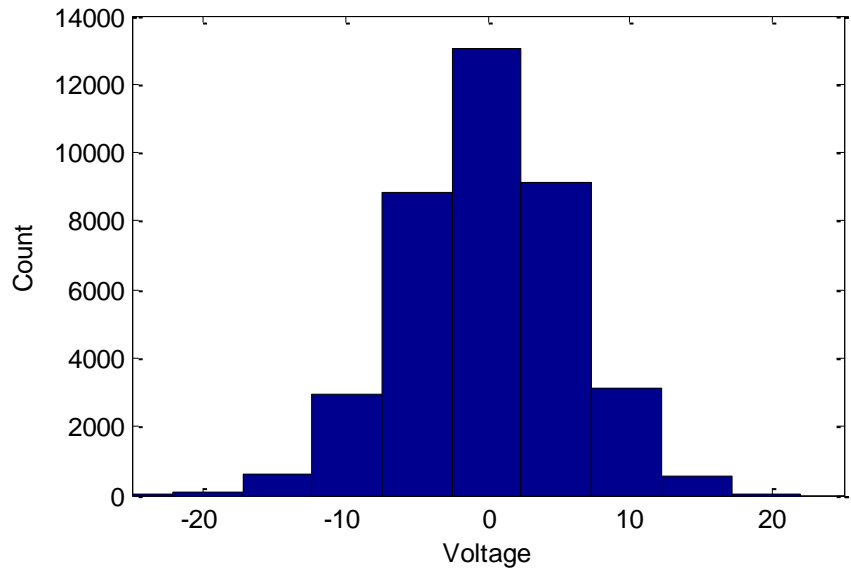

Figure 2.18: An example of probability density function showing normality distribution of EEG data features.

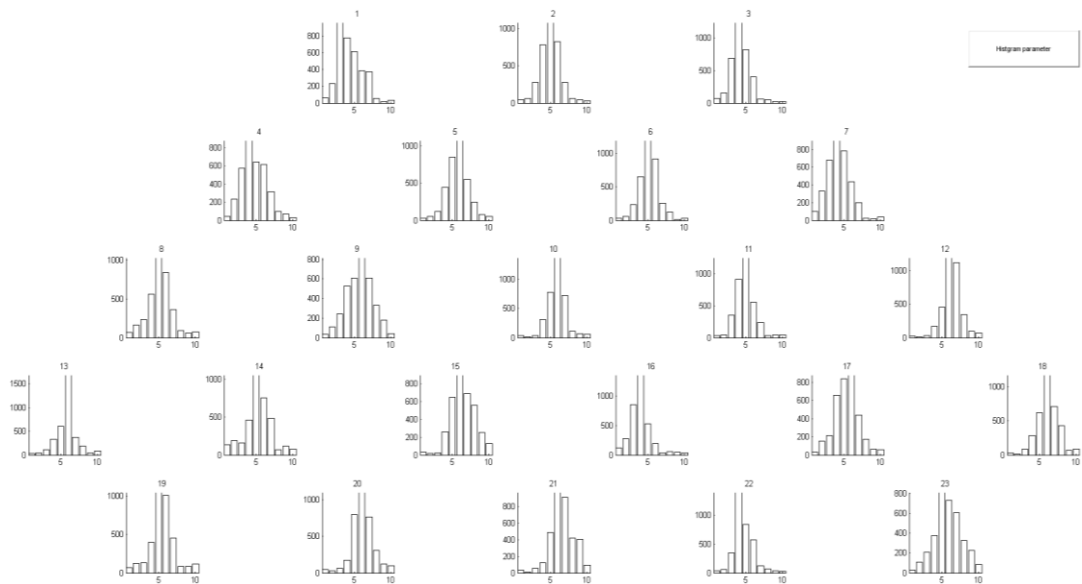

Figure 2.19: An example of probability density function showing normality distribution of fNIRS data features.

## 2.9 Classification using support vector machine (SVM)

Previous neuroimaging studies described in Table 1 and Table 2 have used SVM for classification, estimation and prediction. The SVM is a supervised machine learning technique widely used for classification, regression and density estimation [247]. The method used as common classifier for the performance assessment of individual modality and of EEG and fNIRS modality. The selection of SVM in most of the previous studied was due to its ability to model linear as well as more complex decision boundaries. The decision boundary hyperplane in the SVM usually estimated based on its training dataset by maximizing the distance between the hyperplane to the nearest data point. LIBSVM software is one of the most commonly library used to build the SVM classifier and employed polynomial or the radial basis function (RBF) kernel as stated in Eq.2.12 to nonlinear map data onto a higher dimension space [248].

$$K(x, y) = \exp(\frac{-\| x - y \|^2}{2\sigma^2}) \tag{2.12}$$

where x and y are the two data points and $\sigma$ is the width of RBF. The SVM is then perform by dividing the data sets into a set of cross-validations. One cross validation used for testing the classifier and the other sets of cross-validations used for training the classifier. The process repeated until each cross-validation being used for testing and training.

### 2.9.1 Classification Evaluation

The performance of the classifiers is determined by computing the classification accuracy, sensitivity and specificity as described in [249]. The accuracy was defined as the ability of the classifier to correctly identify the positive and the negative results and can be evaluated using Eq 2.13.

$$Accuracy = \frac{TP + TN}{TP + TN + FP + FN} \times 100 \tag{2.13}$$

where the true positive (TP) are data points correctly labeled as 'stress' in the corresponding dataset and the true negative (TN) are data points correctly labeled as 'not stress' in the corresponding dataset. The sensitivity measures the classifier ability to correctly identify the positive result and was calculated using Eq 2.14.

$$Sensitivity = \frac{TP}{TP + FN} \times 100 \qquad (2.14)$$

where, the false negative (FN) refers to data points incorrectly labeled as ' no stress'. Specificity gives a measure of the classifier ability to identify the negative results defined in Eq.2.15.

$$Specificity = \frac{TN}{TN + FP} \times 100 \qquad (2.15)$$

where, the true negative (TN) are data points correctly labeled as 'stress' in the corresponding dataset, while the false positive (FP) refers to data points incorrectly labeled as 'not stress'.

Additionally, the studies analyzed the classification performance using receiver operating characteristic (ROC) curves that plot the sensitivity versus 1 minus the specificity. The study also evaluated the area under the ROC curve (AUC) as a measure of a classifier's discriminatory power, which is insensitive to class distributions and the cost of misclassification (i.e. AUC=1 indicating perfect classification, whereas AUC=0.5 indicating that the classifier's result is not better than random guess).

## 2.10 Summary

Stress is a growing problem in our society, and nowadays job issues, including high workloads and need of adaptation to constant changes, only serve to worsen the problem. Stress measuring methods based on hormonal techniques and subjective questionnaires are not suitable for real time monitoring and require people to get out of their routine activities. Up to date, the most accurate stress detection systems developed in the state of the art show that stress detection using physiological signals is much more accomplished than using psychological and behavioural responses. Well-researched direct methods electroencephography (EEG) and magnetoencephalography (MEG) rely on the electrical and magnetic field respectively emanating directly from neurons in the brain. These techniques are characterized by high temporal resolution which is advantageous for cognitive processing. However, low spatial resolution makes it difficult to determine the source of activation and the corresponding activity pathway. Indirectly measure by changes in blood volume and oxygenation usually referred to as hemodynamics. These methods of monitoring brain activity include positron emission tomography (PET) and functional magnetic resonance imaging (fMRI) which are both non-optical techniques, and functional near-infrared spectroscopy (fNIRS) an optical technique. The techniques of fMRI, MEG and PET are expensive not only in terms of their manufacture cost but also the cost of the physical infrastructure to house the equipment. Additionally, a disadvantage especially when working with infants is that the techniques do require the subjects to be still and are restrictive not only with head movement but require the subjects to be lying down. Thereby, a multimodal technique that has the complementary nature may be considered. There have been some difficulties for fusing data of different physiological responses until now, but nowadays most of them are solved. Feature-level fusion and voting systems could be used for merging data of different nature, allowing at the same time to reduce the amount of information to work with. This study proposed joint independent component analysis technique to fuse temporal EEG components and spatial fNIRS components in the feature level. Additionally, the study proposed canonical correlation analysis technique to identify a linear relationship between the two sets of variables (EEG and fNIRS) by determining the inter-subject co-variances.

CHAPTER 3

STRESS ASSESSMENT USING FUSION OF EEG AND FNIRS FEATURES

This chapter introduces the proposed methods for preprocessing, feature extraction, statistical analysis, EEG and fNIRS data fusion, classification and cortical connectivity within the inter- and intra-hemisphere. The process of the EEG data involved the artifacts and noise removal using the independent components analysis technique (ICA), the band-pass filter in the range of 0.5-30 Hz, the feature extractions using wavelet transform, and the selection and classification of the features using the support vector machine classifier. Similarly, the fNIRS data process involved removing the motion artifacts, the heartbeat, the Mayer wave, the blood pressure and the breathing noises. Baseline correction was achieved by way of the linear regression, the data moving average, the block averaging, and the extraction and classification of the features. The cortical connectivity was investigated based on the average squared coherence within the inter- and intra-hemisphere for control and stress conditions at seven frequency intervals. Finally, it describes the proposed two feature-level fusion approaches based on joint independent component analysis and canonical correlation analysis. The summary of the overall process of this study is as demonstrated by the flow chart in Fig 3.1.

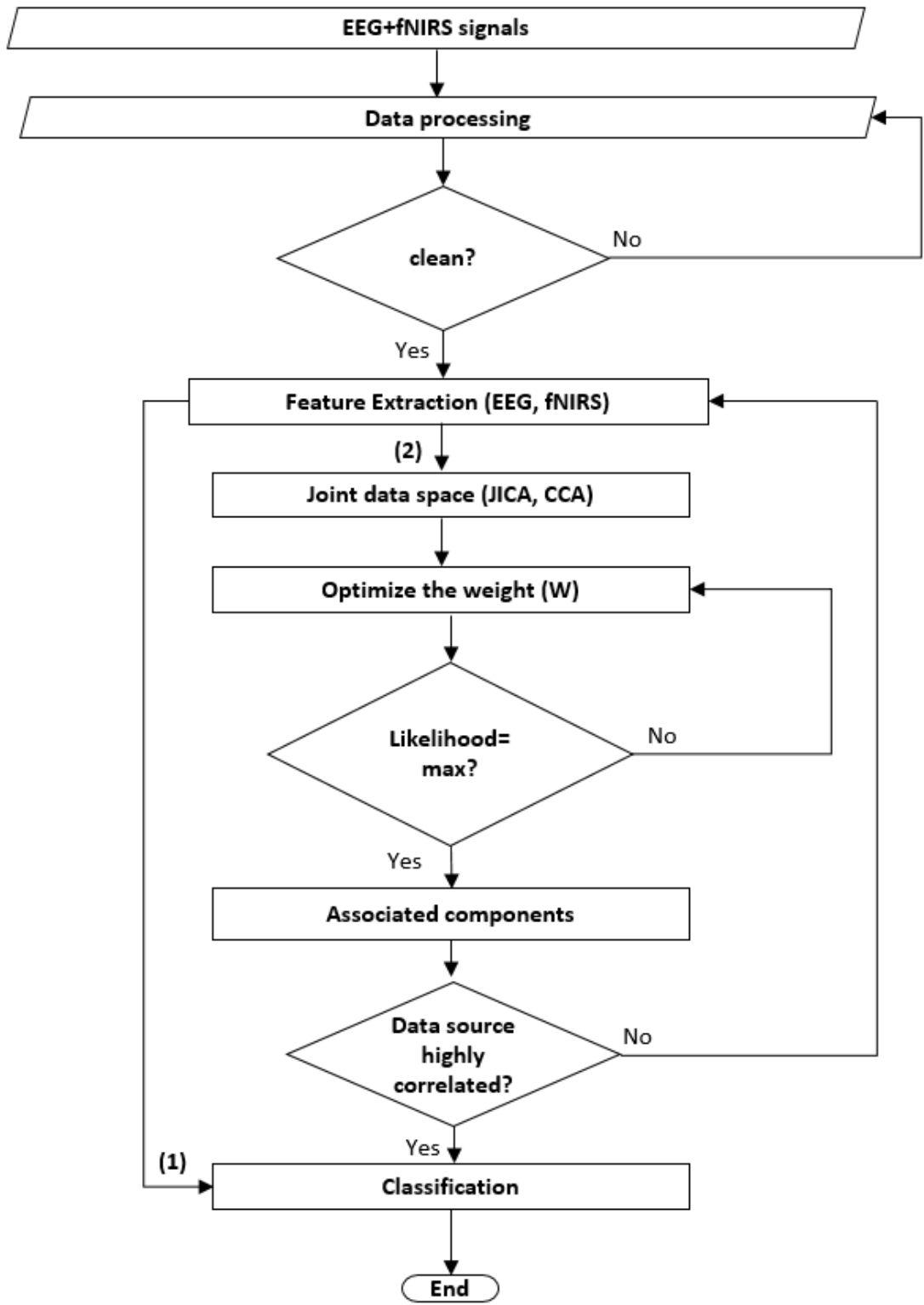

Figure 3.1: overall flow chart of the proposed study.

## 3.1 Subjective measurements

The subjective assessment of mental stress was conducted using questionnaires, NASA-TLX [250]. Participants were asked to evaluate their mental workload before beginning with their task as baseline, after the simultaneous measurements of the EEG and the fNIRS under the control condition and after the simultaneous measurements of the EEG and the fNIRS under the stress condition. As for other subjective measures of workload, NASA-TLX relies on subjects' conscious perceived experience with regards to the effort produced and difficulty of task. NASA-TLX has the advantage of being quick and simple to administer. The index is a multidimensional method with various evaluation degrees, which provides a self-evaluation model to estimate workload through use of six subscales including (MD, mental demand; PD, physical demand; TD, temporal demand; OP, own performance; EF, effort; and FR, frustration). Each of these six subscales scores from 1 to 20 based on performance of participants. To analysis the data collected by the questionnaire, the overall scores were then weighted to 100 and used for evaluation. Two sample t-test was then applied to test the differences between condition with confident interval of 95% corresponding to $p<0.05$.

## 3.2 Objective measurements and data processing

The Objective measurements proposed in this study are based on EEG and fNIRS signals. The method of acquisitions are addressed in details in the experiment part in Chapter 4. In this Chapter, several preprocessing methods applied to the acquired data in order to eliminate the systemic noise and to remove the artifacts from the signals. The preprocessing was performed individually for each modality. The following subsections elaborate the overall preprocessing of the acquired signals.

### 3.2.1 Data acquisition

In this study, data were acquired simultaneously by measuring the EEG signal using the Discovery 24E system (BrainMaster Technologies Inc, Bedford, OH) as well as the hemodynamic responses using the multi-channel fNIRS system (OT-R40, Hitachi Medical Corporation, Japan). The electrodes and channels of the modalities were combined in one single cap custom designed in our lab, more details on the design can be find in Chapter 4 experiment part. The EEG system was equipped with seven active electrodes and two reference electrodes attached to the ear lobes. Similarly, the fNIRS system, was equipped with 16 optical fibres; eight sources (two wavelengths, 695 nm and 830 nm, combined in one source) and eight detectors. The sampling frequency for the EEG was set to 256 Hz and the fNIRS system was sampled at 10 Hz. More details on the data acquisition and experimental set-up are presented in Chapter 4.

### 3.2.2 EEG data processing

EEG data were preprocessed offline using MATLAB (Version R2013a, The MathWorks, Inc., Massachsetts, USA) with custom scripts as well as the plug-in EEGLAB 2013a toolbox [251]. Raw EEG data were bandpass filtered between 0.5 Hz and 30 Hz using the commonly used third order Butterworth filter. The Butterworth filter proposed here due its ability to smooth monotonically decreasing frequency response in the transition region. Independent component analysis (ICA) was applied to remove eye blink and movements artifacts. The channels were decomposed into a number of independent components (by default the number of components is equal to the number of recorded channels). The component corresponding to artifacts was removed. The signals were further analyzed using wavelet transform (WT) [252]. WT constitutes a suitable method for multi-resolution time-frequency analysis. WT decomposed the EEG signals into a set of functions to obtain their approximation and the corresponding coefficients at different levels. In this study the wavelet family of Dubechies-8 (db8) was used in order to decompose the EEG signals into four frequency bands (delta, theta, alpha, and beta). The mean power values were then extracted from these coefficient as described in the following sub-section.

*3.2.2.1 EEG feature extraction*

From the wavelet coefficients, the mean absolute values of the wavelet coefficients in each sub-band and the average power and energy were extracted within the time of the activation period (task condition). The activation period was defined from the onset of the task to the end of the task in each block. The five active blocks were then averaged into a single block of 30 s to facilitate the process and to smooth the overall acquired signals. In this work, a window of 500 ms moving-time interval was used to calculate the features of the EEG signals from the wavelet coefficients. The power spectral density values were calculated using Eq.3.1.

$$P_j = \frac{1}{N}\sum_{n=1}^{N}|x_j(n)|^2, j = 1,2 \qquad (3.1)$$

where $P_j$ constitutes the mean absolute value of the EEG signals, $x_j(n)$ represents the segmented EEG signal in the alpha band at j=1 and the beta band at j=2, $N$ being the length of the clean EEG signal. The energy of the EEG frequency bands was defined and calculated using Eq. 3.2:

$$E_j = \frac{1}{N}\sum_{-\infty}^{\infty}|x_j(n)|^2, j = 1,2. \qquad (3.2)$$

Each feature set was then normalized to the scale of '0' and '1' before being fed into the classifier using Eq.3.3.

$$Feature_{norm} = \frac{x - \min(x)}{\max(x) - \min(x)} \qquad (3.3)$$

where $x$ is the entire feature set, *min(x)* is the minimum value in the feature set and *max(x)* is the maximum value in the feature set. As the mean of the spectral power was highly significant, this study limited the analysis to the mean power features. The entire flow of the processing and feature extraction of EEG data is as shown in Fig 3.2. The statistical analysis was performed to select the dominant region and the most significant electrode responded to the stress task. This study used t-test to study the differences between the brain responses during the control and stress condition. The study assumes that, the differences are significant only if p<0.05. Then support vector machine classifier was applied to classify the data into control or stress.

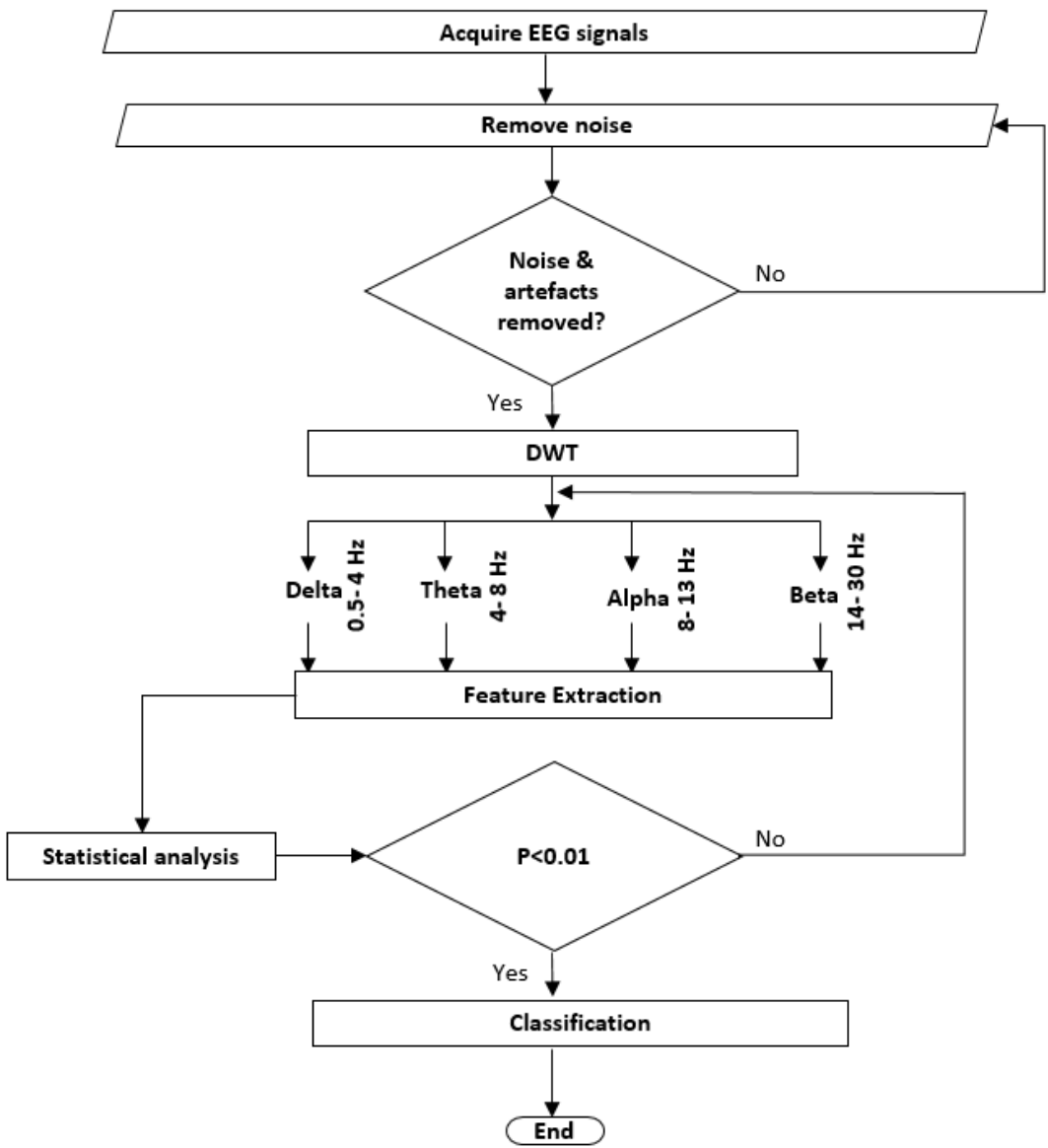

Figure 3.2: EEG flow of data processing and features extraction.

### 3.2.3 fNIRS data processing

The fNIRS signals were transformed into the concentration changes of $O_2Hb$, $HHb$ and total hemoglobin Ht using the modified Beer-Lambert law. In order to reduce noise and artifacts, the fNIRS signals were passed through several pre-processing steps using MATLAB (Version R2013a, The MathWorks, Inc., Massachsetts, USA) with custom scripts as well as the plug-in analysis software *Platform for Optical*

*Topography Analysis Tool* [253]. The process involved filtering the signal in the range of 0.012 to 0.8 Hz using a fifth order Butterworth filter, moving average, baseline correction and epoch extraction. The Butterworth was selected due to its strength in smoothing the overall signal at lower frequency. In order to exclude high frequency artifacts of the signal, a moving average was calculated using a time window of 5 s. In the baseline correction, the study defined a period starting from the onset of the task condition to the end of each task condition as one analysis block. Then linear regression by least mean squares was applied to determine the linear trend of its baseline. This is was done only to include the pure signal of hemodynamic in responding to the arithmetic tasks either under the control or the stress condition. Finally, the data were averaged in all the analysis blocks into a single block to further to reduce the computational time and enhance the overall process. As responses were more pronounced in $O_2Hb$, the analysis in this study was limited to the $O_2Hb$ signals.

### 3.2.3.1 fNIRS features extraction

For each analysis block, features were extracted by calculating the mean concentration, variance, skewness, kurtosis and peak of $O_2Hb$ over the analysis block using a moving time-window of 500 ms. The mean of the signal was calculated as:

$$O_2Hb = \frac{1}{N}\sum_{n=1}^{N}(\Delta O_2Hb)_n \qquad (3.4)$$

where $\Delta O_2Hb$ represents the segmented $O_2Hb$ signal and $N$ is the length of $O_2Hb$ signal.

The variance was calculated as follows:

$$VAR(O_2Hb) = \frac{\sum_{n=1}^{N}(O_2Hb - \mu)^2}{N} \qquad (3.5)$$

where *var* is the variance, $\mu$ is the mean value of $O_2Hb$.

The skewness is computed as follows:

$$Skew(O_2Hb) = E\left[\left(\frac{O_2Hb - \mu}{\sigma}\right)^3\right] \qquad (3.6)$$

where *skew* is the skewness, $E$ is the expected value of $O_2Hb$ at selected time window and $\sigma$ is the standard deviation of $O_2Hb$.

The kurtosis was computed as follows:

$$Kurt(O_2Hb) = E\left[\left(\frac{O_2Hb - \mu}{\sigma}\right)^4\right] \tag{3.7}$$

The signal peak is estimated using the common built-in function in Matlab, the *max* function. The signal slope was determined by fitting a line to all the data points during the mental arithmetic and rest using the built-in function, the *polyfit* function in Matlab. These features were calculated for the rest state, mental arithmetic at control and at stress condition across all the 23 channels from all subjects. All of the feature values were scaled between 0 and 1 before feeding them into the classifier using the Eq.3.8:

$$O_2Hb' = \frac{O_2Hb - \min(O_2Hb)}{\max(O_2Hb) - \min(O_2Hb)} \tag{3.8}$$

where $O_2Hb \in R^n$ denotes the original feature vales, $O_2Hb'$ denotes the rescaled feature values between 0 and 1, max ($O_2Hb$) is the largest value, and min ($O_2Hb$) is the smallest value in the entire signal set. However, in order to facilitate the process of this study, the analysis was limited to the feature mean of the $O_2Hb$ signals due to their highest significant responses to stress stimuli. The overall flow of the fNIRS data processing and features extraction is shown in Fig 3.3.

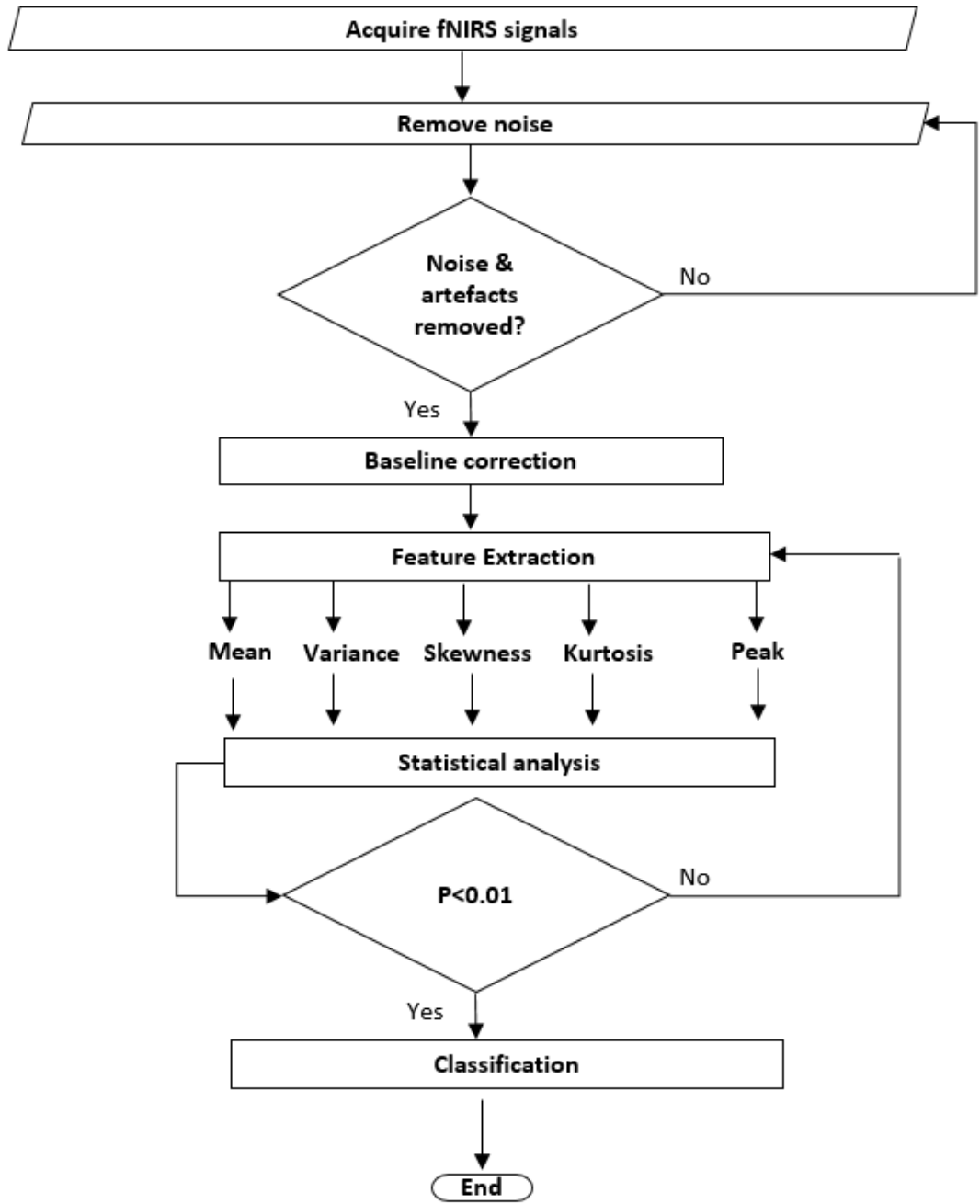

Figure 3.3: fNIRS flow of data processing and features extraction.

## 3.3 Fusion of EEG and fNIRS data

Fusion of multimodality data is an especially challenging problem since brain imaging data types are intrinsically dissimilar in nature, making it difficult to analyze them together without making a number of assumptions. Among these assumption is the unrealistic about the nature of the data. Unlike data integration methods, which tend to use information from one modality to improve the other modality, in this study data fusion techniques incorporate both modalities in a combined analysis. Thus, the combined analysis allows for true interaction between the different data types of EEG and fNIRS. Instead of entering the entire data sets into a combined analysis, this study used an alternate approach to reduce each modality (EEG and fNIRS) to a feature, which is a lower-dimensional representation of selected brain activity. The fusion was then applied to explore associations across these feature data sets through variations across individuals, in this study 25 healthy and 25 with mental stress.

Investigating variations across subjects or between stressed people and non-stressed people at the feature-level provides a natural way to find multimodality associations and also alleviates the difficulty of fusing data types of different dimensionality and nature as well as those that have not been recorded simultaneously. Feature-level analysis has been successfully used in data-driven fusion techniques such as joint-ICA [254]. Given two feature data sets of X, the jICA approach involves concatenating the data sets alongside each other and then performing ICA on the concatenated data set. Note that, this is the first EEG and FNIRS study propose feature fusion based on joint component analysis technique. Joint-ICA assumes that the sources have a common modulation profile A across subjects [254]. This is a strong constraint considering that the data come from two different modalities.

Recently, it has been shown that CCA can allow a more flexible approach to the fusion problem [255]. Due to that, this study propose it for the first time to fuse the EEG and fNIRS features. The fusion method also adopts a feature-based approach and similarly models the feature data set from each modality as a linear mixture of components with varying levels of activations for different subjects. Thus, the relationship between modalities is based on intersubject covariations. The following sub-sections clearly elaborate on the implementations of the proposed fusion technique.

### 3.3.1 Fusion Based on Joint Independent Components Analysis (jICA)

In this method, the EEG and the fNIRS data were fused using two consecutive methods. Firstly, the advantages of the spatial resolution of the fNIRS was used in order to identify the spatial locations of interest (regions of interest). The selection of the EEG electrodes was then based on the significance of their neighboring fNIRS channel response to the mental stress task. Statistical analysis based on t-test was performed to study the significant in brain responses between control condition and stress condition. The higher the t-value, the great is the significant differences between the brain responses under the two conditions. Only channels with the high t-values were considered for fusion and their neighboring electrodes were selected. Secondly, the spatial and temporal advantages of each modality were explored. The EEG and fNIRS sources were transformed via ICA into the temporal tICA components and spatial sICA components. This performed in oreder to extract excellent features with their strength so that they can complement each other. Then joint independent component analysis (jICA) was applied in order to compute the mixing matrix *A,* between the tICA components of the EEG and the sICA components of the fNIRS using the following generative model:

$$X^{EEG} = AS^{EEG}, X^{fNIRS} = AS^{fNIRS} \qquad (3.9)$$

Note that, the main advantages of jICA is that, it maximizes the likelihood between the data features from the two modalities. In this case, the data from both modalities were concatenated and joint independent component analysis just applied to maximize the likelihood between the two data feature sets. Assuming there are two sources per modality (EEG and fNIRS) and two subjects, the mixed data for the EEG are $X^{EEG} = [X_1^{EEG}, X_2^{EEG}]^T$ and the mixed tICA components or sources of the EEG are:

$$S^{EEG} = [S_1^{EEG}, S_2^{EEG}]^T, \qquad (3.10)$$

Each electrode of EEG represent individual component. Similarly, the mixed data for the fNIRS modality of the two subjects is presented by $X^{fNIRS} = [X_1^{fNIRS}, X_2^{fNIRS}]^T$ and the fNIRS sICA components are:

$$S^{fNIRS} = [S_1^{fNIRS}, S_2^{fNIRS}]^T, \qquad (3.11)$$

and the shared linear mixing matrix A is given by:

$$A = \begin{bmatrix} a_{11} & a_{12} \\ a_{21} & a_{22} \end{bmatrix}, \tag{3.12}$$

In order to form a data and source vector for each subject, the two data and sources set are concatenated together into a single vector. The concatenation forms single equation linked between the sources from both modalities as demonstrated in Equation 3.13.

$$\begin{bmatrix} X_1^{EEG} & X_1^{fNIRS} \\ X_2^{EEG} & X_2^{fNIRS} \end{bmatrix} = \begin{bmatrix} a_{11} & a_{12} \\ a_{21} & a_{22} \end{bmatrix} \begin{bmatrix} S_1^{EEG} & S_1^{fNIRS} \\ S_2^{EEG} & S_2^{fNIRS} \end{bmatrix}, \tag{3.13}$$

Additionally, to estimate the mixing matrix $A$, this study employed the infomax algorithm [256]. The algorithm is based on linear ICA model. This means that, the mixing matrix is considered to be time-invariant. However, for long time acquisition, the time invariant assumption is no longer valid. In this case, the method deals with time-invariant ICA problem by dividing the long time sequence into several short segment. The algorithm then applied to these segments batch by batch. In this study, the infomax algorithm uses a natural gradient ascent technique to maximize the output entropy of a neural network. In this context, entropy refers to the independence between the ICA components. The weight matrix of the neural network, $W$ refers to the inverse matrix of the shared mixing matrix, A. The optimization of the weight matrix is achieved by weight updating rule using the neural network as given:

$$\Delta W = \eta \{ I - 2y^{EEG}(S^{'EEG})^T - 2y^{fNIRS}(S^{'fNIRS})^T \} W, \tag{3.14}$$

$$S^{'EEG} = WX^{EEG}, \tag{3.15}$$

$$S^{'fNIRS} = WX^{fNIRS}, \tag{3.16}$$

$$y^{EEG} = g(S^{'EEG}) \tag{3.17}$$

$$y^{fNIRS} = g(S^{'fNIRS}), \tag{3.18}$$

$$g(x) = \frac{1}{1 + e^{-x}}, \tag{3.19}$$

where $I$ is the identity matrix, $S^{'EEG}$, $S^{'fNIRS}$ are the estimated independent sources of the EEG and the fNIRS, $y^{EEG}$ and $y^{fNIRS}$ are the regenerated EEG and fNIRS data respectively. From Eq. 3.19, $g(x)$ is the nonlinear transfer function in the neural network. The initial value for $W$, $W(0)$ is a matrix composed of random vectors. In this fusion approach, the study assumed that the sources associated with the EEG and the fNIRS data modulated the same way across all subjects. A normalization technique was applied to normalize the feature data between the range of 0 and 1 in both modalities. This assumption of common linear covariation for both modalities presents a parsimonious way to link multiple data types and has led to improved results in fMRI studies [254]. Exploring this property in a new type of imaging modality is an emerging approach and strongly needed for future studies. Furthermore, unlike the Calhoun model [254], this fusion technique is performed at the feature level in which features were extracted from each component of the EEG and fNIRS data using small shared-time-moving window of 500 ms.

Note that, the optimization of the weight across the data sources in both modalities helps in reducing the redundant of the features in each modality and maximize the likelihood across the two modality by minimizing the variance within each data set and maximize it in between. This process helps in estimating the data type later on in the classification process. The overall output of the fusion technique is a matrix consists of the regenerated data features from the associated components in which their likelihood were maximized. It could be concluded that, when associated components/sources are highly correlated, the task will be highly localize into that particular region. Exploring the source- correlation map matric across all individuals could help in the diagnosis of mental stress. Clear indication of the fusion process is as demonstrated in the flow chart in Fig 3.4. Note that, only highly correlated sources were used for classification performance of the data feature-fusion.

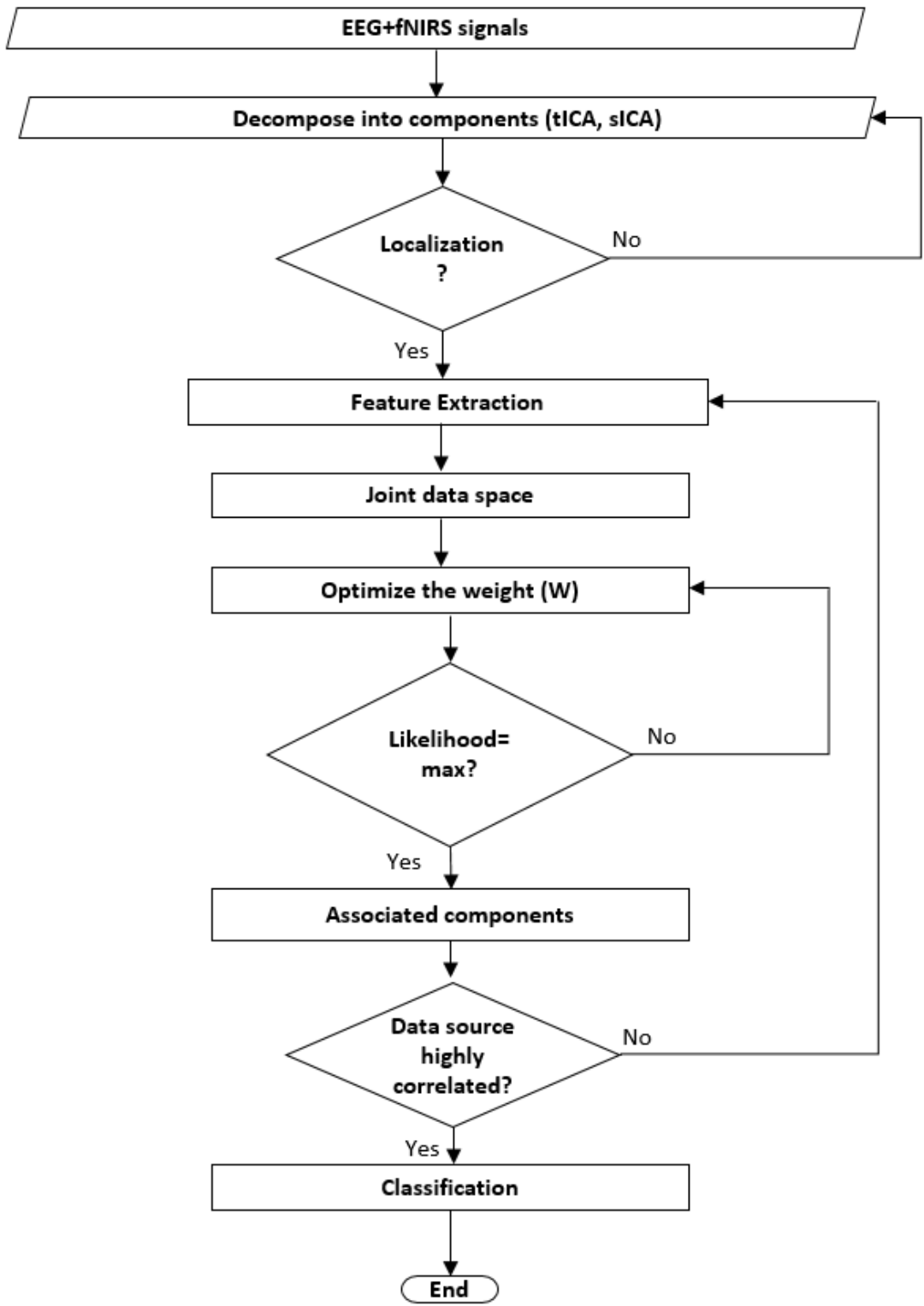

Figure 3.4: Flow chart of fusion based on joint independent component analysis.

### 3.3.2 Fusion Based on Canonical Correlation Analysis (CCA)

Canonical correlation analysis allows a different mixing matrix for EEG and fNIRS modality and used to find a transformed coordinate system that maximizes inter-subject covariation across the two data sets (one from EEG and the other one from fNIRS). This method decomposes each dataset into a set of components (such as spatial areas for fNIRS/or temporal segments for EEG) and their corresponding mixing profile, called canonical variates. The canonical variates have varying levels of activations for different subjects and are linked if they modulate similarly across subjects. After decomposition, the canonical variates correlate to each other only on the same indices and their corresponding correlation values are called canonical correlation coefficients. Compared to joint independent analysis that constrains two features to have the same mixing matrix, canonical correlation analysis is flexible in that it allows common as well as distinct level of connection between two features.

In this method, the EEG dataset comprised seven signal components (each corresponds to signal from one of the seven EEG electrodes in the alpha frequency band) and the fNIRS dataset had 23 signal components (each corresponds to single fNIRS channel). The main advantage in this method was that it does not give any priority to any modality as in the case of jICA method. In jICA, the functional near infrared was used to construct the spatial location for the EEG due to its fine spatial resolution. However, in canonical correlation analysis, the entire dataset or components were considered for evaluations. In this method, both datasets were preprocessed to extract features using a sliding window of 1 s, as described in Sections 3.2.2 (in this case the study used the wavelet family of Dubechies-2 (db2)) and 3.2.3. The canonical correlation analysis (CCA) of the EEG and the fNIRS data was performed at the feature level. Supposed that $X \in \mathbb{R}^{n \times p}$ and $Y \in \mathbb{R}^{n \times q}$ are two matrices, each contains $n$ observations with $p$ and $q$ feature-dimensions from the two modalities, respectively .The variable n here stands for the datapoint or features from each modality while the variable p and q are the number of electrodes in EEG and the number of channels in fNIRS data. Let $S_{xx} \in \mathbb{R}^{p \times p}$ and $S_{yy} \in \mathbb{R}^{q \times q}$ represent the within-sets covariance matrices and $S_{xy} \in \mathbb{R}^{p \times q}$ represent the between-set covariance matrix in which $S_{xy} = S_{yx}^{T}$. This study proposed the CCA method to derive a linear combination of canonical variates $X^* = W_x^T X$ and $Y^* = W_y^T Y$ that maximizes the pair-wise correlation across the two feature sets according to:

$$r(X^*, Y^*) = r(W_x^T X, W_y^T Y) = \frac{W_x^T S_{xy} W_y}{\sqrt{(W_x^T S_{xx} W_x)(W_y^T S_{yy} W_y)}} \tag{3.20}$$

where the canonical coefficients $W_x \in \mathbb{R}^p$ and $W_y \in \mathbb{R}^q$ are two arbitrary non-zero vectors and the solution involves constraining the two terms in the denominator to be equal to 1:

$$W_x^T S_{xx} W_x = W_y^T S_{yy} W_y = 1, \tag{3.21}$$

Note that the canonical variates are uncorrelated within each data set and have zero mean and unit variance. Additionally, these variates have nonzero correlation only in their corresponding indices. The maximization was achieved using Lagrange multipliers that solve the following optimization model [257]:

$$Model \begin{cases} \max \rho(X^*, Y^*), \\ W_x^T S_{xx} W_x = W_y^T S_{yy} W_y = 1, \\ W_x \in \mathbb{R}^p, W_y \in \mathbb{R}^q \end{cases} \tag{3.22}$$

Applying the Lagrange multiplier to the canonical correlation at Eq.3.22, the transformation can be obtained as:

$$L(X^*, Y^*) = L(W_x^T X, W_y^T Y) = W_x^T S_{xy} W_y - \frac{\lambda_1}{2}(W_x^T S_{xx} W_x - 1) - \frac{\lambda_2}{2}(W_y^T S_{yy} W_y - 1)$$
$$\tag{3.23}$$

where $\lambda_1$ and $\lambda_2$ are the Lagrange multipliers. Setting the partial derivatives of $L(X^*, Y^*)$ with respect to $W_x$ and $W_y$ equal to zero gives:

$$\frac{\partial L}{\partial W_x} = S_{xy} W_y - \lambda_1 S_{xx} W_x = 0 \tag{3.24}$$

$$\frac{\partial L}{\partial W_y} = S_{yx} W_x - \lambda_2 S_{yy} W_y = 0 \tag{3.25}$$

Lagrange multiplier optimization technique proposed in this study due to its simplicity and flexibility in building the optimizer network. Additionally, the technique has the advantages of finding local maxima or local minima with less constrains. Multiplying both sides of Eq. 3.24 and Eq. 3.25 with $W_x^T$ and $W_y^T$ respectively and taking into consideration the constrains in Eq.3.21, the equations can be simplified into:

$$\begin{cases} W_x^T S_{xy} W_y = \lambda_1 W_x^T S_{xx} W_x = \lambda_1 \\ W_y^T S_{yx} W_x = \lambda_2 W_y^T S_{yy} W_y = \lambda_2 \end{cases} \qquad (3.26)$$

Let $\lambda_1 = \lambda_2 = \lambda$ then

$$\rho(X^*, Y^*) = W_x^T S_{xy} W_y = W_y^T S_{yx} W_x = \lambda \qquad (3.27)$$

This shows that the Lagrange multipliers $\lambda_1$ and $\lambda_2$ are equal to the correlation coefficient of $W_x^T$ and $W_y^T$. Substitute to Eq.3.24 and Eq.3.25, the transformation matrices $W_x$ and $W_y$ can be found using the eigenvalue equations as proposed by [258]:

$$\begin{cases} S_{xx}^{-1} S_{xy} S_{yy}^{-1} S_{yx} W_x = \lambda^2 W_x \\ S_{yy}^{-1} S_{yx} S_{xx}^{-1} S_{xy} W_y = \lambda^2 W_y \end{cases} \qquad (3.28)$$

where $W_x$ and $W_y$ are the eigenvectors and $\lambda^2$ is a vector of eigenvalues or the square of the canonical correlations. The number of non-zero eigenvalues in each equation is $d = rank(S_{xy}) \leq \min(n, p, q)$, sorted in decreasing order, $\lambda_1 \geq \lambda_2 \geq \cdots \geq \lambda_d$. Eventually, fusion was performed by the concatenation of the transformed feature vectors within the associated components according to [255], using the following equation:

$$F = \begin{pmatrix} X^* \\ Y^* \end{pmatrix} = \begin{pmatrix} W_x^T X \\ W_y^T Y \end{pmatrix} = \begin{pmatrix} W_x & 0 \\ 0 & W_y \end{pmatrix}^T \begin{pmatrix} X \\ Y \end{pmatrix} \qquad (3.29)$$

where $F$ represents the canonical correlation discriminant features. The final form of fusion is given by a matrix that correlated the associated sources or components across all subjects. The higher the correlation is, the localized is the stress to that particular sources/components. From the overall matrix of associated components, classification accuracy, sensitivity, specificity and area under ROC were obtained using support vector machine describe in the next section. The overall process involved in the canonical correlation analysis fusion method is as demonstrated in the flow chart in Fig 3.5. Note that, small components were neglected and considered as a noise in the feature dimension. This confirms the feasibility of reducing the redundancy and eliminating unwanted signals/noise.

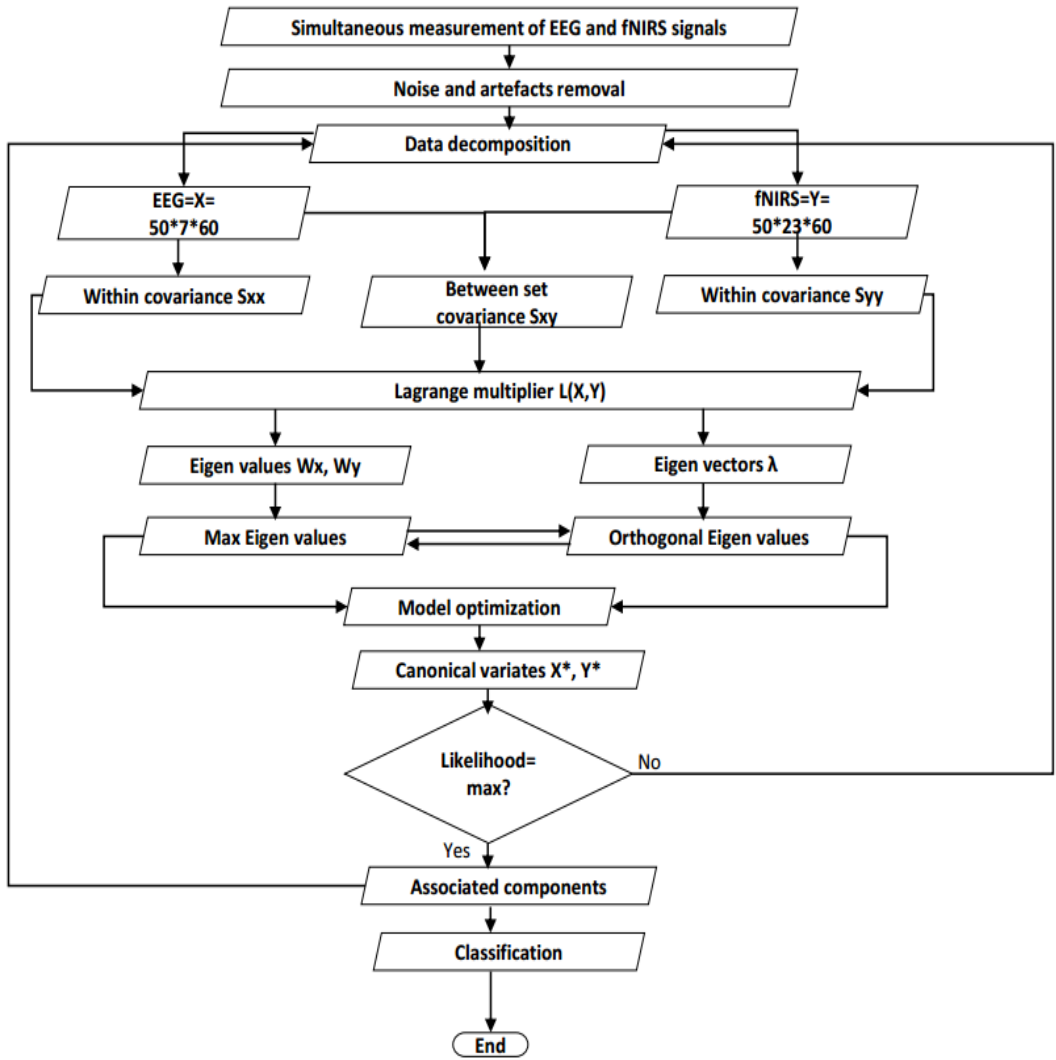

Figure 3.5: Flow chart of fusion based on canonical correlation analysis.

### 3.3.3 Feature selection and classification

The feature selection was done using statistical analysis of t-test. The t-test is one of the most used technique in feature selection and localization. In order to better understand the statistical significance of the stress effects on PFC activities, this study used a two sample t-test to measure the differences in task response (NASA-TLX, alpha amylase level, EEG and fNIRS signal features) between the control and the stress conditions. The differences in all the aforementioned responses (control versus stress) were considered statistically significant if *p value < 0.01*. Additionally, for the fNIRS responses, topographical maps of T-values were used to study the effects of mental stress on different PFC subregions. The use of T-value here is due to its reputability as a good indicator to local activation/deactivation. Unlike other studies in literature, this study defined empirically a threshold value of t > 3 to be considered as the significant change in the $O_2Hb$. The selection of t>3 is to obtain good focality point of the oxygenated hemoglobin to the stress on the PFC subregions.

In this study, support vector machine (SVM) is proposed as common classifier for the performance assessment of individual modality and the fusion of EEG and fNIRS modality. As described in Chapter 2, the SVM is a supervised machine learning technique widely used for classification, regression and density estimation. The selection of SVM in this study was due to its ability to model linear as well as more complex decision boundaries as the case of mental stress. The decision boundary hyperplane in the SVM was estimated based on its training dataset by maximizing the distance between the hyperplane of the control and stress data features to the nearest data point. LIBSVM software was used to build the SVM classifier in this study and employed the radial basis function (RBF) kernel to nonlinear map data onto a higher dimension space.

The EEG and fNIRS signals were computed at 0.5/1 Hz in features extraction stage as demonstrated in the flow charts of the EEG and fNIRS data. This gave 60 samples of EEG/fNIRS per analysis block. These samples were then fed as input features for the SVM classifier. This study, adopted leave-one-out approach for cross validation among the 25 subjects. The performance of the classifiers was then determined by computing the classification accuracy, sensitivity and specificity as described in. The accuracy was defined as the ability of the classifier to correctly identify the positive/stress and the negative/control results and can be evaluated using Eq 2.13.The sensitivity measures the classifier ability to correctly identify the positive result/stress and was calculated using Eq 2.14. Specificity on the other hand gives a measure of the classifier ability to identify the negative results/no-stress defined in Eq.2.15.

Additionally, this study analyzed the classification performance using receiver operating characteristic (ROC) curves that plot the sensitivity versus 1 minus the specificity. It also evaluated the area under the ROC curve (AUC) as a measure of a classifier's discriminatory power, which is insensitive to class distributions and the cost of misclassification (i.e. AUC=1 indicating perfect classification, whereas AUC=0.5 indicating that the classifier's result is not better than random guess).

## 3.4 EEG and fNIRS connectivity

In order to discover the network or functional connectivity within and between each PFC subregion, this study analyzed the EEG signals using the coherence function based on Fourier transform. The study investigated the effects of mental stress on the functional connectivity of intra-hemispheres (the pairs are FP1-F3, FP1-F7 and F7-F3; FP2-F4, FP2-F8 and F8-F4) and inter-hemispheres (the pairs are FP1-FP2, F3-F4 and F7-F8).The coherence function allows to find common frequencies and to evaluate the similarity of signals. However, it does not give any information about time. There are two often-used methods to calculate the coherence function, namely the Welch method and the MVDR (Minimum Variance Distortionless Response) method. This study used Welch's averaged modified periodogram technique to estimate the cross-spectral and power spectral density to obtain the squared coherence. The functional connectivity was then mapped based on the squared coherence threshold of 0.6. The coherence function is calculated as derived in Eq (2.10) in Chapter 2 in which, $S_{xx}$ and $S_{yy}$ are the power spectra of the EEG signals from the right and the left hemisphere, $S_{xy}$ is the cross-power spectrum of the signals, $\langle . \rangle$ indicates window averaging in the frequency band of 8 Hz- 13 Hz. The output of the coherence ranges between '0' and '1'. The higher the coherence value, the higher the connectivity.

Similarly, this study also investigated the effects of mental stress on the functional connectivity of inter-hemispheres (The pairs are Ch [1-3, 4-7, 5-6, 8-12, 9-11, 13-18, 14-17, 15-16, 19-23 and 20-22]) and the intra-hemispheric (Ch-[1,4,5,8,9,13-15,19,20] within the right hemispheric alone and Ch-[3,6,7,11,12,16-18,22,23] within the left hemispheric alone) PFC by calculating the squared coherence of the $O_2Hb$ between all channel pairs of participants. As described previously in the part of EEG, the study also applied Welch's averaged modified periodogram technique on the fNIRS signals to estimate the cross-spectral and power-spectral density in order to obtain the squared coherence on the time point of 3 minutes task (sampled at 10 Hz per second). The duration of time period in obtaining the functional connectivity is within the significant acquisition/duration time according to recent study in which significant rest-state functional connectivity can be obtained with short acquisition time of one minute and can be reproduced at less than seven to ten minutes [231]. More details on the functional properties has been discussed in details in Chapter 2. The functional connectivity in this study was then mapped based on the squared coherence threshold of ≥0.6 similarly as proposed by [212].

This study investigated the PFC functional connectivity within seven frequency intervals: I (0.009-0.02 Hz), II (0.02- 0.04 Hz), III (0.04-0.06 Hz),  IV (0.06- 0.08Hz), V (0.08-0.10 Hz), VI (0.0.009-0.1 Hz) and VII (0.1-0.8 Hz) [212]. This is due to that, the typical frequency band used in the fMRI field to assess functional connectivity is in the range of 0.01-0.1 Hz because other bands are contaminated by noise and physiological artifacts such as respiratory- and cardiac-related fluctuations in oxygen supply [259].

Different colors were used to indicate the various connectivity strengths in the coherence maps. All channel pairs in inter and intra-hemispheric were classified into three connection groups: (1) connectivity between the DLPFC regions, (2) connectivity between the VLPFC regions; and (3) connectivity between the Frontopolar area. The average coherence values were calculated for each connectivity type. To reveal the differences in connectivity between control and stress groups in inter and intra-hemispheric PFC with all the frequency intervals, the study analyzed the differences between them in channel pair basis. The differences were considered statistically significant if p value < 0.01.

### 3.5 Summary

This chapter describes in details the experimental protocol, the stress inducement procedure, the control of the simultaneous measurements of EEG and fNIRS systems, the data processing, the features extraction, statistical analysis and classifications. It also presents the proposed fusion technique that combines the EEG and the fNIRS signals for improving the detection rate of mental stress. Additionally, it also presents a new sight in evaluating the stress based on connectivity within inter and intra-hemispheric PFC areas. The work consists of five parts; first part deals with developing a technique to induce stress on the participants at the workplace. The stress inducement was based on established mental arithmetic task with time pressure and negative feedback (i.e. message of "correct', "incorrect" and "time's up") of peer performance. The stress inducement procedures was confirmed by measuring alpha amylase level from all the participants. In the second part, the control of simultaneous measurement was implemented in MATLAB and triggers were sent to both Discovery 24E system and OT-R40 system through parallel and serial ports to mark the start and the end of the task in each block. In the third part, EEG data were preprocessed using EEGLAB toolbox as well custom script-coding. The data were bandpass filtered in the range of 0.5- 30 Hz and eye movements and artefacts were removed using independent component analysis technique. The data were further processed using wavelet transform. Power features were then extracted from the wavelet coefficient at four frequency bands; delta, theta, alpha and beta. Two sample t-test was used to evaluate the significant differences between control and stress subjects. Similarly, fNIRS data were preprocessed using POTATo toolbox as well custom script-coding. Significant features were extracted from the mean oxygenated hemoglobin using time-window intervals.

In the fourth part, new algorithms for fusion were proposed to fuse EEG and fNIRS data in the feature-level fusion. One of the feature-level fusion was based on joint independent component analysis technique (jICA). The other one is based on canonical correlation analysis technique. In both techniques, EEG and fNIRS sources were transformed via ICA into components within their time record. The study then applied joint independent component analysis (jICA) to compute the mixing matrix A, between ICA components of EEG and ICA components of fNIRS using linear generative model. The mixing matrix A was then estimated using infomax algorithm which used a natural gradient ascent technique to maximize the output entropy of the neural network. The output of the fusion technique was then evaluated using support vector machine (SVM). The CCA is a statistical method to identify a linear relationship between two sets of variables by determining the inter-subject co-variances. The CCA works as a linear mixing model which maximizes the correlation between pairs of canonical variates, in this case, the features of brain response recorded by each modality (EEG/fNIRS) for individual subject. Based on these covariations, the study assess the association between EEG and fNIRS data to study the effects of mental stress on our working memory. Similarly as the first method, this method used SVM to classify the task and evaluate the performance of the algorithm. Lastly, the magnitude square coherence was implemented to study the cortical connectivity within inter and intra-hemispheric PFC areas under the control and stress conditions respectively.

CHAPTER 4

RESULTS AND DISCUSSION

This chapter presents the results and discusses in detail their significance to the field of study as well as to the research community. It discusses the results obtained by alpha amylase, EEG and fNIRS modalities separately and after fusion for the assessment of mental stress. It also discusses in detail the effects of stress on the cortical connectivity within the inter- and intra-hemispheric parts of the brain.

## 4.1 Experimental procedures

In the present study, the experiment was conducted in four phases. In the first phase, the inclusion and exclusion criteria for the samples of the study and the time for the measurements were identified. In the second phase, psychological stress stimuli developed in order to induce stress on the participants at the experimental workplace. In the third phase, an integrated cap that combine EEG electrodes and fNIRS optodes or fibres was developed, tested and validated before it can be used in the actual experiment. In the fourth phase, a technique to simultaneously record the EEG and the fNIRS signals while the participants completing a specific task under control and stress conditions was developed. The following subsections describe in detail the overall process.

### 4.1.1 Study sample

A total of twenty-five males, all of them being right-handed adults (aged 22±3, head size 54±2 cm), participated in the simultaneous measurement process of the EEG and fNIRS measurements. Smokers were excluded and measurements were limited to males subjects in order to avoid the increase in base amylase level [86]. All participants were informed prior to the experiment and gave written consent. None of these participants had a history of psychiatric or neurological disorders. Each was seated in a comfortable chair placed in a room with adequate air conditioning in order to avoid the influence of environmental stress. Each was asked to minimize his head movements and to keep calm throughout the entire experiment. The experiment protocol had been approved by a local ethical committee and was performed in accordance with the Declaration of Helsinki.

### 4.1.2 Stress stimuli

The stress task was designed based on the Montreal Imaging Stress Task (MIST) and demonstrated using a graphical user interface [260]. The experiment protocol was performed in four steps: Step 1: A brief introduction was given to the participants to familiarize them with the proposed tasks; Step 2: The participants were trained for five minutes in a mental arithmetic (MA) task in order to estimate the time it took each individual to answer each question. The task involved 3 one-digit integers (ranging from 0 to 9) and the operators were limited to + or – (for example 7-3+1). The answer for each question was displayed on a computer monitor in the sequence of '0' to '9' (as shown in Fig. 4.1) and each participant had to select the right answer by a single left-click on the mouse; Step 3 (i.e. the control phase): Simultaneous EEG and fNIRS measurements were performed for a total of five minutes without time limit per question. The participants were instructed to answer the questions as quickly and as accurately as possible and received no feedback on whether or not their answer was correct; Step 4 (i.e. stress phase): The average time recorded during the training phase was reduced by 10% and was set as a time limit. If answering wrongly or failing to answer a question within the time limit, the participant received a negative feedback in the form of "incorrect" or "time is up" being displayed on the monitor. Furthermore, the average peer performance set to 90% was displayed on the screen to further increase the stress on the participants. Actually, the participants were expected to score 40-50% when the time given to answer each question was reduced by 10%. The entire recording (control phase and stress phase) took a total duration of nearly 25 minutes, each phase consisting of five blocks of EEG and fNIRS recordings. Prior to each recording phase (control and stress), the baseline was measured for a total duration of 20 seconds. During the baseline recording, the participants were instructed to look at a fixation cross on the computer monitor and to get ready for the next task. Fig. 4.1 gives an overview of the block design of the task. In each block, a mental arithmetic task was introduced for 30 seconds followed by 20 seconds of rest. The red dashed-line marked the start of the task and the green dashed-line marked the end of the task (the marker was presented in every block). The stressors were based on the time pressure and the negative feedback of each individual performance as demonstrated in Fig 4.1(b). note that, five samples (S1-S5) of alpha amylase were collected; S1, five minutes before the control condition as baseline for control, S2 immediately after control condition, S3 five minutes before the stress condition, S4 immediately after stress condition, and S5 five minutes after stress condition as marked in the figure with the yellow rectangle.

During the experiment, all participants were instructed to answer each question correctly and not to guess the answer. In order to evaluate if the participants were paying due attention to the task, their accuracy in answering the questions was calculated. The average score was 90% accuracy in the control phase and 40% in the stress phase, as expected based on the original MIST article [260].

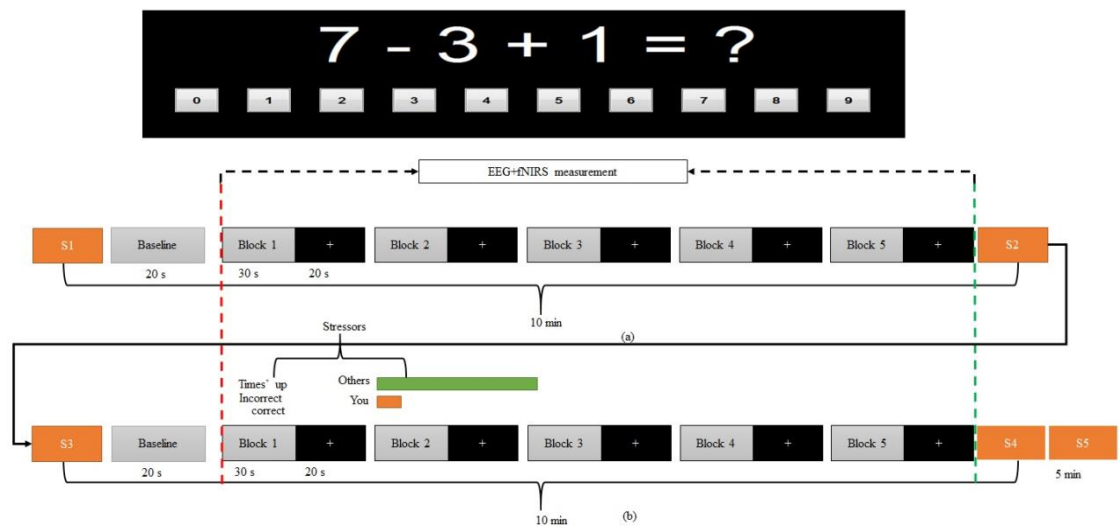

Figure 4.1: Experiment block design. A total of five active blocks existed for each of the (a) control and (b) stress condition.

### 4.1.3 Development of integrated cap

An integrated cap contains eight fNIRS sources or optodes and eight fNIRS detectors with seven active EEG electrodes designed and fabricated with AutoCAD software and printed using a 3D printer. The design was in line with the international 10-20 system of electrode placement. This research adopted the current optical topography (OT) holders that come with the OT-R40 machine and integrated the EEG electrodes. The holders consisted of a rubber base fitted with optode sockets. The rubber base had been designed such as that the distance between the source and detector was fixed at 30 mm. This ensured the optimal recording of the OT signals. Figure 4.2 shows the final printed layout of the designed integrated EEG+fNIRS probe holder. The red socket signifies where to place the OT optode-source and the blue socket where to insert the optode-detector of the OT system. The EEG electrodes were placed according to the 10-20 system placement as shown in Fig. 4.2, outer view. Fig 4.3 and Fig. 4.4 shows the final layout of the probe-holder within outer and inner views after placing the sources and detectors. As shown in these figures, the cap was designed ergonomically in which participants felt comfortably when placing the holder over their heads.

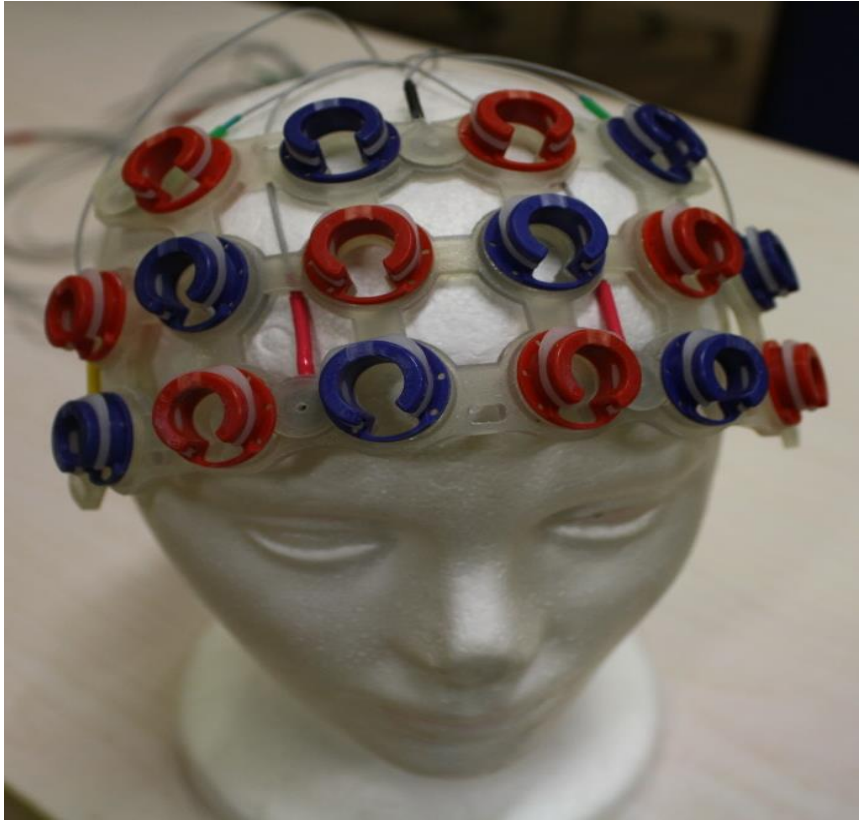

Figure 4.2: Outer-view of the designed integrated probe holder.

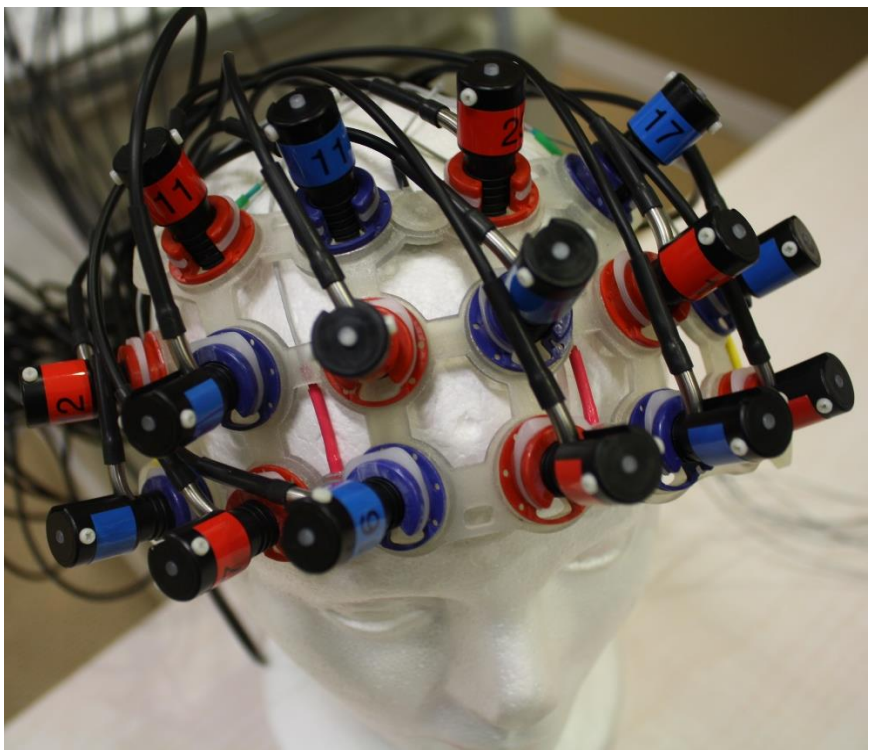

Figure 4.3: EEG+fNIRS electrodes and source-detector layout arranged
in the 10-20 system.

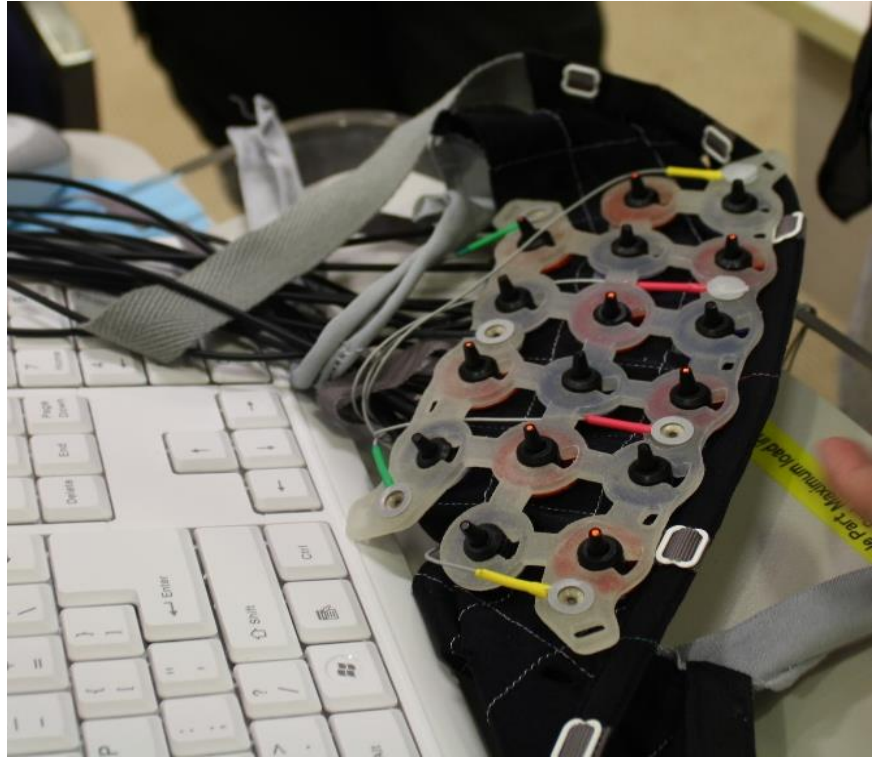

Figure 4.4: Inner view of the integrated probe holder.

### 4.1.4 Control of simultaneous measurement

The control of the simultaneous measurement was implemented in MATLAB in which triggers were sent to the Discovery 24E (BrainMaster Technologies Inc., Bedford, OH) system and the OT-R40 (Hitachi Medical Inc, Japan) system through parallel and serial ports in order to mark the start and the end of the task in each block of the mental arithmetic task. This was achieved by three connections. The first connection was the communication link between the EEG amplifier and the computer (Desktop used in this study) which was straightforward by using a 3m USB cable with choke for interface. The second communication link was established via the EEG amplifier. The Discovery 24E had two channels (Channel 23 and 24) available to receive markers from the external source, in this case the main measurement controller. The markers here referred to the signs used to indicate the start or the end of an experiment session. In order to create the link, the EEG amplifier was connected through a special cable (381-071) to a PC printer port. The cable was provided by BrainMaster and was optically isolated for safety and noise immunity. In order to send the marker to the EEG amplifier, the study used the MATLAB data acquisition

Toolbox to write a signal (either '1' as Start or '0' as End) to Channel 23/24 of EEG amplifier. The sample MATLAB code is as follows:

Figure 4.5 shows an example of the EEG recording by the BrainMaster acquisition software. The markers for 'Start' and 'End' are visible and distinguishable from each other, as highlighted within the sample 'A1' or 'A2'. These markers were essential to the subsequent processing of the EEG data in certain activity blocks. The other communication link was established between the OT-R40 and the main measurement controller. This was achieved using the RS232 interface of the OT-R40 and a USB-to RS232 converter.

The communication link was tested in advance using the OT-R40 interface. In this study, 'F9' was chosen to mark the start of a block and 'F7' to mark the end of a block. Similar to the EEG, these markers are important to allow users averaging blocks of brain signals for a particular task. Figure 4.6 shows an example of the OT-R40 markers within a single channel record while performing five blocks of the arithmetic task.

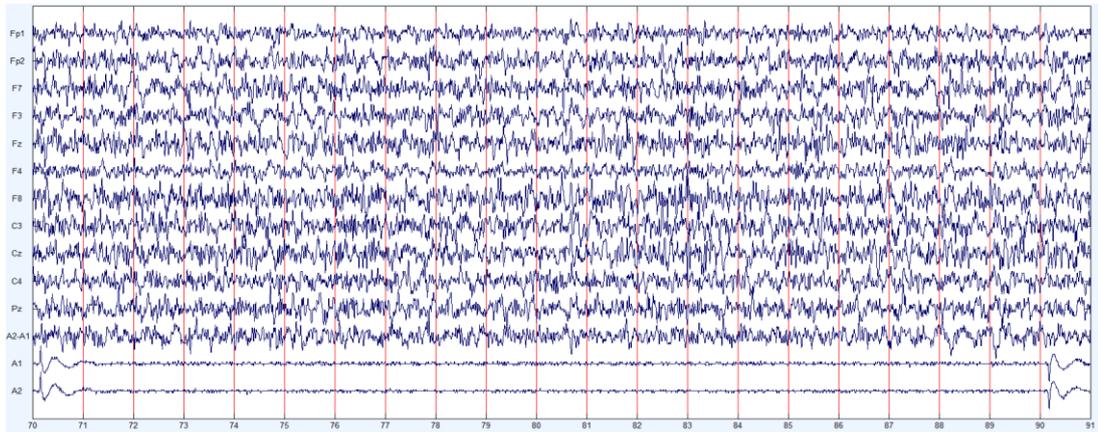

(a)

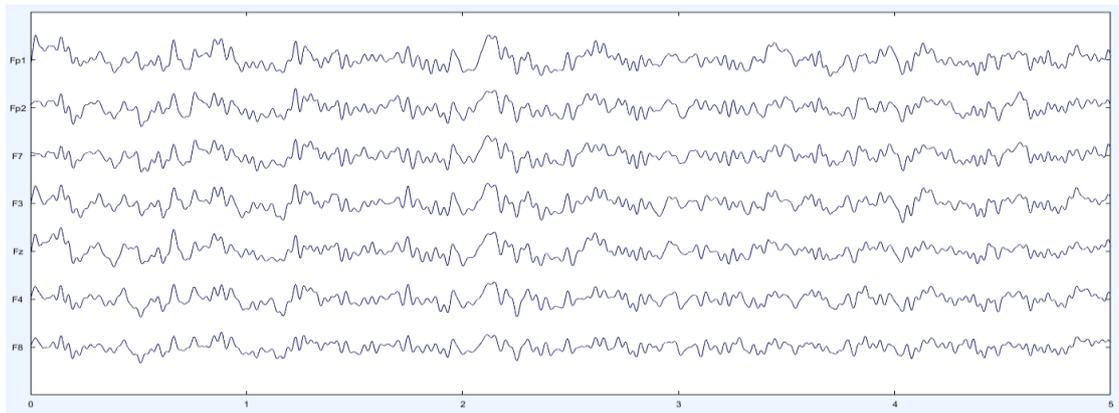

(b)

Figure 4.5: EEG markers (a) showing more channels, (b) 7-active EEG electrodes (clean signal).

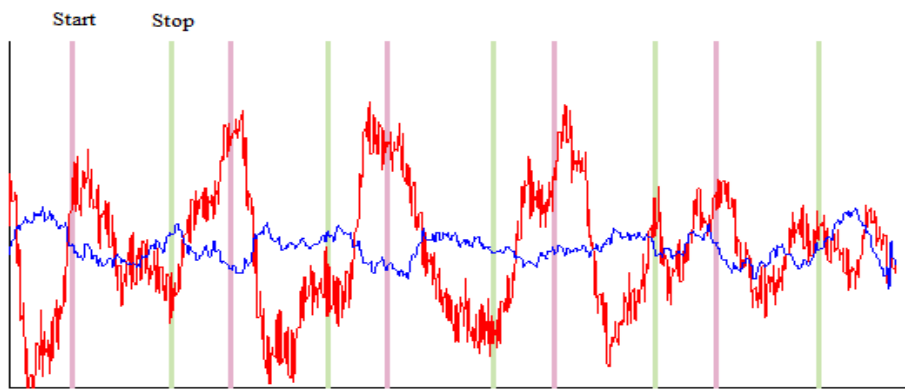

Figure 4.6: OT-R40 markers for the start and end of the task at each block.

## 4.2 Subjective measurements

The subjective assessment of mental stress was conducted using questionnaires, NASA-TLX [250]. Participants were asked to evaluate their mental workload before beginning with their task as baseline, after the simultaneous measurements of the EEG and the fNIRS under the control condition and after the simultaneous measurements of the EEG and the fNIRS under the stress condition. As for other subjective measures of workload, NASA-TLX relies on subjects' conscious perceived experience with regards to the effort produced and difficulty of task. NASA-TLX has the advantage of being quick and simple to administer. The index is a multidimensional method with various evaluation degrees, which provides a self-evaluation model to estimate workload through use of six subscales including (MD, mental demand; PD, physical demand; TD, temporal demand; OP, own performance; EF, effort; and FR, frustration). Each of these six subscales scores from 1to 20 based on performance of participants. The overall scores were then weighted to 100 and used for evaluation.

The results of subjective ratings of workload measured by NASA-TLX across control and stress conditions for all subscales and overall workload of the NASA-TLX have been summarized in Table 4.1. The weighted workload in the control and stress conditions were 22.2 and 75.4, respectively. Based on the participant's subjective responses, in control condition OP and in stress condition FR, TD and PD had dominant importance and in control and stress conditions EF had the lowest importance. Overall, there was a significant differences between the subjective score from control to stress condition. The p and t values are summarized in Table 4.1.

Table 4.1: Comparison of subjective variables Means ± SE across control and stress conditions

| NASA TLX | Control | Stress | T-test result | |
|---|---|---|---|---|
| | | | p-value | t-value |
| MD | 20±7.6 | 75±4.7 | <0.0001 | 6.30 |
| PD | 25±8.4 | 77±6.3 | <0.0001 | 6.41 |
| TD | 23±6.6 | 78±4.5 | <0.0001 | 6.31 |
| OP | 85±6.3 | 20±3.1 | <0.0001 | 7.41 |
| EF | 30±4.5 | 68±2.3 | <0.0001 | 5.81 |
| FR | 32±3.2 | 88±1.2 | <0.0001 | 7.62 |
| WWL | 22.2±2.5 | 75.4±3.7 | <0.0001 | 6.21 |

Abbreviations: MD, mental demand; PD, physical demand; TD, temporal demand; OP, own performance; EF, effort; FR, frustration; WWL, weighted workload; MW, mental workload;  T-test analysis; SE, standard error.

## 4.3 Objective measurements

This research measured mental stress using the three techniques of COCORO meter (Nipro, Osaka, Japan), EEG and fNIRS modalities. Due to the popularity of salivary alpha amylase in measuring stress levels (measured by COCORO meter) in clinical practice and behavioural studies, it used as a reference to confirm the induction procedure of stress on the participants. Similarly, the EEG and the fNIRS are proposed as a new approach to assess mental stress objectively as they complement each other. The procedures involved in the measurements are described in the following sub-sections.

### 4.3.1 Salivary alpha amylase collection

A hand-held monitor COCORO meter (Nipro, Osaka, Japan) was used to measure the salivary alpha amylase activity as shown in Fig 4.7 (a). Five samples of salivary alpha amylase were collected from each participant during the entire experiment. The first sample was collected five minutes before the beginning of the experiment as a baseline sample and labeled as (S1) as shown in Fig 4.8. The second sample was collected immediately at the end of control task and labeled as (S2). The third sample was collected five minutes after the control task as recovery of the control phase and labeled as (S3). The fourth sample was collected immediately after the stress task and labeled as (S4) as shown in Fig 4.9. The fifth sample was collected five minutes after the stress task and labeled as (S5). Note that in the analysis section, the third sample (S3) was also used as the baseline for the stress phase. The entire sequence of the salivary sample collection is as illustrated in Fig.4.7 (b).

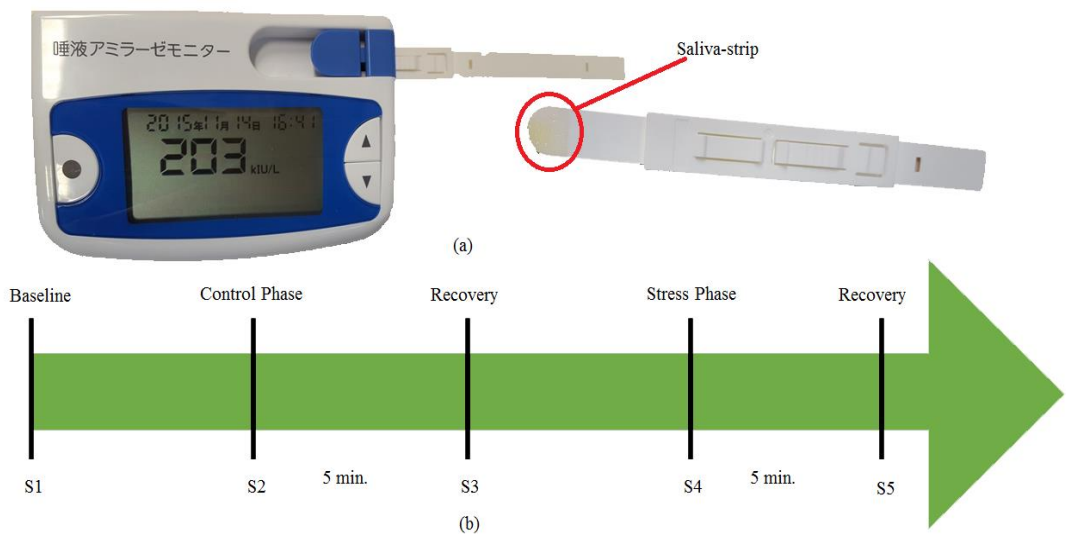

Figure 4.7. (a) COCORO meter interface and saliva-strip, (b) Graphical summary of salivary alpha amylase samples collection.

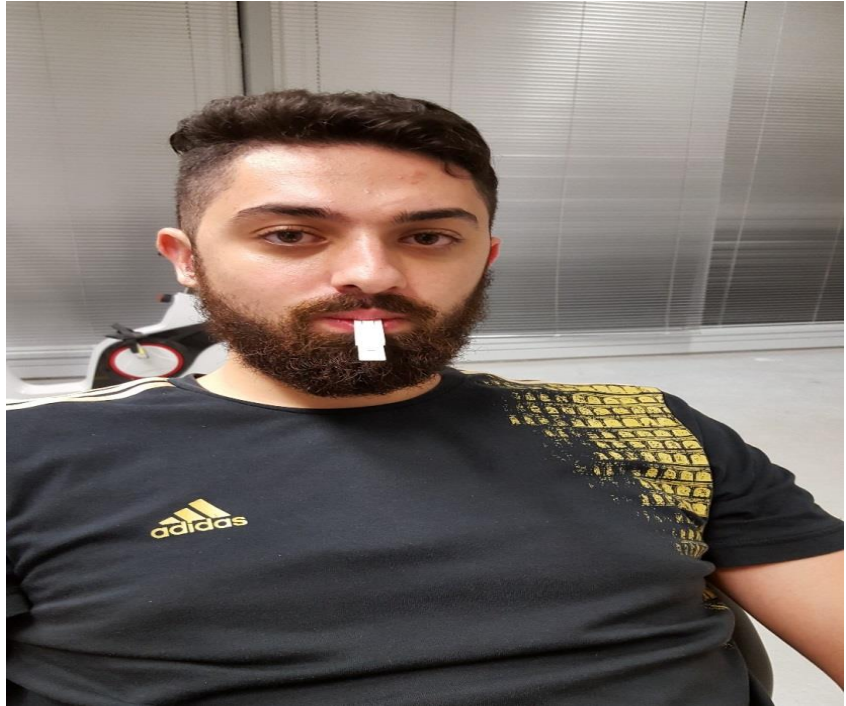

Figure 4.8: Alpha amylase collection at baseline stage.

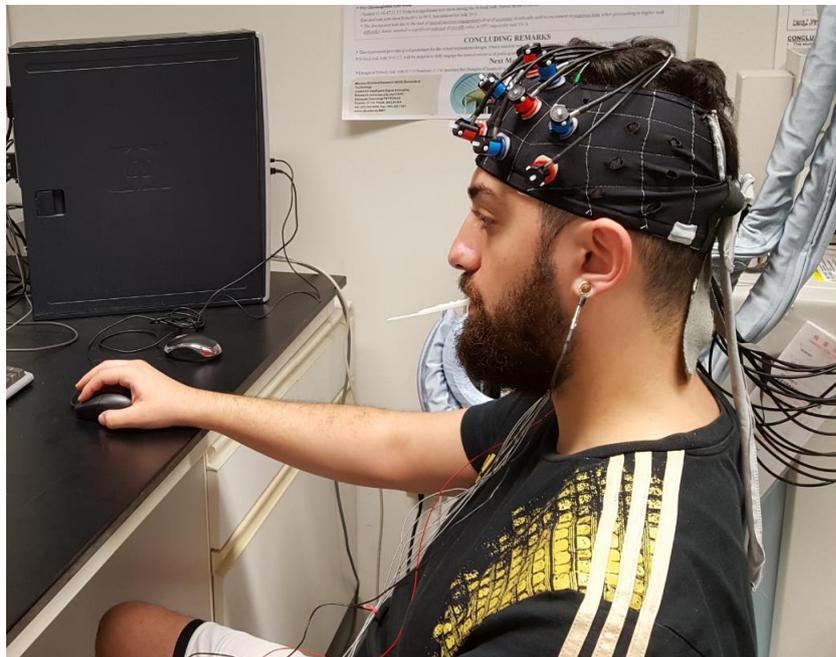

Figure 4.9: Alpha amylase collection immediately after the task condition.

The results of the alpha amylase level demonstrated a significant increase during the stress condition compared to control and rest conditions and return to the baseline after five minutes of rest. The increase in alpha amylase was noticeable consistently in all the subjects. At the baseline, the highest and lowest amylase level values were measured as 20 (kIU\L) and 3.0 (kIU\L) with a standard deviation of 4.6 (kIU\L) respectively. Similarly, under the control condition, the highest and lowest amylase level values were measured as 61 (kIU\L) and 45 (kIU\L) with a standard deviation of 6.56 (kIU\L) respectively. At the recovery after control and stress conditions (sample 3 and sample 5), the highest and lowest amylase level values were measured as 33 (kIU\L) and 19 (kIU\L) with a standard deviation of 4.01 (kIU\L) respectively. However, under the stress condition, the highest amylase level was measured as 120 (kIU\L), and the minimum value was measured as 71 (kIU\L) with a standard deviation of 13.6 (kIU\L). Figure 4.10 illustrates the overall results of the amylase levels of the five collected samples during the entire experiment. The study further analyzed the salivary alpha amylase responses using the two-sample t-test. The increase in the alpha amylase level under the stress condition was found to be significant as compared to the control condition, with a mean p-value of $p<0.001$. The significant differences in the alpha amylase level between the stress and the control condition confirmed the inducement of stress using the proposed task.

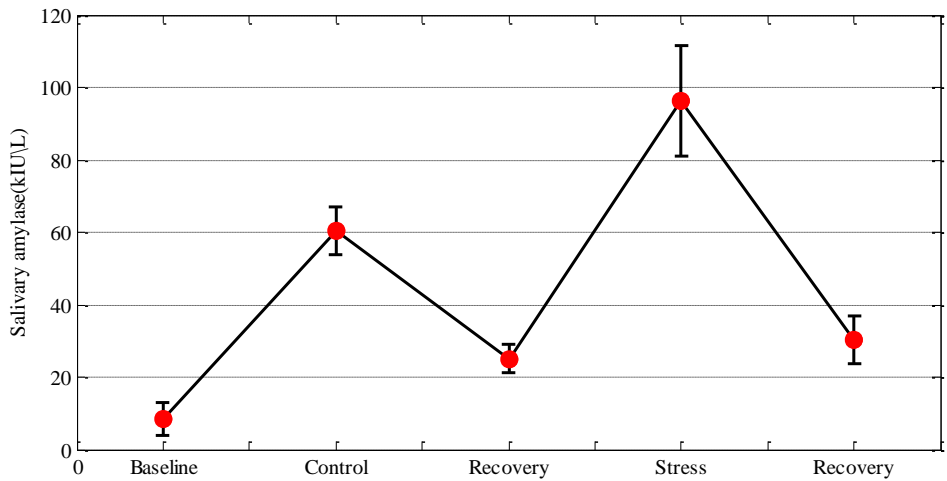

Figure 4.10: Alpha amylase samples collection.

### 4.3.2 EEG+fNIRS data acquisition

Simultaneous measurements of the EEG and the fNIRS were performed on the participants while they were solving mental arithmetic tasks under control and stress conditions. The data were simultaneously recorded using the Discovery 24E system (BrainMaster Technologies Inc, Bedford, OH) and the multi-channel fNIRS system (OT-R40, Hitachi Medical Corporation, Japan). The EEG system was equipped with seven active electrodes  placed on the positions FP1, F7, F3, Fz, FP2, F8, F4 and reference electrode 'A1-A2' attached to the earlobe according to the international 10-20 system. The sampling frequency for the EEG was set to 256 Hz and the impedance was reduced to 5K ohms by applying small amounts of gel directly to the scalp. The fNIRS system, on the other hand, was equipped with 16 optical fibres; eight sources (two wavelengths, 695 nm and 830 nm, combined in one source) and eight detectors. The distance between the pairs of the source and the detector probes was set to 3cm. The measurement area between a pair of source-detector probes was defined as channel (Ch). A total of 23 channels were recorded in this study as shown in Fig 4.11. The sampling frequency for the fNIRS system was fixed at 10 Hz.

The EEG electrodes and the fNIRS channels were co-registered into three right and three left PFC scalp quadrants (Frontopolar area (FPA): FP1, FP2, Ch-[9, 10, 11, 15, 16, 20, 21, and 22], the Ventrolateral prefrontal area (VLPFC): F7, F8, Ch-[8, 13, 14, 19, 12, 17, 18, and 23] and the Dorsolateral prefrontal area (DLPFC): F3, F4, Ch-[1-7]). The Placements of the EEG electrodes and the fNIRS channels are shown in Fig 4.10.  The character 'R' within the black box in the bold line stands for the right PFC area and the other character 'L' stands for the left PFC area. Figure 4.12 shows the overall experiment set-up.

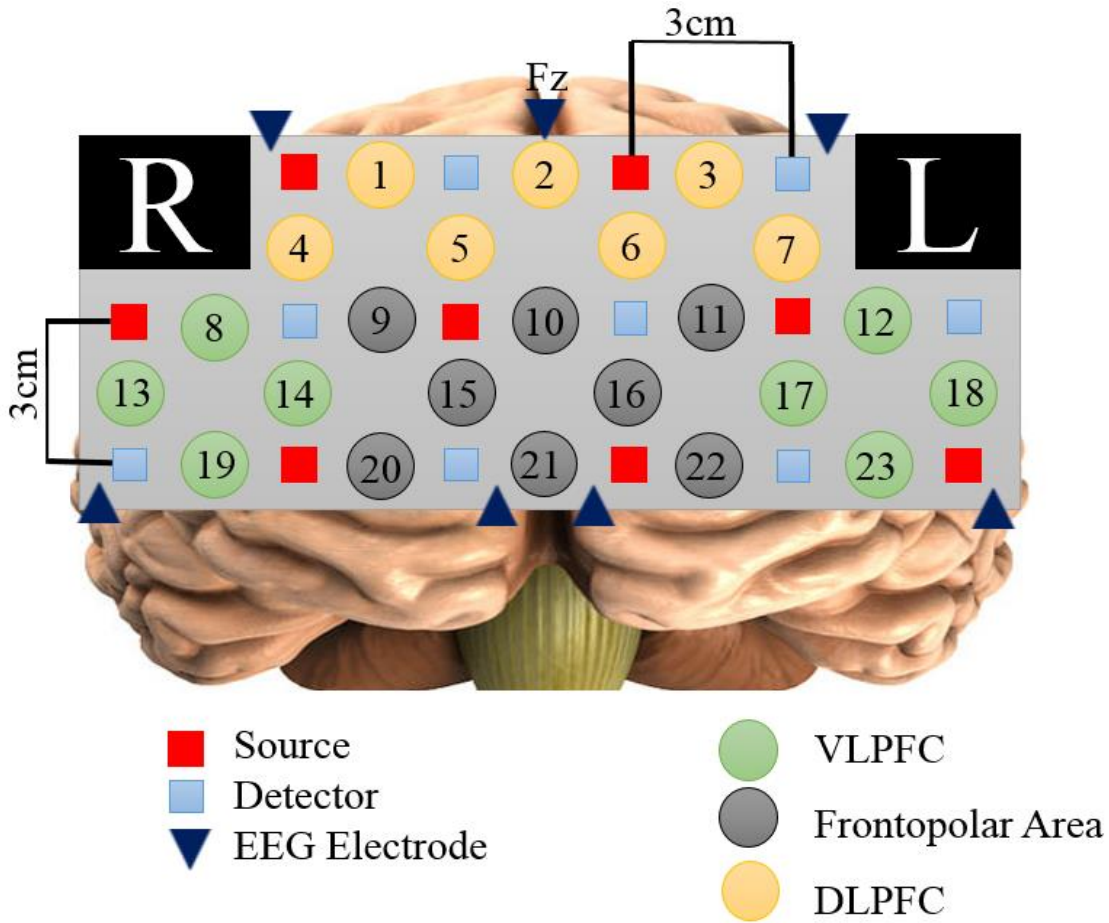

Figure 4.11: EEG+fNIRS electrode/channel Placement based on the international 10-20 system.

From the configurations of EEG electrodes and fNIRS channels, the subregions were highlighted based on channel locations. The Dorsolateral PFC highlighted with yellow circles, the Ventrolateral PFC highlighted with green circles and the Frontopolar area highlighted with grey circles.

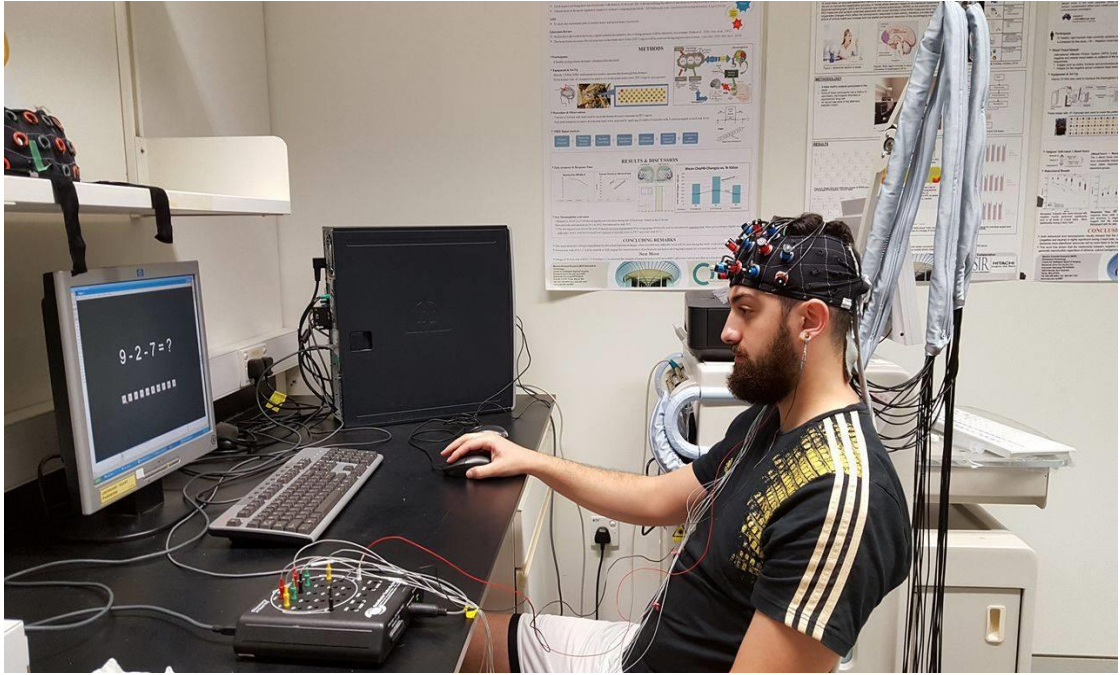

Figure 4.12: EEG+fNIRS experimental set-up.

## 4.4 Fusion of EEG+fNIRS using jICA

### 4.4.1 EEG result

In order to investigate the effects of the stress task on the brain fluctuations, the power values of the alpha and the beta rhythm for all the selected electrodes were calculated. The cortical activation of the brain during the stress task revealed an increase in the beta rhythm power and a decrease in the alpha rhythm power on the PFC. Figure 4.13 shows the boxplot representing the normalized mean power values of alpha and beta under the control and stress condition in all the 25 subjects. The statistical analysis showed that the brain response was significantly different under the stress condition from that of the control state as represented by features in beta and alpha rhythm. In order to show the contribution of each PFC subregion represented by the individual electrode in the beta and alpha band, the t-values and p-values were measured accordingly. For the beta rhythm, the measured t-value and p-value for FP1 was (t=1.8, p=0.067), for FP2 was (t=4.1, p=0.0013), for F3 was (t=2.5, p=0.020), for F4 was (t=3.8, p=0.002), for Fz was (t=3.1, p=0.005), for F7 was (t=3.0, p=0.005), and for F8 was (t=1.7, p=0.070) respectively. Similarly, the measured t-value and p-value in the alpha rhythm for FP was (t=2.89, p=0.008), for FP2 was (t=3.3, p=0.003), for F3 was (t=5.3, p=0.000), for F4 was (t=6.4, p=0.000), for Fz was (t=2.9, p=0.007), for F7 was (t=2.9, p=0.001), and for F8 was (t=5.4, p=0.000) respectively. The overall result demonstrated that the alpha rhythm responded more significantly to mental stress as compared to the beta rhythm. This can be clearly seen as highlighted in Fig 4.13. The result suggests that the alpha rhythm may be a better indicator of mental stress. The overall statistical analysis is shown in Table 4.2.

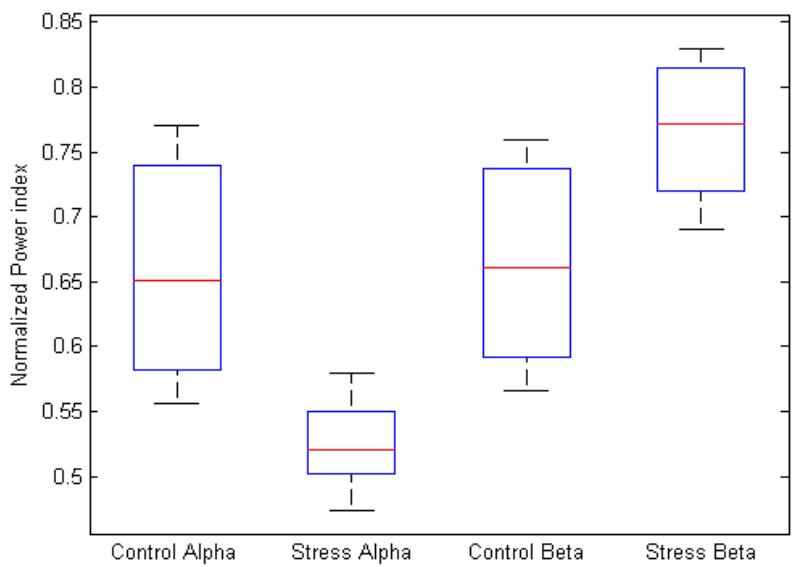

Figure 4.13: The normalized Alpha and Beta rhythm power values in two mental states: control and stress for average of 25-subjects.

### 4.4.2 fNIRS result

The fNIRS result demonstrated a dramatic increase in the oxygenated hemoglobin concentration $O_2Hb$ during control condition as compared to the baseline and stress condition in most PFC regions. The grand average time-course of oxygenated hemoglobin concentration $O_2Hb$ within all the channels' locations during the control condition is shown in Figure 4.11. The red line indicates the grand average time-course of oxygenated hemoglobin concentration $O_2Hb$, and the blue line indicates the grand average time-course of deoxygenated hemoglobin concentration HHb respectively. As expected, the increase in the oxygenated hemoglobin concentration $O_2Hb$ is associated with a slight decrease in the deoxygenated hemoglobin concentration. The association in the hemodynamic responses (increase in oxygenated and decrease in the deoxygenated hemoglobin) confirmed the cerebral activities related directly to the task, and the less noisy were the collected signals. From the time-course, it is also noticeable that the oxygenated hemoglobin concentration returned to the baseline at the 20 second rest condition after the task. In contrast, less cortical activation and in some cases a strong decrease in the oxygenated hemoglobin concentration $O_2Hb$ from the baseline was observed under the stress condition. The strong decrease in the oxygenated hemoglobin concentration under the stress condition was found on the right PFC region. Obviously, there was an increase in the oxygenated hemoglobin concentration during stress in some locations on the left PFC. The channels labeled as 10, 11, 16, 17, 21 and 22, for example, show an increase in the oxygenated hemoglobin concentration under stress, yet not significantly so. The grand average time-course of the oxygenated hemoglobin concentration $O_2Hb$ (red line) and the deoxygenated hemoglobin concentration HHb (blue line) is shown in Fig 4.14, the control condition and Fig 4.15, the stress condition. The vertical red dash-line marks the start of the task and the vertical green dash-line marks the end of the task.

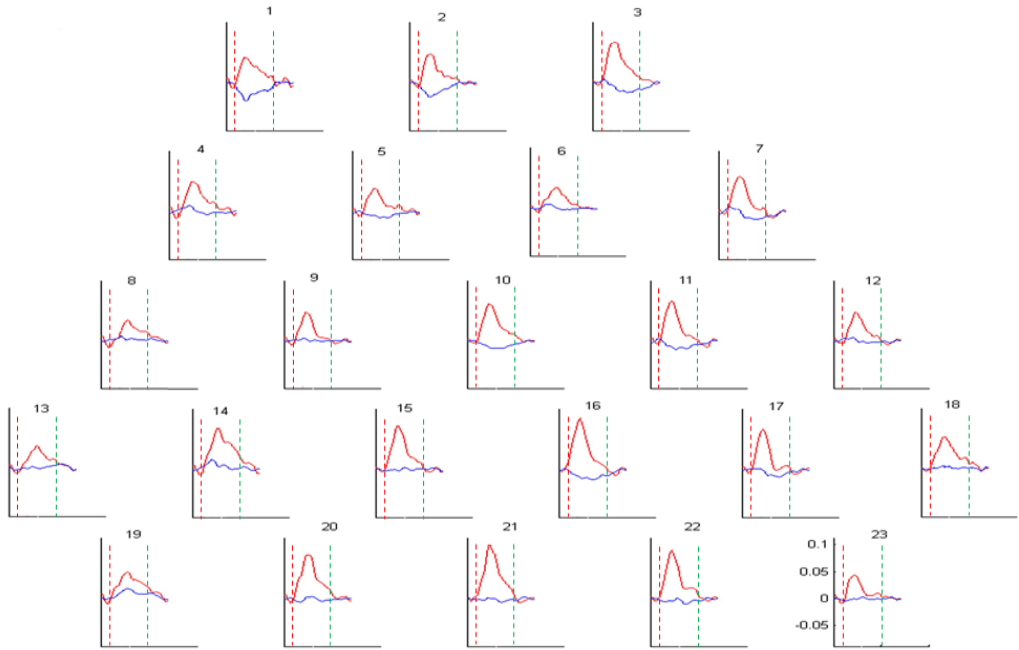

Figure 4.14: Mean time-courses of $O_2Hb$ (red line) and HHb (blue line) under the control condition.

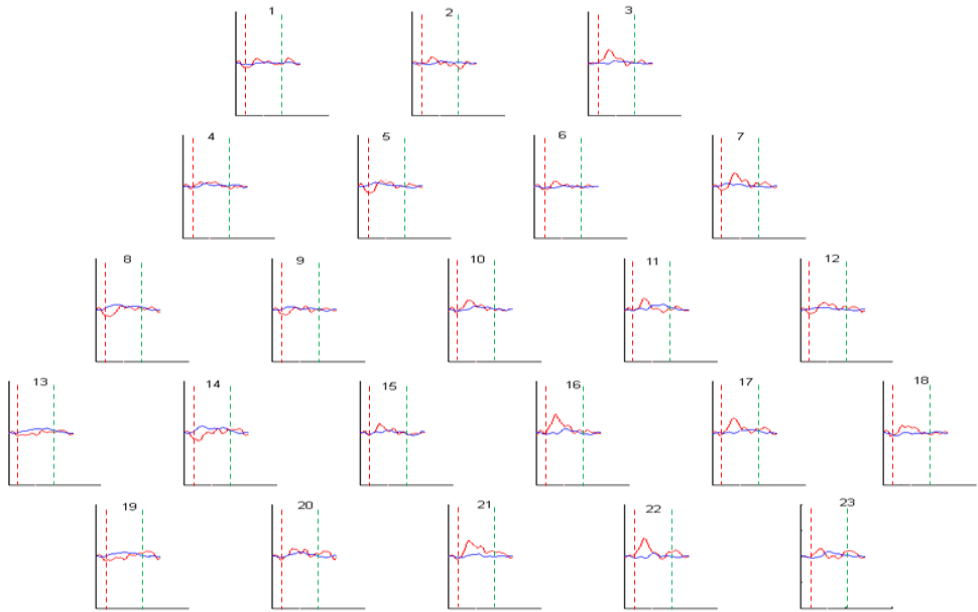

Figure 4.15: Mean time-courses of O2Hb (red line) and HHb (blue line) under the stress condition.

The overall behaviour of the PFC regions under control and stress conditions is presented in the topographical maps of Fig 4.16 and Fig 4.17 respectively. The topographical maps of oxygenated hemoglobin concentration O2Hb response measured as the averaged mean concentration from all the subjects across all channels with channel number labeling from '1' to '23'. When examined closer, the study found, on average, the decrease in the oxygenated hemoglobin concentration O2Hb response was highly localized to the right PFC. The topographical map of oxygenated hemoglobin and deoxygenated hemoglobin activation under control the condition are reported for the average of 25-subjects. The red coloring in the topography maps indicate higher activation and the blue coloring indicates less activation in cortical activities.

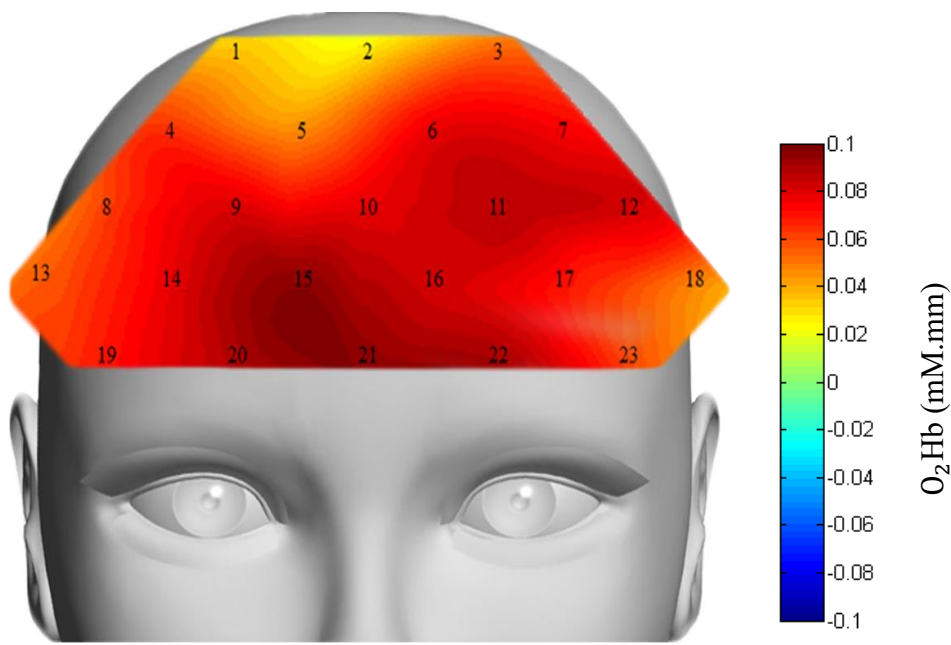

Figure 4.16: Topographical map of O$_2$Hb under control condition.

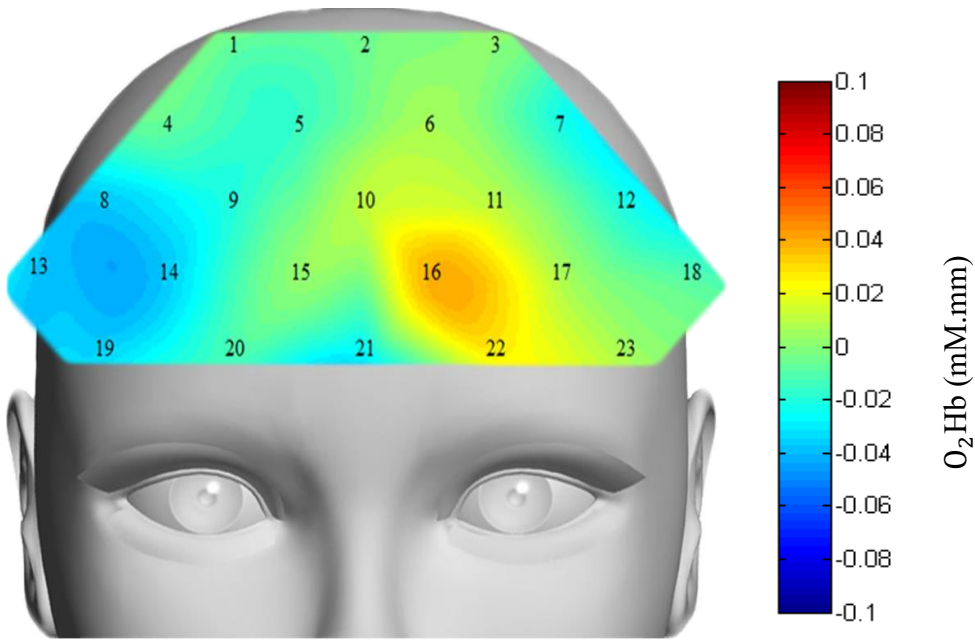

Figure 4.17: Topographical map of HHb under the stress condition.

The statistical analysis demonstrated a significant reduction in the oxygenated hemoglobin concentration from the control condition to the stress condition in all the subjects across most of the channels in the PFC. This thesis studied the contribution of each region on the entire PFC by measuring their t-value and p-values as means for statistical evaluation. The calculated t-value and p-value for all the channels. The overall statistical analysis for the EEG and the fNIRS signal modalities is summarized in Table 4.2.

The statistical analysis reported in terms of electrode naming, F4A which represents the EEG electrode of F4 in the Alpha band, and F4B represents the F4 EEG electrode in the Beta band. Similar sequence used for the rest of electrodes in which the character 'A' is used to represent the statistical name in the alpha band and the character 'B' used to represent the statistical name at beta band respectively.

Table 4.2: Statistical analysis of EEG alpha and beta and O2Hb of fNIRS measurements based on the two-sample t-test.

| Channel No | t-value | p-value | Channel No | t-value | p-value | Electrode | t-value | p-value |
|---|---|---|---|---|---|---|---|---|
| 1 | 4.3 | 0.0016 | 14 | 5.5 | 0.0000 | Fz Alpha | 2.9 | 0.0073 |
| 2 | 2.2 | 0.0370 | 15 | 5.3 | 0.0000 | Fz Beta | 3.1 | 0.0058 |
| 3 | 3.8 | 0.0022 | 16 | 2.1 | 0.0496 | F3 A | 5.3 | 0.0000 |
| 4 | 5.6 | 0.0000 | 17 | 3.0 | 0.0070 | F3 B | 2.5 | 0.0200 |
| 5 | 3.35 | 0.0033 | 18 | 5.2 | 0.0000 | F8 A | 5.4 | 0.0000 |
| 6 | 4.8 | 0.0002 | 19 | 5.0 | 0.0000 | F8B | 1.7 | 0.0700 |
| 7 | 2.6 | 0.0231 | 20 | 2.8 | 0.0011 | FP2 A | 3.3 | 0.0030 |
| 8 | 3.3 | 0.0038 | 21 | 1.2 | 0.2500 | FP2 B | 4.1 | 0.0013 |
| 9 | 5.5 | 0.0000 | 22 | 0.7 | 0.4193 | FP1 A | 2.89 | 0.0080 |
| 10 | 3.1 | 0.0059 | 23 | 5.1 | 0.0001 | FP1B | 1.8 | 0.0670 |
| 11 | 1.1 | 0.2904 | F4 A | 6.4 | 0.0000 | F7 A | 2.9 | 0.0010 |
| 12 | 2.3 | 0.0355 | F4 B | 3.8 | 0.0021 | F7 B | 3.0 | 0.0051 |
| 13 | 5.1 | 0.0000 | | | | | | |

From the overall statistical analysis on all the EEG electrodes and all the fNIRS channels summarized in Table 4.2, the EEG electrodes and the fNIRS channels of interest were identified to be used for performance evaluation and classification. For the EEG modality, six electrodes with the highest t values were selected; F4, FP2, F8, Fz, F3 and F7. Since the alpha rhythm responded more significantly to stress than the beta rhythm, only the alpha rhythm features considered for further analysis. For the fNIRS modality, since most of the channels responded significantly to the task, a threshold value of $t \geq 5$ was considered for the channel selection criteria. Eight channels responded above the threshold value, with six channels located on the right PFC and two channels located on the left PFC, namely Ch4, Ch9, Ch13,Ch14, Ch15, Ch19, Ch18 and Ch23. The six channels with the highest t-values were then selected for the fNIRS performance evaluation and classification.

For the EEG and the fNIRS fusion, the number of features for the performance evaluation was maintained the same as individual modality. Three fNIRS channels with the highest t values were selected, together with the EEG electrodes located in the closest proximity to these three fNIRS channels. The three fNIRS+EEG pairs are:

Ch-4 with F4, Ch-15 with FP2, and Ch-18 with F7, and their signals were used as inputs to the proposed jICA model. Note that, the lateralization was considered when selecting the channels and electrodes for fusion due to its dominant in stress studies. The fusion model was then evaluated using SVM in the same manner as the unimodal EEG and fNIRS.

### 4.4.3 Joint independent component analysis result

The result from the joint independent component analysis (jICA) is presented in a matrix map for all the 25-subjects as shown in Fig 4.18. The map clearly shows the potential of the fNIRS in constructing source-localization for the EEG in estimating stress. The obtained map indicates that the jICA emphasizes the consistency of the activation region almost across all subjects.

The classification accuracy obtained using the SVM classifier is shown in Figure 4.19. The classification results are illustrated as boxplot for all the three modalities, namely EEG, fNIRS and EEG+fNIRS. The boxplot demonstrates the distribution of SVM accuracies across all subjects. The accuracy is reported in term of its occurrence within the maximum values, at third quartile (75%), at median value (position at the center), at its first quartile (25%) and at its minimum value. For the sole EEG, the average classification accuracy was measured as 91.7±5.3%, for the sole fNIRS the average classification accuracy was measured as 84.1±6.8%, and under the fusion of EEG+fNIRS the measured accuracy was 95.1±3.9% for the average of the 25 subjects respectively. Note that, the maximum accuracy value achieved by the three modalities are the same at 100%. At third quartile, median, first quartile and minimum accuracy value, fusion outperforms sole EEG and sole fNIRS with p<0.001. From the overall classification accuracy result, it is clearly seen that the fusion of EEG+fNIRS increased not only for the average of the 25 subjects but also for each individual subject as demonstrated by the small error in the boxplot. The fusion increased the accuracy by an average of +3.4% compared to the sole EEG and +11.0% compared to the sole fNIRS. The improvement was found to be significant, p<0.001 as measured by the t-test.

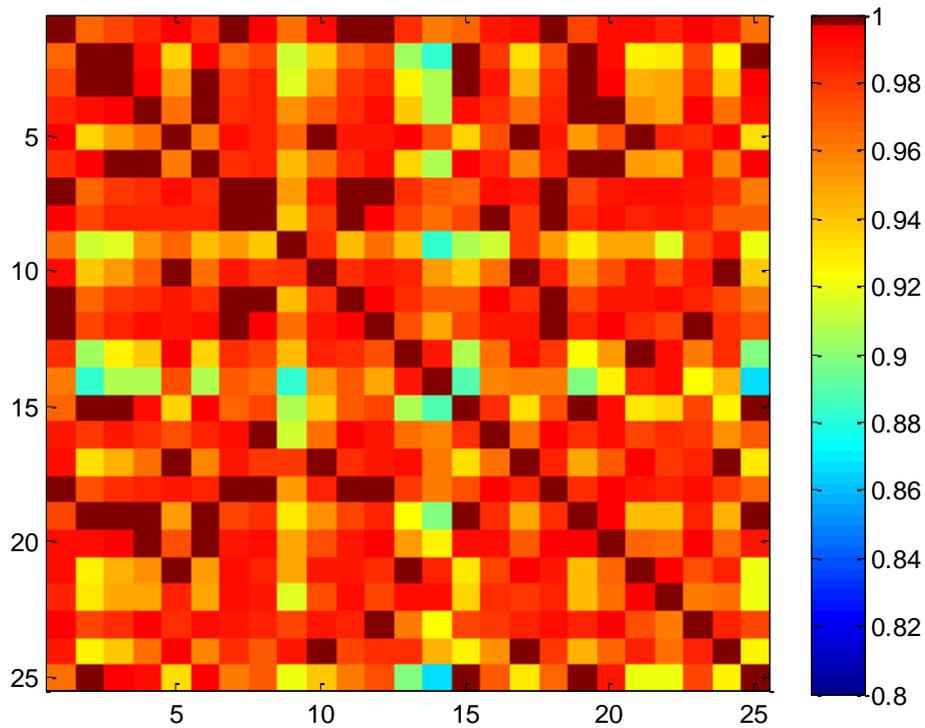

Figure 4.18: Source matrix across all subjects obtained through joint independent
component analysis.

Similarly, the classification results on the sensitivity and the specificity for each individual as well as the fusion of both modalities are shown in Figure 4.20 and Figure 4.21 respectively. The average classification sensitivities of the sole EEG, the sole fNIRS and the fusion of EEG+fNIRS were 90.4±5.7%, 82.4±6.3% and 94.2±4.3%, respectively. There were significant improvements (p<0.001) in the sensitivity of the fusion approach (EEG+fNIRS) compared to the unimodals with an average improvement of +3.8% compared to the sole EEG and +11.8% compared to the sole fNIRS, respectively. The average classification specificities of the sole EEG, the sole fNIRS and the fusion of EEG+fNIRS, on the other hand, were 93.4±4.4%, 86±7.2% and 96.6±2.8%, respectively. Significant improvements due to the fusion over the individual modality were observed, as expected, with mean a p-value of p<0.001.

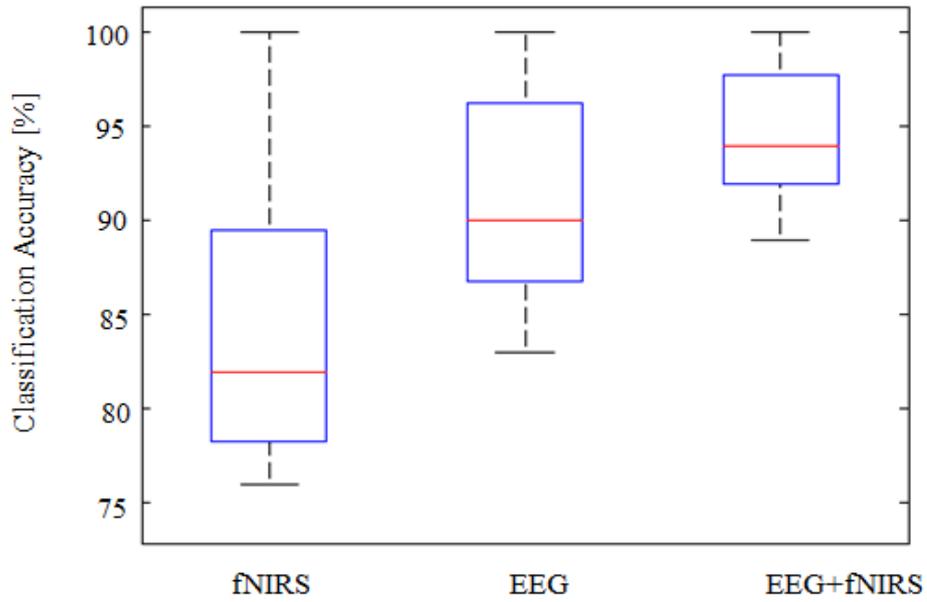

Figure 4.19: Boxplots representing the classification accuracy measured by SVM for 25 subjects.

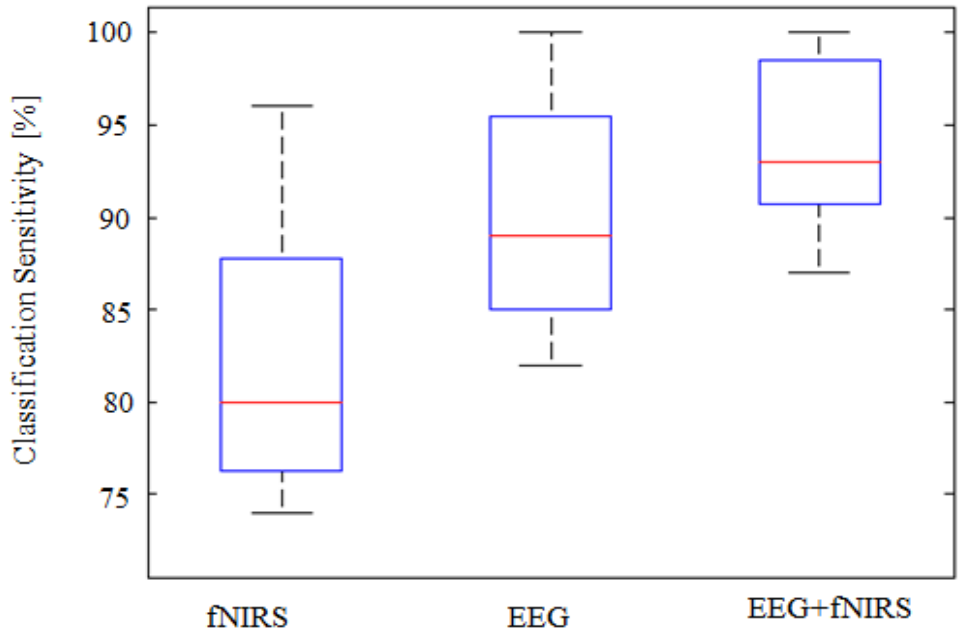

Figure 4.20: Boxplots representing the classification sensitivity calculated for 25 subjects.

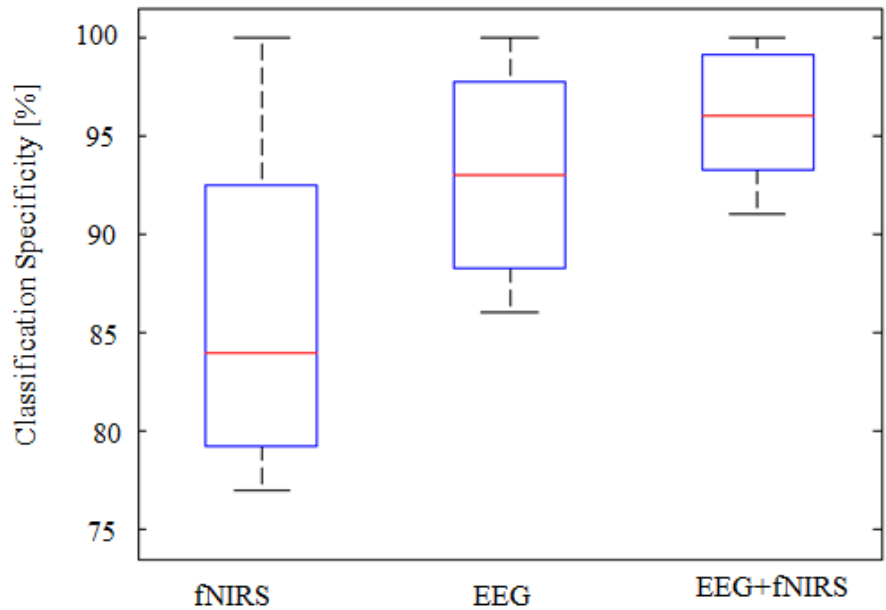

Figure 4.21: Boxplots representing the classification specificity calculated for 25 subjects.

### 4.4.4 Discussion on the results

The experiment results revealed that significant improvements were achieved through the joint independent component analysis fusion approach in all three performance metrics, with p<0.001. This confirmed our hypothesis that the fusion of the EEG and the fNIRS components can improve the mental stress detection. Undoubtedly, the proposed feature-fusion model enabled us to take the advantages of the strengths of both modalities in the unified analysis. The improvements in the classification accuracy achieved in this study is also consistent with the existing EEG and fNIRS hybrid studies but not in all cases [77-79]. The inconsistency in the hybrid results may due to the level of integration or fusion being adopted. Both Fazli and Putze applied fusion at the decision level, i.e. using a meta-classifier to integrate the outputs from one EEG classifier and one fNIRS classifier. It is likely that the outputs from the EEG classifier and the fNIRS classifier were highly correlated with little complementary information. In contrast, Yin was able to consistently improve the decoding of the motor imagery tasks when considering the feature level fusion of the bimodal EEG and fNIRS [233].

## 4.5 Fusion of EEG+fNIRS using CCA

### 4.5.1 Results of individual modality

The EEG result shows a decrease in the alpha rhythm from the control condition to the stress condition in all seven electrodes for all the participants. Figure 4.22 shows the averaged normalized alpha rhythm under the control and the stress conditions of all subjects at all EEG electrodes. The statistical analysis demonstrated significant differences in alpha rhythms between control and stress conditions in all the electrodes with mean p-values of <0.01 as the case in 'F7'; p-value of <0.001 as the case in 'FP1','FP2', 'F3' and 'Fz'; and p-value of <0.0001 as the case in 'F4', and 'F8', respectively. The changes in alpha suppression indicates that, the PFC response differently to situations where arithmetic task is presented with the time pressure and negative feedback. This could be due to the fact that, stressful task with the time pressure and feedback induced negative emotion and impairs the performance of working memory in which firing rate of neurons reduced.

Similarly, the results from the fNIRS show a decrease in the concentration change of oxygenated hemoglobin from the control condition to the stress condition, and the decrease was found consistently across all subjects with mean *p-values* < 0.01. Figure 4.23 to Figure 4.26 shows the topographical maps of four subjects where the first subject exhibits a higher cortical activation at the left DLPFC region as compared to the other three on the right FPA, under the control condition. Note that the label of channels on the PFC subregions is similar to the label in Fig 4.11. On the other hand, Figures 4.27 to 4.30 show the topographical maps of the same subjects under the stress condition. The right VLPFC is consistently the region with the least oxygenation under the stress condition among the participants. Figures 4.31 to 4.33 show the topographical maps for the average of the 25 subjects under the control and the stress condition with their corresponding t-map respectively. The calculated t-values were reconstructed to generate T-map using bicubic interpolation function developed by Sutoko et. al., and is built in the Platform for Optical Topography Analysis Tool [33]. From the overall topographical maps of all the subjects in this study, it can be clearly seen that the reduction in brain activation under stress is localized to a specific PFC subregion – right VLPFC, instead of being distributed across the entire PFC.

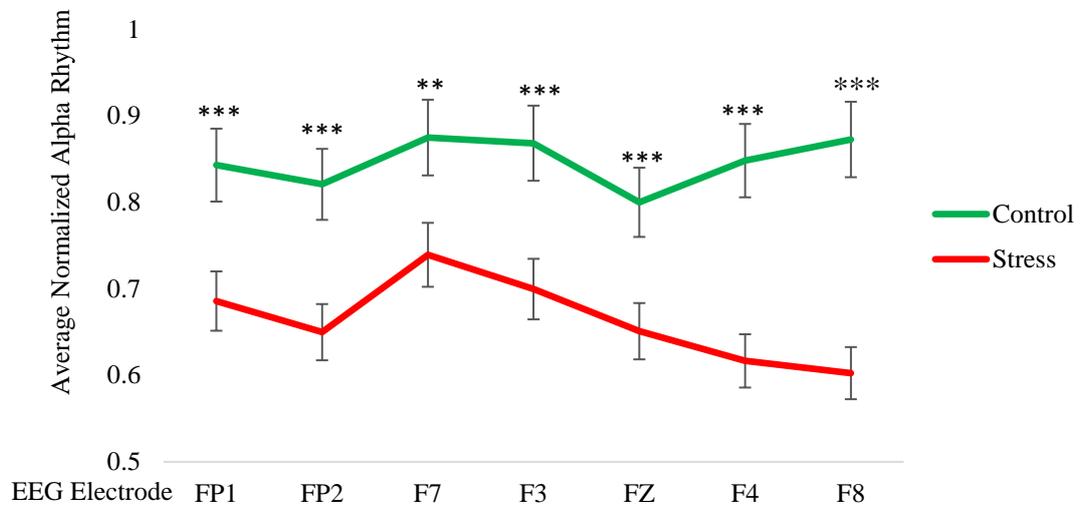

Figure 4.22: Normalized alpha rhythm under the control (green line) and the stress condition (red line) with seven corresponding EEG electrodes located at the PFC area. The marks '**', '***' and '****' indicate that, the task is significant with p<0.01, p<0.001 and p<0.0001 respectively.

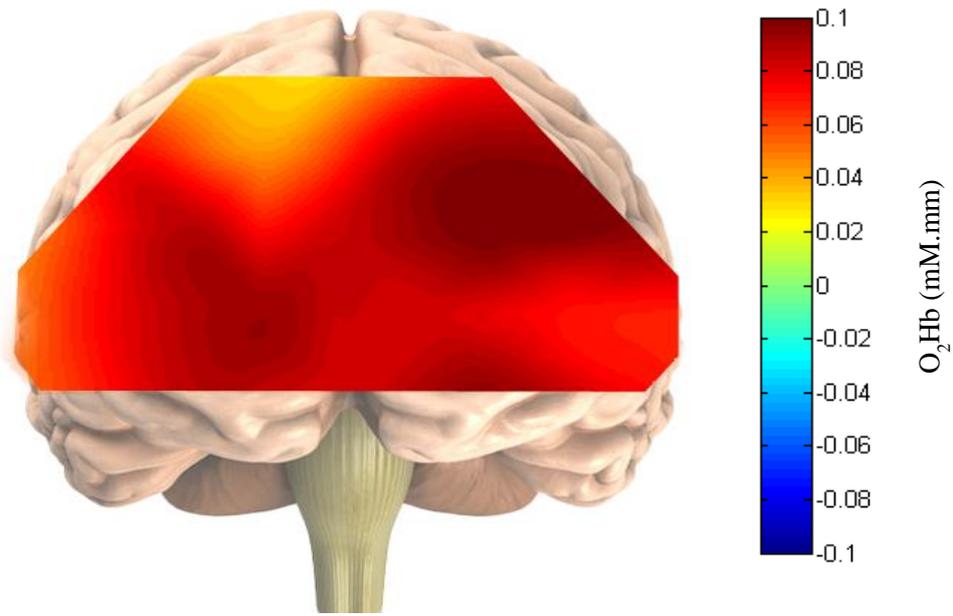

Figure 4.23: Mean oxygenated hemoglobin under the control condition of the first subject.

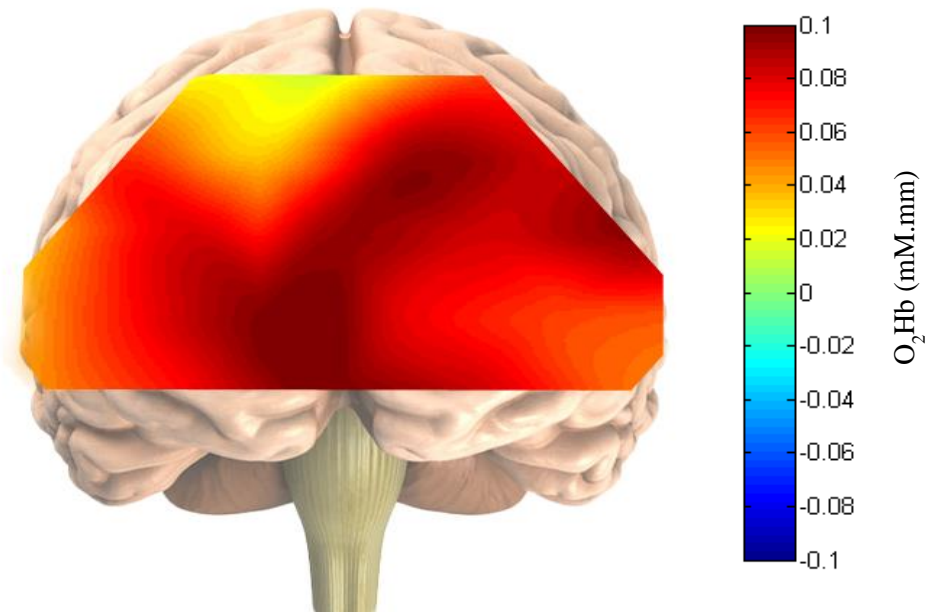

Figure 4.24: Mean oxygenated hemoglobin under the control condition of the second subject.

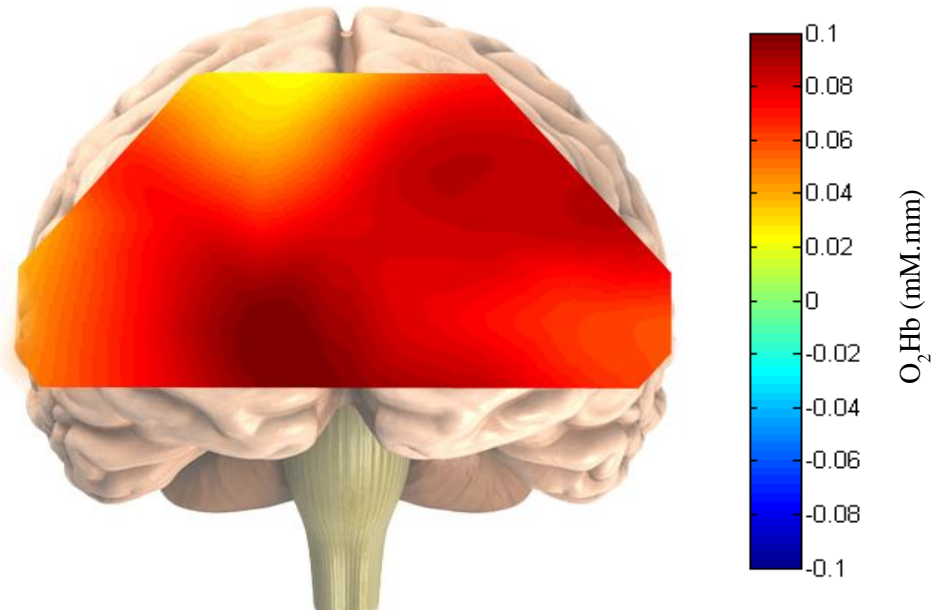

Figure 4.25: Mean oxygenated hemoglobin under the control condition of the third subject.

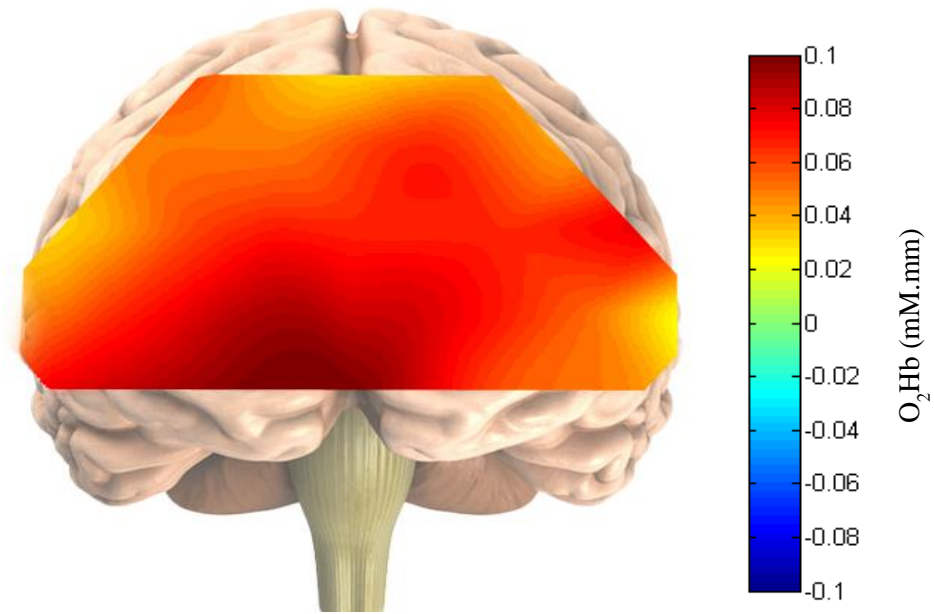

Figure 4.26: Mean oxygenated hemoglobin under the control condition of the fourth subject.

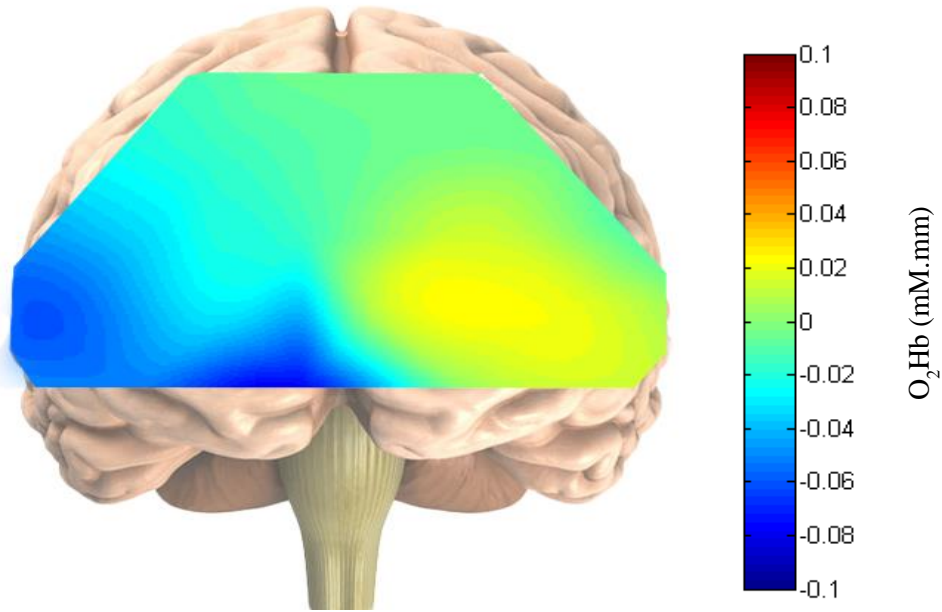

Figure 4.27: Mean oxygenated hemoglobin under the stress condition of the first subject.

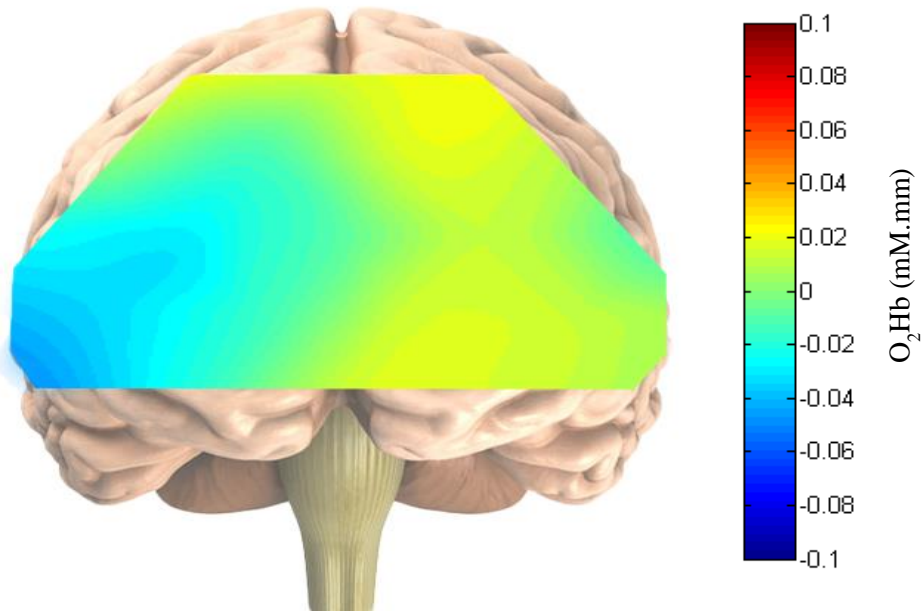

Figure 4.28: Mean oxygenated hemoglobin under the stress condition of the second subject.

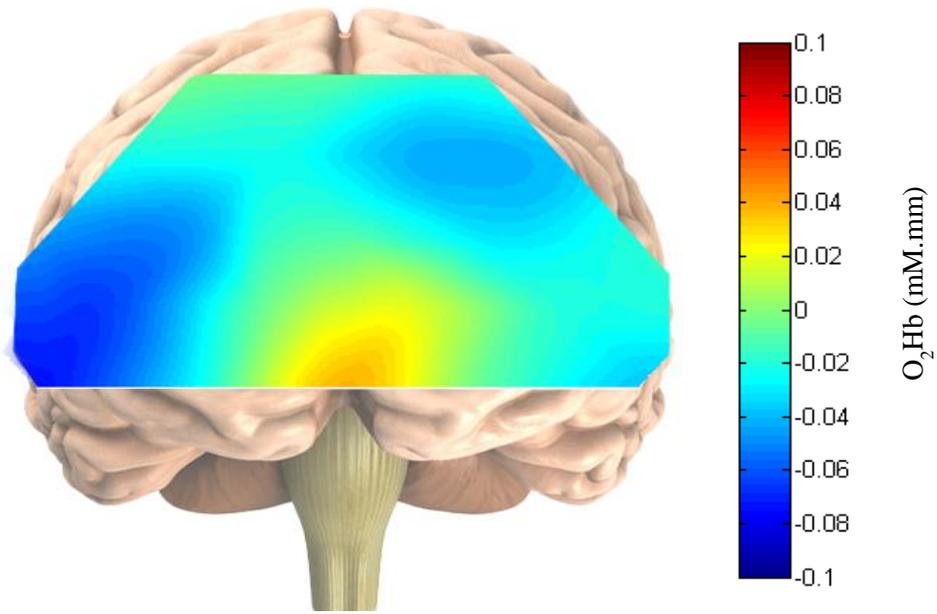

Figure 4.29: Mean oxygenated hemoglobin under the stress condition of the third subject.

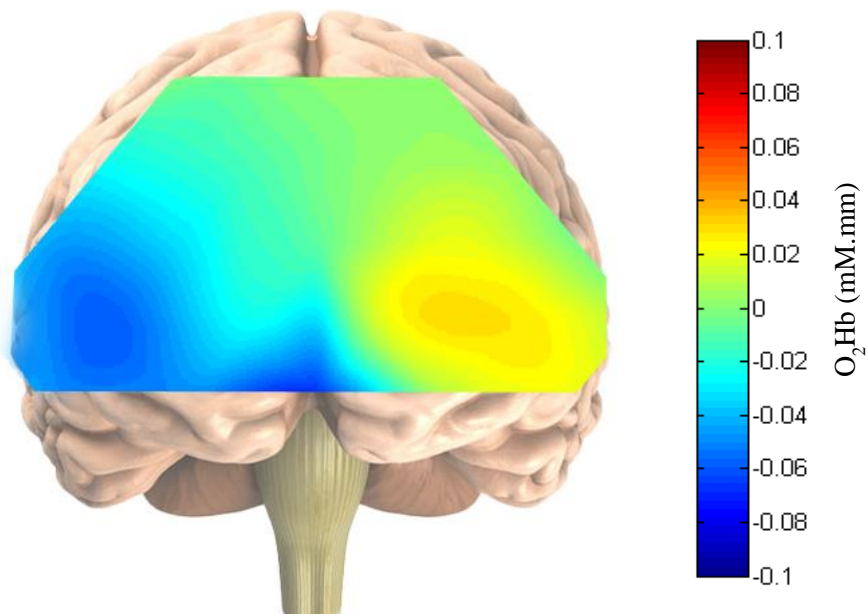

Figure 4.30: Mean oxygenated hemoglobin under the stress condition of the first subject.

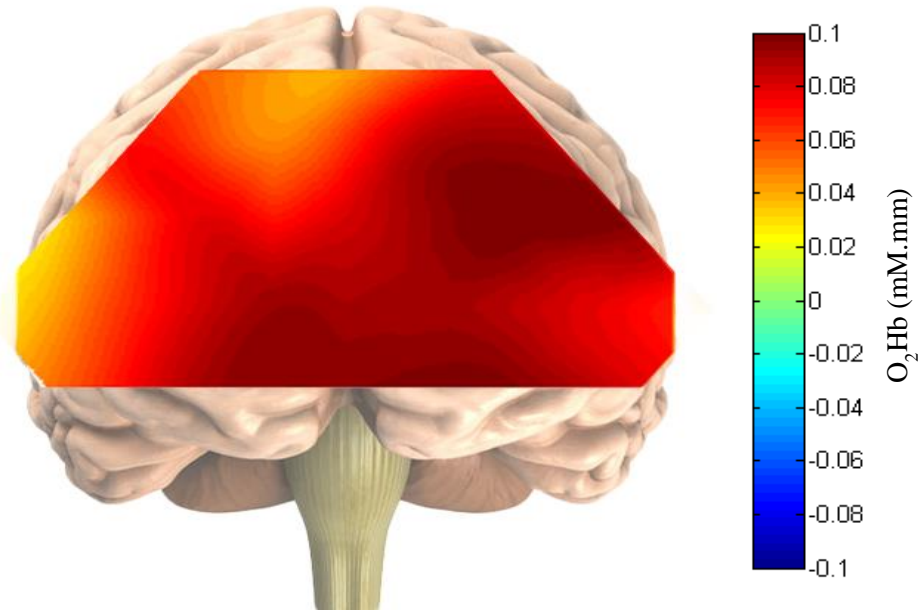

Figure 4.31: Topographical map of the oxygenated hemoglobin concentration under the control condition for the average of the 25 subjects.

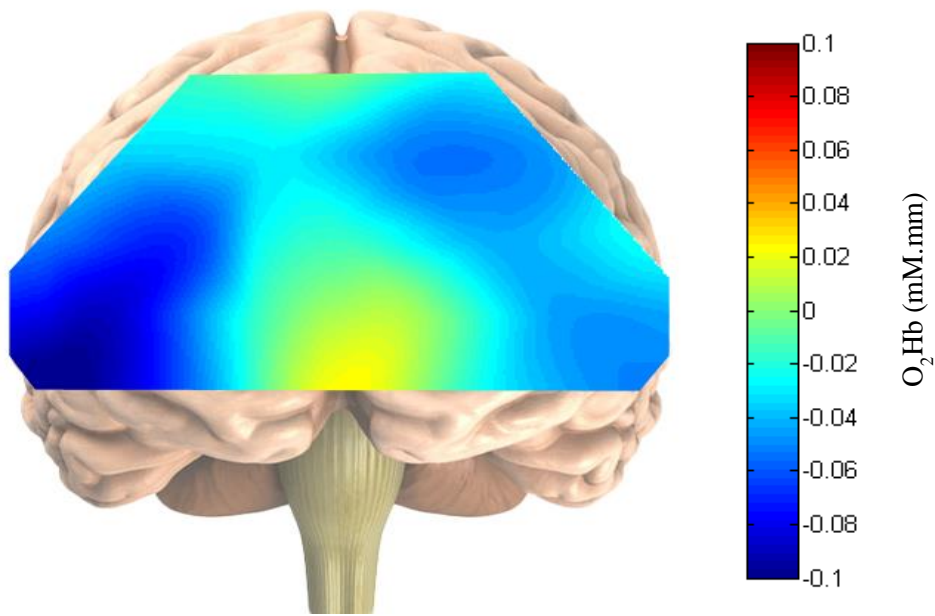

Figure 4.32: Topographical map of the oxygenated hemoglobin concentration under the stress condition for the average of the 25 subjects.

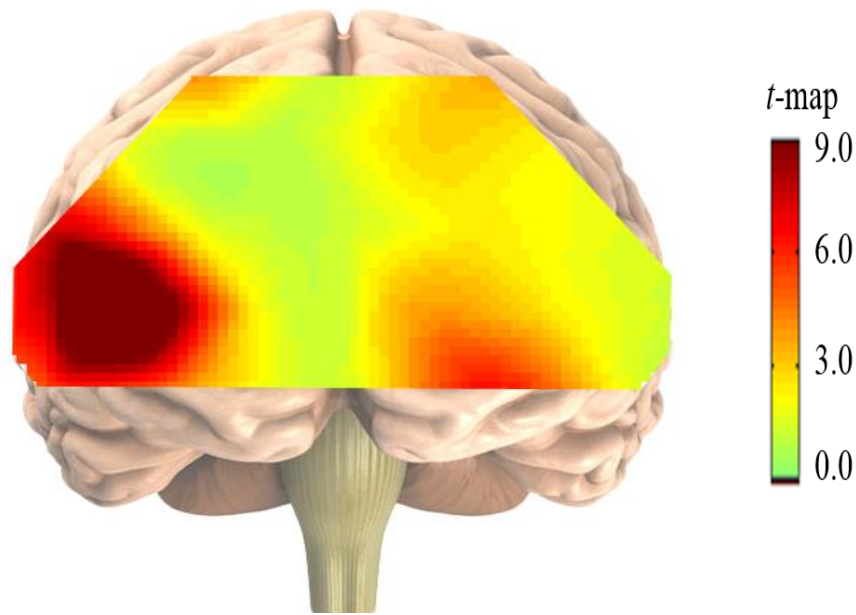

Figure 4.33: Topographical map representing the statistical analysis of the 25 subjects based on their averaged t-values.

## 4.5.2 Result of fusion based on CCA

This study analyzed the covariance matrices of the feature sets from the two modalities on the entire PFC region with the six subregions of interest. Each of the six subregions was being represented by a single EEG/fNIRS channel (VLPFC [right: F8 and Ch19; left: F7 and Ch23], DLPFC [right: F4 and Ch1; left: F3 and Ch3], and FPA [right: FP2 and Ch20; left: FP1 and Ch22]), refer to Table 4.3 for more details. The criteria of selection were based on the correlation level of the components from the transformed feature vectors, discarding those with small canonical correlations. Figure 4.34 shows the canonical correlations of the alpha rhythm and the oxygenated hemoglobin of the stress features obtained by applying the CCA to the entire datasets arranged in descending rank order. These canonical correlations were computed from the estimated joint covariance matrix. As observed from the correlation variants among subregions with the highest correlation coefficient of 0.95 at the right VLPFC, 0.88 at left DLPFC, 0.81 at right DLPFC, 0.72 at left VLPFC, 0.61 at the right FPA and the sixth highest correlation coefficient of 0.48 located at the left FPA respectively. The higher the correlation value, the more focal or localized the stress is. It is also observed that, the precision in localizing stress to the right VLPFC across all subjects is high as demonstrated by the low standard deviation at component '1'. The overall standard deviation across the components was obtained as 0.025 for the first component, 0.028 at the second component, 0.0455 at the third component, 0.054 at the fourth component, 0.062 at the fifth component and 0.084 at the sixth component. The pair of components showing the strongest correlation ($r$=0.95) across the two data sets demonstrates the highest significant difference ($p$<0.00001) between the control and the stress subjects. The overall cross-subject source correlation matrix (map) is displayed in Figure 4.35. The matrix shows the consistency in inter-subject correlation between the modalities.

Table 4.3: CCA pairs of EEG alpha rhythm and $O_2Hb$ of fNIRS.

| No | Region | Component pair | |
|----|--------|------|------|
|    |        | EEG | fNIRS |
| 1 | Right VLPFC | F8 | Ch19 |
| 2 | Left DLPFC | F3 | Ch3 |
| 3 | Right DLPFC | F4 | Ch1 |
| 4 | Left VLPFC | F7 | Ch23 |
| 5 | Right FPA | FP2 | Ch20 |
| 6 | Left FPA | FP1 | Ch22 |

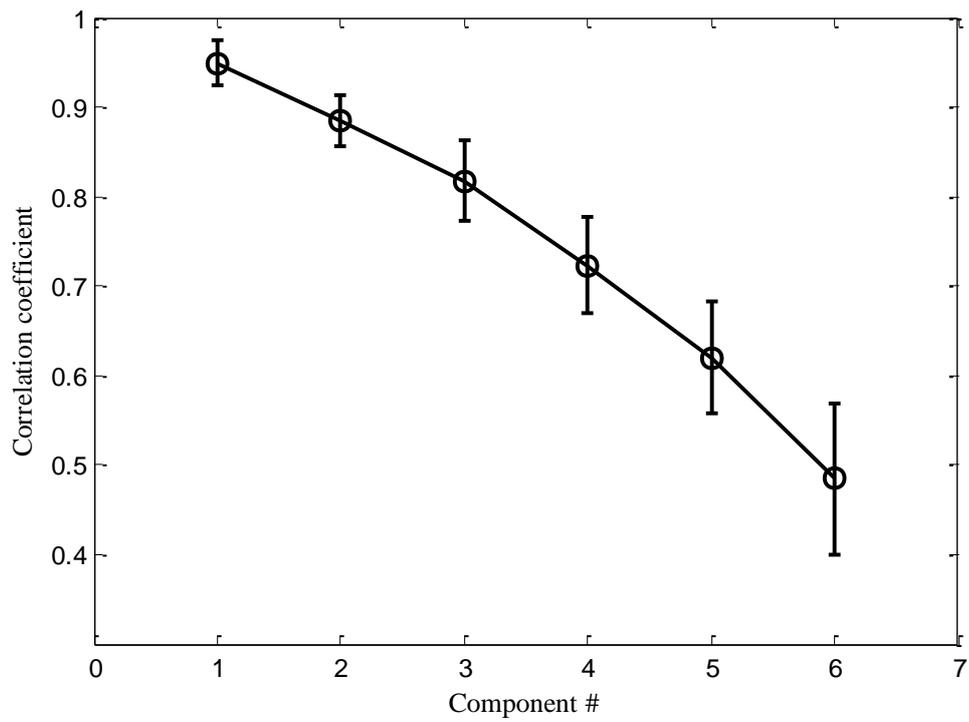

Figure 4.34: Correlation coefficients resulting from the EEG+fNIRS CCA (sorted in decreasing order).

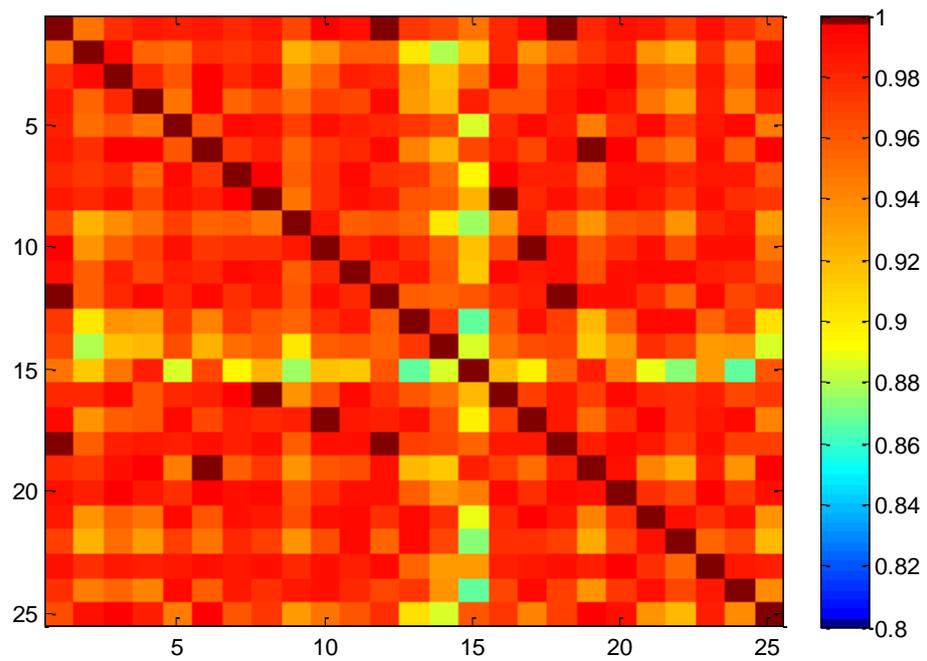

Figure 4.35: Cross-subject source correlation matrix using the CCA technique. The x and y labels represents the number of subjects on the entire correlation matrix.

### 4.5.3 Classification evaluation

The classification results for the individual modality and after the fusion are presented by their ROC curves in Figure 2.36 (a), (b) and (c). Figure 4.36(a) shows the classification evaluation of the sole EEG modality within all the PFC subregions (based on combination of FP1, FP2, F3, F4, F7 and F8), VLPFC subregion (based on combination of F8 and F7), DLPFC subregion (based on combination of F4 and F3) and FPA subregion (based on combination of FP1 and FP2). The labels in term of colors were in accordance with the; using six-bilateral electrodes use red-line, VLPFC with blue-line, DLPFC with green-line and FPA with black-line in all the two modalities. For the fusion of the EEG and the fNIRS channels in the different combinations of two-electrodes and two-channels within: the VLPFC labelled with red-line, DLPFC labelled with blue-line, FPA labelled with green-line and six-bilateral electrodes labelled with black-line and the six-channels of fNIRS labelled with cyan line.

The average classification accuracy, sensitivity, specificity and AUC values calculated for each PFC subregion are (0.899, 0.858, 0.837 and 0.831), (0.875, 0.870, 0.866 and 0.860), (0.920, 0.845, 0.808 and 0.801), and (0.957, 0.903, 0.900 and 0.925) respectively. Similarly, Fig 4.36(b) shows the classification evaluation of the sole fNIRS modality within all the PFC subregions (based on combination of Ch22, Ch20, Ch19, Ch23, Ch1 and Ch3), VLPFC subregion (based on combination of Ch19 and Ch23), DLPFC subregion (based on combination of Ch1 and Ch3) and FPA subregion (based on combination of Ch20 and Ch22). The average classification accuracy, sensitivity, specificity and AUC values calculated are (0.856, 0.829, 0.808 and 0.793), and (0.823, 0.795, 0.833 and 0.829), (0.880, 0.862, 0.783 and 0.758), and (0.927, 0.897, 0.863 and 0.863). Looking at the performance of the individual modality, the EEG outperforms the fNIRS in all the classification measurements.

On the other hand, Figure 4.36(c) shows the classification evaluation under the fusion of the EEG+fNIRS signal modality within the VLPFC subregion, the DLPFC subregion and the FPA subregion as compared with the excellent results obtained by the individual modality from the entire PFC area. The average classification accuracy, sensitivity, specificity and AUC values in the format of entire PFC area (at six-bilateral electrode/channel), VLPFC, DLPFC and FPA were calculated and presented as shown in Table 4.3. Additionally, the overall classification performance and improvements of fusing each subregion over the entire PFC area of each modality and over each subregion by individual modality is presented in Table 4.4. The results of the classification supports the dominant of right VLPFC to stress as obtained by CCA method.

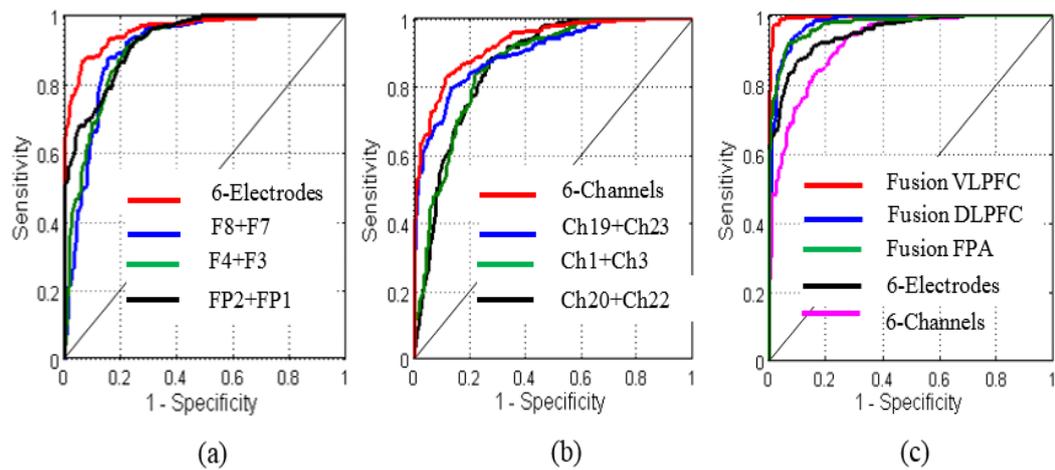

Figure 4.36: ROC curves (a) EEG modality, (b) fNIRS modality and (c) Fusion of EEG and fNIRS.

Table 4.4: Overall Classification Performance and fusion improvements

| Region | | EEG | fNIRS | EEG+fNIRS | ++REEG | ++RfNIRS | ++EEG | ++fNIRS |
|---|---|---|---|---|---|---|---|---|
| Six-Bilateral | Accuracy % | 89.8 | 85.6 | | | | | |
| | Sensitivity % | 87.5 | 82.3 | | | | | |
| | Specificity % | 92.0 | 88.0 | | | | | |
| | AUC % | 95.7 | 92.7 | | | | | |
| VLPFC | Accuracy % | 85.8 | 82.9 | 97.7 | 11.9 | 14.8 | 7.9 | 12.1 |
| | Sensitivity % | 87.0 | 79.5 | 96.6 | 9.6 | 17.1 | 9.1 | 14.3 |
| | Specificity % | 84.5 | 86.2 | 98.7 | 14.2 | 12.5 | 6.7 | 10.7 |
| | AUC % | 90.3 | 89.7 | 99.5 | 9.2 | 9.8 | 3.8 | 6.8 |
| DLPFC | Accuracy % | 83.7 | 80.8 | 92.5 | 9.0 | 11.7 | 2.7 | 6.9 |
| | Sensitivity % | 86.6 | 83.3 | 92.9 | 6.3 | 9.6 | 5.4 | 10.6 |
| | Specificity % | 80.8 | 78.3 | 92.0 | 11.1 | 13.7 | 0 | 4.0 |
| | AUC % | 90.0 | 86.3 | 97.6 | 7.6 | 11.3 | 1.9 | 4.9 |
| FPA | Accuracy % | 83.1 | 79.3 | 92.5 | 9.4 | 13.2 | 2.7 | 6.9 |
| | Sensitivity % | 86.0 | 82.9 | 92.0 | 6.0 | 9.1 | 4.5 | 9.7 |
| | Specificity % | 80.1 | 75.8 | 92.9 | 12.8 | 17.1 | 0.9 | 4.9 |
| | AUC % | 92.5 | 86.3 | 97.0 | 4.5 | 10.7 | 1.3 | 4.3 |

Note that, the ++REEG represents the improvements of fusing EEG and fNIRS in each subregion over EEG modality in that subregion; ++RfNIRS represents the improvements of fusing EEG and fNIRS in each subregion over fNIRS modality in that subregion; ++EEG represents the improvements of fusing EEG and fNIRS in each subregion over optimum EEG modality; ++fNIRS represents the improvements of fusing EEG and fNIRS in each subregion over optimum fNIRS modality, respectively.

### 4.5.4 Discussion on the results

The overall results from the EEG signals across all the subjects revealed a decrease in the alpha rhythm on the entire PFC area. Specifically, the electrodes from the right PFC subregions; F4 and F8 were highly sensitive to stress as reported by their p-values, mean $p < 0.0001$. On the other hand, the electrode 'F7' from the left PFC subregion responded differently with less significant difference in the cortical activities with the p-value reported as $p < 0.01$. Compared to the other electrodes, it shows a hemispheric difference. This difference in hemispheric activities is thought to be associated with frontal-alpha asymmetry. Thus, the decrease in the right frontal alpha rhythm constitutes evidence of a negative response due to the stressors. The decrease in the alpha rhythm on the PFC is consistent with previous emotional and anxiety studies and with studies that administered cortisol to human subjects [41, 42, 261, 262]. Additionally, the difference in activities of the right and the left PFC in this study is in line with previous EEG studies that showed hemispheric differences under stress conditions [263]. Therefore, the obtained results suggest that the cortical activities shifted from diffused to focal cortical activities under the stress condition.

Likewise, the fNIRS signals showed higher activation and increase in the oxygenated hemoglobin concentration on the entire PFC area while solving the arithmetic task under the control condition. Specifically, a higher activation was found within the left PFC area which can be explained as the left PFC being sensitive to the arithmetic task. In particular, the higher activations of oxygenated hemoglobin were observed consistently in individual subjects as well as in the average of all subjects. The results reported in Figures 4.31 to Fig 4.33 showed the overall oxygenated hemoglobin concentration of the 25 subjects under the control and the stress conditions with their corresponding t-map respectively. Similarly as with each individual subject, the activities on the right VLPFC subregion were significantly reduced under stress, $p<0.0001$. It is evidenced that the oxygenated hemoglobin shifted from diffused to focal under the maintained psychological stress condition. This result is in line with our hypothesis stating that the PFC functions under the maintained psychological stress shifts from diffuse to focal cortical activities.

Interestingly, the reduction of the oxygenated hemoglobin concentration on the PFC subregions in our study is consistent with previous fMRI human and animal stress studies [26, 32, 264, 265]. Furthermore, a similar reduction of the oxygenated hemoglobin on the DLPFC have been seen in subjects suffering from post-traumatic stress disorder [266, 267]. Additionally, reduced cortical activities have been reported on the ventromedial PFC while inhibiting a fear response [268]. Thus, showing similar dysfunctions in the PFC subregions in healthy individuals under maintained psychological stress may provide the insight necessary to treat neuropsychiatric conditions that afflict many people. However, in light of the present results, it is important to consider the methodological and neuroimaging modality differences between studies. This study could thus suggest that our approach may have greater sensitivity to detect discriminable patterns on the PFC subregions to stress under naturalistic settings.

The classification result of the EEG signals demonstrated a high accuracy in discriminating stress in the control state with 89.8% and area under the curve of 95.7% using 6-bilateral electrodes over the entire PFC area. Similarly, the fNIRS classification accuracy result was 85.6%, the area under the curve being 92.7% using six-bilateral channels over the PFC. Due to the excellent temporal resolution of the EEG modality, it outperforms the fNIRS modality by +4.2% in accuracy and +3.0% in AUC. On the other hand, the fNIRS outperforms the EEG modality by +1.7% in terms of specificity over the VLPFC. Additionally, studying each subregion individually, this study found that the VLPFC subregion outperforms other subregions in terms of accuracy in both modalities. As reported in Table 4.4, the VLPFC outperformed other PFC subregions by +2.7% accuracy in the EEG modality and by +3.6% in the fNIRS modality. The highest accuracy on the right VLPVC obtained through the fNIRS confirmed its dominance in mental stress.

The fusion of the EEG+fNIRS features using the proposed CCA method discovered the associations across the two modalities and estimated the components responsible for these associations. It jointly analysed the two modalities to fuse information without giving preference to either modality. The method identified the relationships based on the natural inter-subject co-variances between the modalities. Six pairs of components were estimated based on their degree of correlation across the modalities as showed in Fig.4.34. The difference in the correlation values across the modalities indicated the level of subregional activities (dysfunction) and the dominance of specific subregions in their susceptibility to mental stress. The first pair of components had a correlation of 0.95 showing the most significant reduction in the oxygenated hemoglobin on the right VLPFC associated with a decrease of the alpha rhythm at the time of stress. These associations across the modalities (given by the highest correlation at the right VLPFC) confirmed the subregion most susceptible to mental stress. The performance evaluation of the CCA also supported the dominance of the right VLPFC to stress as measured by the classifier accuracy, sensitivity, specificity and AUC of each subregion when compared with the entire PFC measuring an excellent EEG/fNIRS modality.

The classification evaluation of the VLPFC subregion resulted in 97.7% accuracy, 96.6% sensitivity, 98.7% specificity and 99.5% AUC respectively. Compared to the sole EEG modality, the fusion of the VLPFC subregion demonstrated improvement of +9.9% in the accuracy, +9.1% in the sensitivity, +6.7% in the specificity and +3.8% in the area under the curve. Similarly, the fusion of the DLPFC subregion showed an improvement of +2.7% in the accuracy, +5.4% in the sensitivity, and +1.9% in the AUC. The fusion of the EEG+fNIRS over the FPA showed an improvement of +2.7% in the accuracy, +4.5% in the sensitivity, +0.9% in the specificity and +1.3% in the AUC. Compared to the sole fNIRS modality, the fusion of the VLPFC subregion demonstrated an improvement of +14.1% in the accuracy, +14.3% in the sensitivity,

+10.7% in the specificity and +6.8% in AUC. Similarly, the fusion of the DLPFC subregion showed an improvement of +6.9% in the accuracy, +10.6% in the sensitivity, +4% in the specificity and +4.7% in the area under the curve. The fusion of the FPA showed an improvement of +6.7% in the accuracy, +9.7% in the sensitivity, +4.9% in the specificity and +4.3% in AUC. Using two sample t-test, the proposed fusion significantly improved the classification accuracy, sensitivity, specificity and area AUC compared to the sole EEG and the sole fNIRS modality, $p<0.01$.

## 4.6 EEG and fNIRS connectivity

The EEG PFC connectivity on the inter-hemisphere and intra-hemisphere within all subjects across all pairs of electrodes under control and stress conditions are as shown in Figure 4.37 (A) and (B) respectively. The results investigated on the right and left; FPA, VLPFC and DLPFC in the form of the inter-hemisphere and intra-hemisphere respectively. The EEG inter-hemisphere connectivity results demonstrated a significant reduction from the control condition to the stress condition on the right and left VLPFC (F8 and F7), the right FPA (FP2), and the right and left DLPFC (F4 and F3) respectively. Only the left FPA at FP1 was not significantly affected by stress. Similarly, the intra-hemispheric connectivity was significantly reduced from the control to the stress condition on the right FPA (FP2), the right VLPFC (F8), and the right DLPFC (F4) respectively. The strength of the connectivity is as demonstrated by lines, the red line indicating high connectivity and the blue line indicating less connectivity.

A) Inter

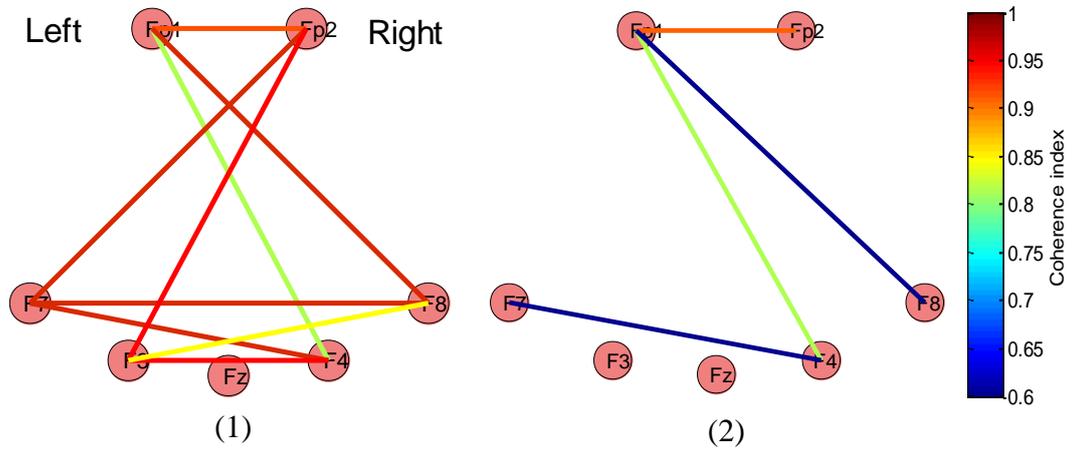

(1)                                    (2)

B) Intra

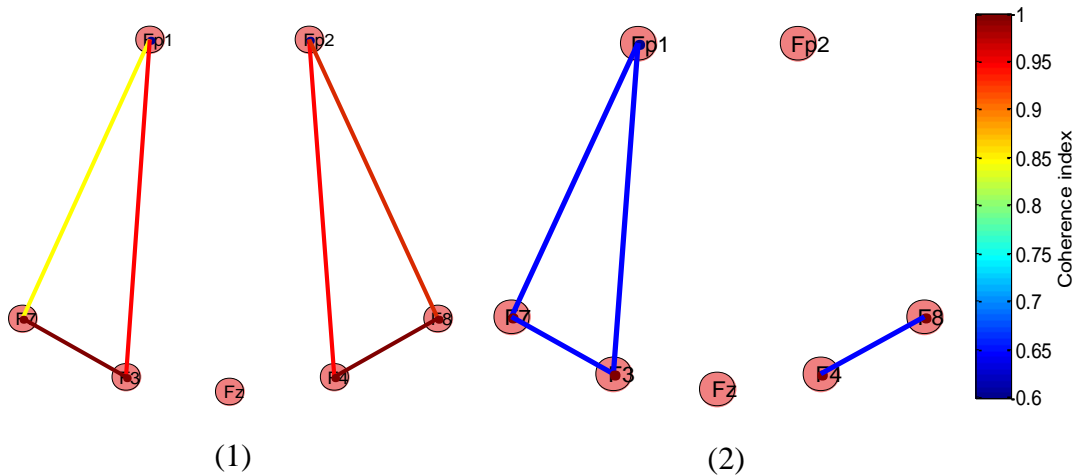

(1)                                    (2)

Figure 4.37: EEG connectivity (A) inter-hemispheric PFC connectivity (1) under the control condition and (2) under the stress condition, (B) intra-hemispheric PFC connectivity (1) under the control condition and (2) under the stress condition.

In contrast, the fNIRS connectivity on the inter-hemispheric and intra-hemispheric PFC demonstrated a significant reduction from the control condition to the stress condition in all the frequency bands. The results of the connectivity at the frequency intervals of I (0.009-0.02 Hz), II (0.02- 0.04 Hz), III (0.04-0.06 Hz), IV (0.06-0.08Hz), V (0.08-0.10 Hz), VI (0.0.009-0.1 Hz) and VII (0.1-0.8 Hz) were highly reduced as shown in Figures 4.38 to 4.41 for the control and the stress condition within the inter-hemispheric and intra-hemispheric PFC areas respectively. A reduced brain connectivity on the inter-hemispheric PFC was mostly found at the right DLPFC (Ch4 and Ch1), the right VLPFC (Ch8, Ch13 and Ch14) and the right and left FPA (Ch9, Ch15 and Ch22), respectively. Similarly, the connectivity within the intra-hemispheric PFC demonstrated a significant reduction from the control to the stress condition, on the right VLPFC (Ch19), on the left and the right DLPFC. The overall statistical analysis for all the EEG electrodes and all the fNIRS channels is presented in Table 4.5 (inter-hemispheric PFC) and Table 4.6 (intra-hemispheric PFC) respectively.

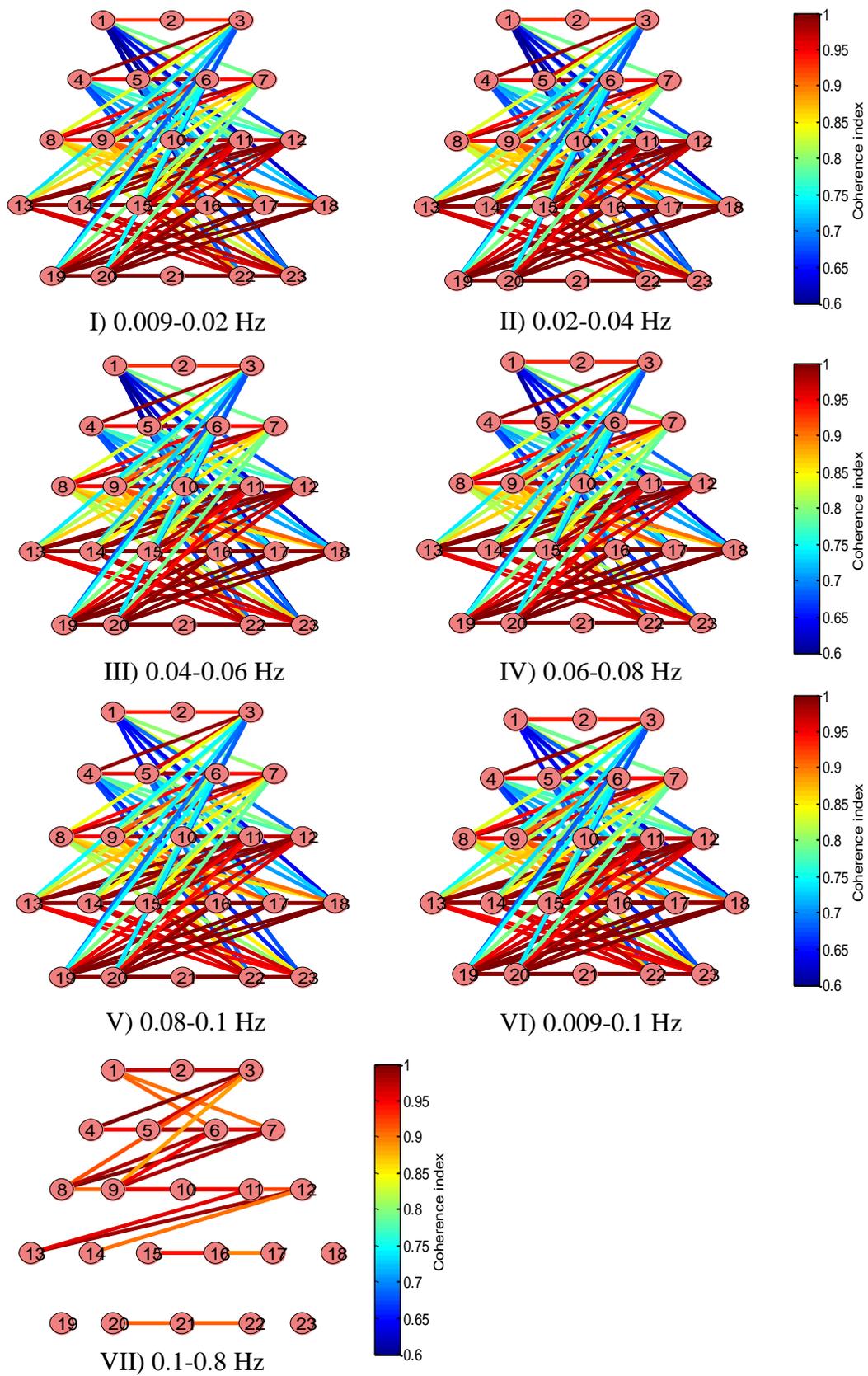

Figure 4.38: fNIRS inter-hemispheric PFC connectivity under the control condition.

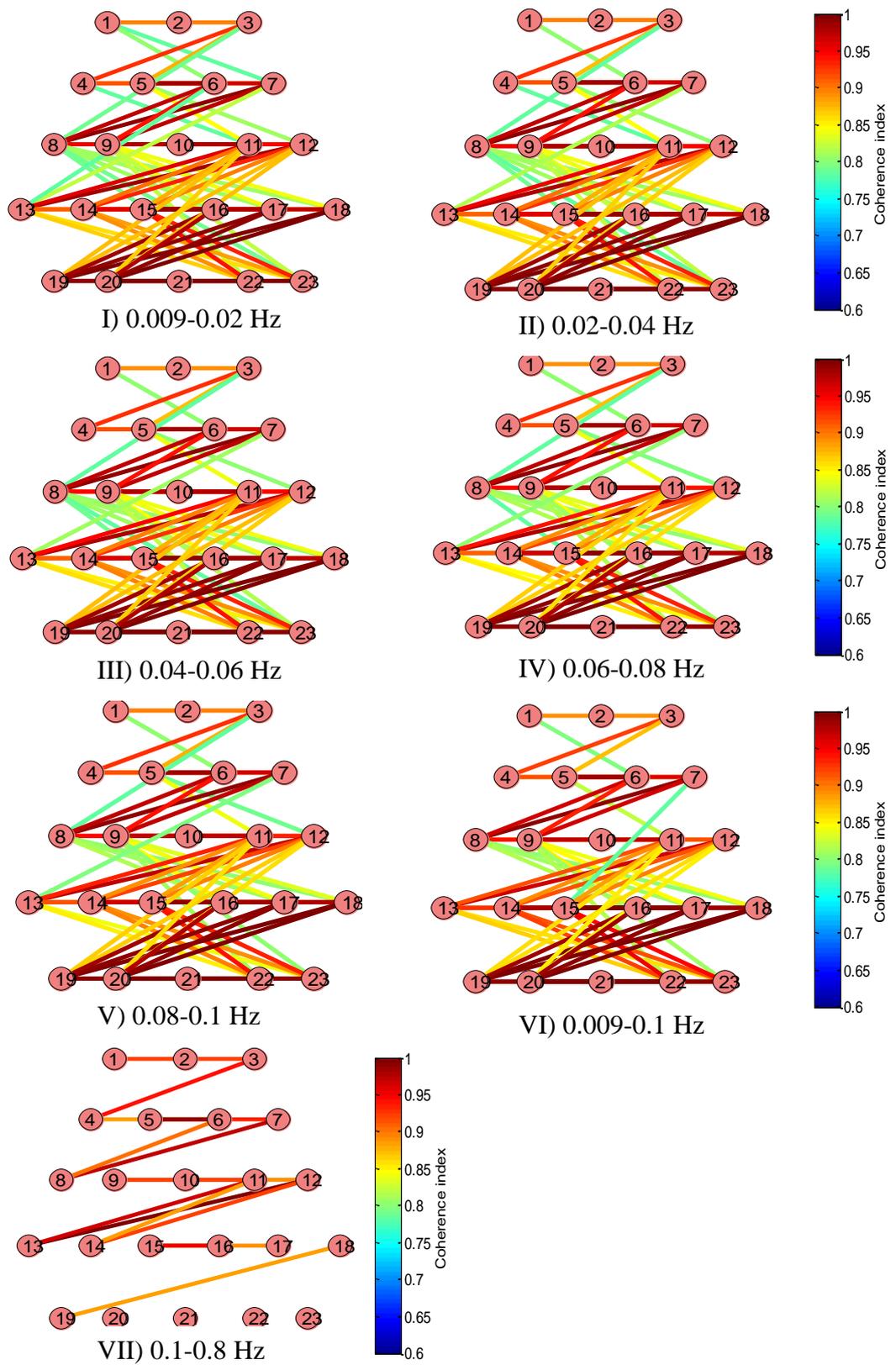

Figure 4.39: fNIRS inter-hemispheric PFC connectivity under the stress condition.

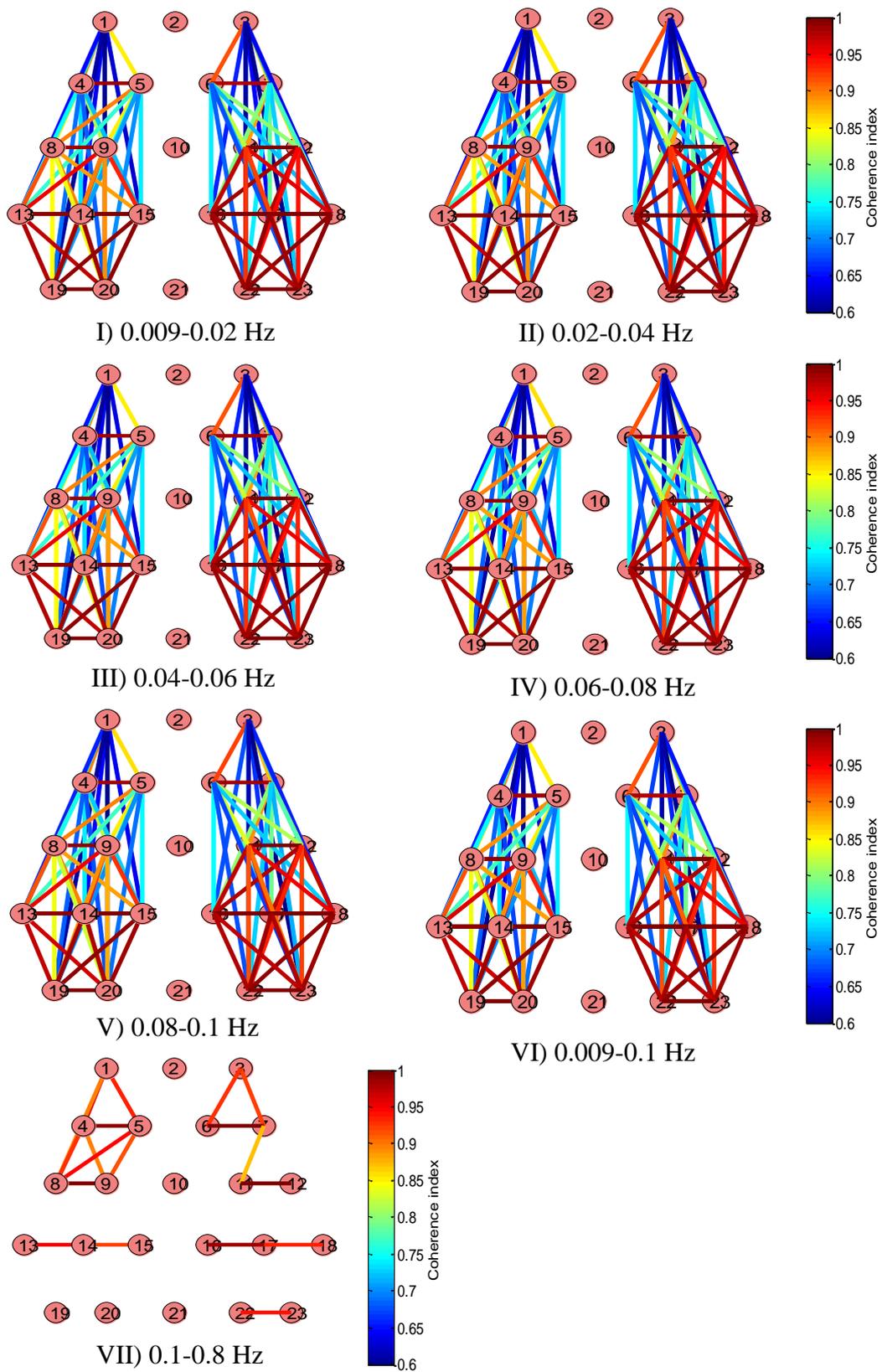

I) 0.009-0.02 Hz

II) 0.02-0.04 Hz

III) 0.04-0.06 Hz

IV) 0.06-0.08 Hz

V) 0.08-0.1 Hz

VI) 0.009-0.1 Hz

VII) 0.1-0.8 Hz

Figure 4.40: fNIRS intra-hemispheric PFC connectivity under the control condition.

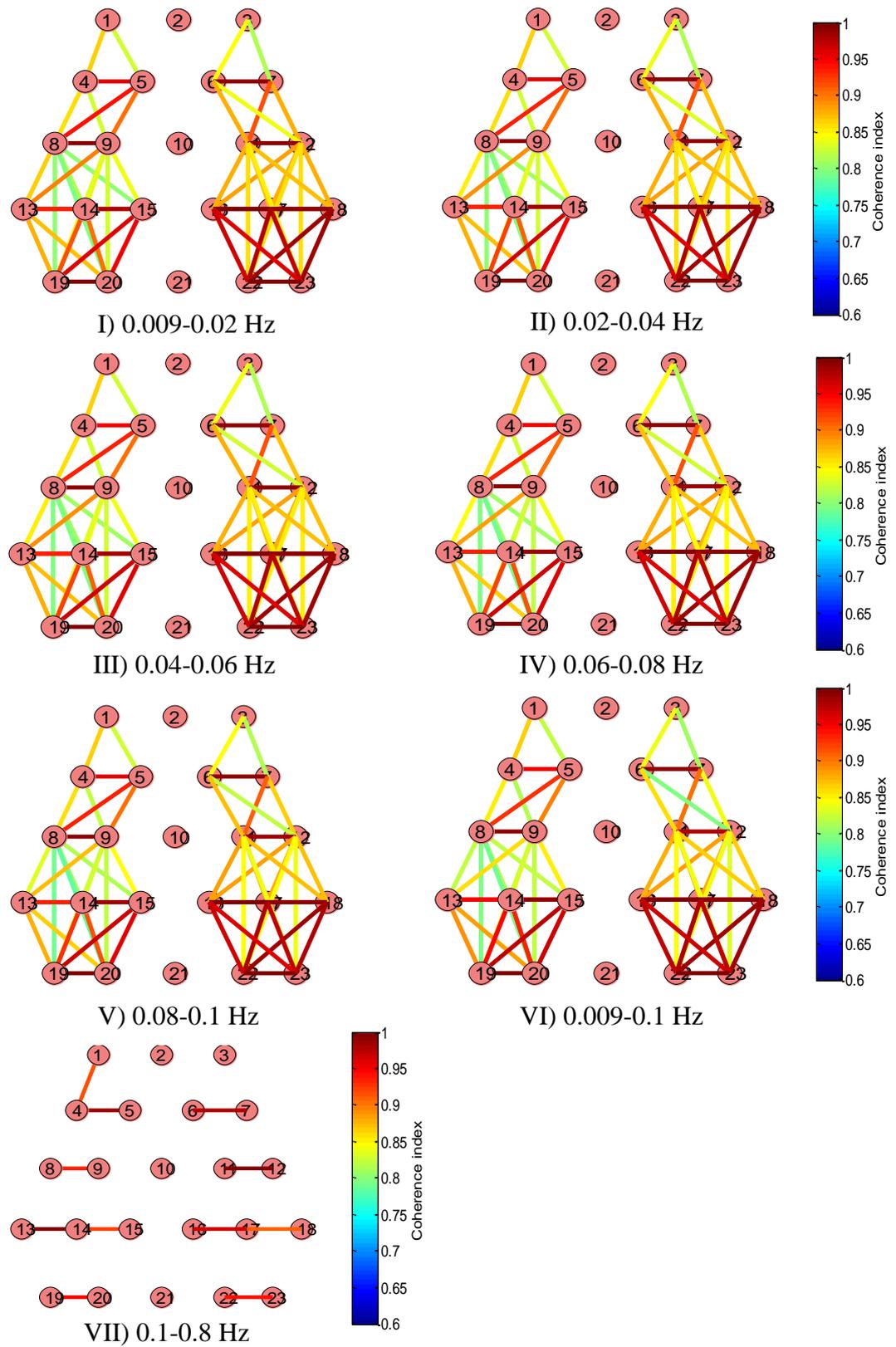

Figure 4.41: fNIRS intra-hemispheric PFC connectivity under the stress condition.

Table 4.5: Statistical analysis of the alpha EEG and O$_2$Hb of the fNIRS connectivity measurements within inter-hemispheric PFC based on the two-sample t-test.

| Channel No | t-value | p-value | Channel No | t-value | p-value |
|---|---|---|---|---|---|
| 1 | 3.95 | 0.0000 | 16 | 3.00 | 0.0085 |
| 3 | 3.38 | 0.0033 | 17 | 1.83 | 0.0874 |
| 4 | 4.60 | 0.0000 | 18 | 1.77 | 0.0982 |
| 5 | 2.78 | 0.0132 | 19 | 1.87 | 0.0787 |
| 6 | 1.85 | 0.0852 | 20 | 1.76 | 0.0974 |
| 7 | 3.00 | 0.0085 | 22 | 2.95 | 0.0092 |
| 8 | 2.24 | 0.0389 | 23 | 1.76 | 0.0970 |
| 9 | 3.00 | 0.0083 | FP1 | 0.41 | 0.9321 |
| 11 | 0.90 | 0.3825 | FP2 | 3.11 | 0.0041 |
| 12 | 1.45 | 0.1639 | F3 | 6.00 | 0.0000 |
| 13 | 2.21 | 0.0430 | F4 | 2.11 | 0.048 |
| 14 | 2.47 | 0.0251 | F8 | 4.31 | 0.0001 |
| 15 | 2.40 | 0.0291 | F7 | 3.51 | 0.0025 |

Table 4.6: Statistical analysis of the alpha EEG and O$_2$Hb of the fNIRS connectivity measurements within intra-hemispheric PFC based on the two-sample t-test.

| Channel No | t-value | p-value | Channel No | t-value | p-value |
|---|---|---|---|---|---|
| 1 | 3.95 | 0.0014 | 16 | 1.87 | 0.0821 |
| 3 | 3.98 | 0.0013 | 17 | 1.36 | 0.1912 |
| 4 | 2.80 | 0.0130 | 18 | 1.87 | 0.0811 |
| 5 | 2.50 | 0.0214 | 19 | 2.23 | 0.0400 |
| 6 | 2.90 | 0.0110 | 20 | 1.52 | 0.1489 |
| 7 | 2.80 | 0.0121 | 22 | 1.45 | 0.1686 |
| 8 | 1.30 | 0.2300 | 23 | 1.36 | 0.1918 |
| 9 | 2.30 | 0.0480 | FP1 | 0.31 | 0.9871 |
| 11 | 0.14 | 0.8800 | FP2 | 6.00 | 0.0000 |
| 12 | 1.36 | 0.1913 | F3 | 0.41 | 0.9683 |
| 13 | 0.14 | 0.8913 | F4 | 3.11 | 0.0041 |
| 14 | 1.12 | 0.2791 | F8 | 3.11 | 0.0041 |
| 15 | 2.21 | 0.0430 | F7 | 0.51 | 0.9560 |

### 4.6.1 Discussion on the EEG+fNIRS connectivity

This study investigated mental stress based on the patterns of the EEG and fNIRS-based functional connectivity using the indices called mean squared coherence. Different colors were used to indicate the various connectivity strengths in the coherence maps. All channel pairs in inter and intra-hemispheric were classified into three connection groups: (1) connectivity between the DLPFC regions, (2) connectivity between the VLPFC regions; and (3) connectivity between the frontopolar area. The average coherence values were calculated for each connectivity type on the frequency intervals of I (0.009-0.02 Hz), II (0.02- 0.04 Hz), III (0.04-0.06 Hz), IV (0.06- 0.08Hz), V (0.08-0.10 Hz), VI (0.0.009-0.1 Hz) and VII (0.1-0.8 Hz). The results demonstrated a significant reduction in functional connectivity from the control to the stress condition in inter- hemispheric and intra-hemispheric PFC areas. The connectivity was highly reduced on the right DLPFC (Ch1, Ch4 and F4) followed by the left DLPFC (Ch3, Ch7) and right VLPFC (Ch19, Ch8, Ch14 and F8) and as reported by both modalities. In summary, this study demonstrated that the EEG and fNIRS-based functional connectivity reveals different patterns for different mental states (control and stress) in all frequency intervals. This recommends the functional connectivity within brain sites as good indices to discriminate the stress state from that of control state.

### 4.7 Summary

In first part of this chapter, the results of alpha amylase level under control and stress conditions are presented and discussed. The results showed significant increase in the alpha amylase level during stress. The results of alpha amylase confirmed that, the proposed task with time pressure and negative feedback significantly induce stress on all the participants. In the second part of this chapter, the results of feature level fusion using joint independent component analysis technique as well as canonical correlation analysis technique are presented and their performance compared with individual modality based on the classification accuracy, sensitivity, specificity and areas under the curve. The results of the feature level fusion significantly improve the detection rate of mental stress, p<0.01 compare to individual EEG and fNIRS modality. Overall, the developed fusion techniques have contributed to the state-of-the art by improving the detection rate of mental stress. Up to date, this is the first study reported fusion at feature level as a promising method in improving the overall detection/system performance. Another novel approach reported in this study is that, the proposed methods localize the stress to the right ventrolateral PFC area. In the last part of this chapter, the results of the cortical connectivity under control condition and under stress condition are presented. The study investigated the connectivity within inter and intra-hemispheric PFC areas at seven frequency intervals and found that, stress reduce the connectivity in the dorsolateral PFC and on the right ventrolateral PFC. The results of connectivity was in line with the alpha rhythm and oxygenated hemoglobin showing right dominant of ventrolateral PFC to stress. It could be concluded that, stress disrupt the PFC connectivity and report functional connectivity on PFC as a good index of stress.

CHAPTER 5

CONCLUSIONS AND FUTURE WORK

This chapter discusses the conclusions derived from the main findings of this thesis and highlights the main contribution of the study to the research field as well as society. Additionally, it highlights some points for future work in order to translate this type of research study to the clinical practice.

## 5.1 Conclusions

This thesis investigated mental stress on 25 healthy subjects based on simultaneous measurements of electroencephalography (EEG) and functional near infrared spectroscopy (fNIRS) over the prefrontal cortex (PFC). The stress stimulus used in this study was based on established mental arithmetic tasks with the level of difficulty and under time pressure with negative feedback. The confirmation of stress elucidation on the participants was assessed by measuring their alpha amylase levels. The specific aims of this study contained herein were: 1) to control the simultaneous measurement of EEG and fNIRS on the PFC while participants solving the arithmetic task under control and stress conditions; 2) to detect the level of stress based on the cortical activity and cortical connectivity induced by arithmetic tasks under control and stress conditions; and 3) to develop a fusion method of EEG and fNIRS features to improve mental stress detection over individual modality.

In this study, mental stress on the participants was induced in the laboratory (workplace) using established mental arithmetic tasks with a certain level of difficulty and under time pressure with negative feedback as described in Chapter 4 (Experiment). The study confirmed the procedures of stress by measuring each participant's salivary alpha amylase level before the task as baseline, immediately after the task in the control and stress conditions. The obtained results confirmed that the applied time- and peer-pressure did increase significantly ($p<0.001$), the stress level of all participants, as compared to the baseline and the control conditions.

The EEG and the fNIRS-based cortical and functional connectivity demonstrated significant reduce under stress as compare to control condition. The main finding of the EEG results showed a significant decrease in the alpha rhythm power *($p<0.01$)* and an increase in the beta rhythm power *($p<0.02$)* in the PFC subregions under the stress condition as compared to the control condition. Additionally, the EEG statistical results of the cortical connectivity showed a significant decrease in the right hemisphere under the stress condition as compared to the control condition, $p<0.05$. The significant differences were found in both: inter- hemisphere and the inter-hemisphere PFC. Likewise, the fNIRS signals showed a higher activation or increase in the oxygenated hemoglobin concentration on the entire PFC area while solving the arithmetic task under the control condition. The statistical analysis of the functional connectivity of the fNIRS showed a significant decrease, $p<0.01$ under stress compared to that under the control condition in the seven frequency intervals. The reduction in the functional connectivity was found in the inter- and intra-hemisphere PFC with much reduction on the dorsolateral PFC a well as right ventrolateral PFC. Up to date, this is the first fNIRS study investigating mental stress on different frequency intervals within different brain areas and identifying the dorsolateral and right ventrolateral PFC as the brain regions mostly sensitive to stress.

Finally this study explored two different feature-level fusions of the EEG and the fNIRS for improving the detection rate of mental stress. The study investigated the feature-level fusion using the joint independent component analysis (jICA) and the canonical correlation analysis (CCA) methods. Prior to the jICA, the study used the advantages of the fNIRS spatial resolution to construct the spatial locations of the EEG after which the features of the EEG and the fNIRS from the selected components were fused. The spatial location was based on t-values in which the most significant channels responded to stress were selected. The optimization of the fused features was based on neural network. The neutral gradient decent was used to maximize the output entropy between the components. The output of the fusion was then assess in term of their accuracy, sensitivity, specificity and area under ROC curves. Overall, the proposed jICA fusion method significantly improves the classification accuracy, sensitivity and specificity on average by +3.46% compared to the EEG and +11.13% compared to the fNIRS. The fusion technique significantly improve the detection rate of mental stress as hypothesis in the early section of this study. Similarly, the CCA method was used in this study maximized the covariance across all the subjects within the EEG and the fNIRS components. In this method, all components were used for analysis without giving priority to any modality. The optimization technique used in this technique was based on Lagrangian multipliers. The output of the CCA fusion technique was then evaluated using SVM classifier. Similarly as the case of jICA, significant improvements achieved in all the evaluation metrics namely; accuracy, sensitivity, specificity and area under the ROC curves. The CCA method improves the classification accuracy, sensitivity and specificity on average by +8.56% compared to the EEG and +13.03% compared to the fNIRS. As expected, the entire experimental results revealed that significant improvements were achieved by the proposed fusions approaches in all the three performance metrics, with *p<0.01*. This confirmed our hypothesis that the fusion of the EEG and the fNIRS can improve the mental stress detection.

## 5.2 Contributions

The list of contributions of this research to the field of study as well as the society is as follows:

- This is the first study investigated the effects of stress on the PFC activities based on EEG alpha rhythm as well as hemodynamic responses of fNIRS. The present findings make important contributions to the field of EEG and fNIRS. Specifically, they show for the first time, that it is possible to detect a negative correlation between activity in key regions of the task-positive network and task-negative network with EEG alpha rhythm and oxygenated hemoglobin of fNIRS.

- The present study showed for the first time, less cortical connectivity within the inter- and intra-hemispheric parts of the brain under stress condition based on EEG and fNIRS (using seven frequency intervals) signals.

- The study presents a novel fusion based on joint independent component analysis in which temporal EEG components and spatial fNIRS components were fused in the feature level and their mixing matrix was obtained using neural network optimization technique. Another novel approach was based on canonical correlation analysis in which the temporal EEG components and the fNIRS spatial components were fused by maximizing their inter-subject covariations.

- The study discovered the right ventrolateral PFC as the dominant region to mental stress which could serve as a base to neurofeedback in the clinical practice.

## 5.3 Future work

The study described in this thesis possesses several limitations, and more research is necessary in order to advance this line of inquiry. First, this study was performed in healthy male subjects. The cortical activities in the PFC are expected to behave differently in young and healthy than in more elderly populations. In order to truly understand and model stress at the workplace, many groups of individuals (young, elderly, females, subjects with neurological impairments, etc.) should be tested. Second, the EEG and the fNIRS analysis in this study have been aimed at improving the detection rate of mental stress rather than distinguishing between different levels of stress. The focus of the present analysis is assessing whether the proposed modalities and the proposed fusion techniques can improve the detection rate of mental stress. In the future, it also may be interesting to grade mental stress into different levels. Additionally, in order to apply the findings of this study in clinical practice, more extensive research considering the aforementioned limitations need be done.

# REFERENCES

[1] H. Selye, "The Stress Syndrome," *AJN The American Journal of Nursing,* vol. 65, pp. 97-99, 1965.

[2] M. I. Ismail and T. Teck-Hong, "Identifying work-related stress among employees in the Malaysian financial sector," *World,* vol. 3, pp. 229-243, 2011.

[3] A. Alberdi, A. Aztiria, and A. Basarab, "Towards an automatic early stress recognition system for office environments based on multimodal measurements: A review," *Journal of biomedical informatics,* vol. 59, pp. 49-75, 2016.

[4] E. A. f. Safety and H. a. Work, "European Opinion Poll on Occupational Safety and Health," ed: EU-OSHA Bilbao, 2013.

[5] N. Sharma and T. Gedeon, "Modeling observer stress for typical real environments," *Expert Systems with Applications,* vol. 41, pp. 2231-2238, 2014.

[6] B. S. McEwen, "Central effects of stress hormones in health and disease: Understanding the protective and damaging effects of stress and stress mediators," *European journal of pharmacology,* vol. 583, pp. 174-185, 2008.

[7] C. Hammen, "Stress and depression," *Annu. Rev. Clin. Psychol.,* vol. 1, pp. 293-319, 2005.

[8] P. Strike and A. Steptoe, "Systematic review of mental stress-induced myocardial ischaemia," *European heart journal,* vol. 24, pp. 690-703, 2003.

[9] A. Vyas, R. Mitra, B. S. Rao, and S. Chattarji, "Chronic stress induces contrasting patterns of dendritic remodeling in hippocampal and amygdaloid neurons," *The Journal of Neuroscience,* vol. 22, pp. 6810-6818, 2002.

[10] S. Cohen, R. C. Kessler, and L. U. Gordon, *Measuring stress: A guide for health and social scientists*: Oxford University Press, 1995.

[11] T. H. Holmes and R. H. Rahe, "The social readjustment rating scale," *Journal of psychosomatic research,* vol. 11, pp. 213-218, 1967.

[12] T.-K. Liu, Y.-P. Chen, Z.-Y. Hou, C.-C. Wang, and J.-H. Chou, "Noninvasive evaluation of mental stress using by a refined rough set technique based on

biomedical signals," *Artificial intelligence in medicine,* vol. 61, pp. 97-103, 2014.

[13]    A. Kaushik, A. Vasudev, S. K. Arya, S. K. Pasha, and S. Bhansali, "Recent advances in cortisol sensing technologies for point-of-care application," *Biosensors and Bioelectronics,* vol. 53, pp. 499-512, 2014.

[14]    D. H. Hellhammer, S. Wüst, and B. M. Kudielka, "Salivary cortisol as a biomarker in stress research," *Psychoneuroendocrinology,* vol. 34, pp. 163-171, 2009.

[15]    N. M. Stephens, S. S. Townsend, H. R. Markus, and L. T. Phillips, "A cultural mismatch: Independent cultural norms produce greater increases in cortisol and more negative emotions among first-generation college students," *Journal of Experimental Social Psychology,* vol. 48, pp. 1389-1393, 2012.

[16]    Y. Noto, T. Sato, M. Kudo, K. Kurata, and K. Hirota, "The relationship between salivary biomarkers and state-trait anxiety inventory score under mental arithmetic stress: a pilot study," *Anesthesia & Analgesia,* vol. 101, pp. 1873-1876, 2005.

[17]    J. A. Bosch, E. J. de Geus, E. C. Veerman, J. Hoogstraten, and A. V. N. Amerongen, "Innate secretory immunity in response to laboratory stressors that evoke distinct patterns of cardiac autonomic activity," *Psychosomatic medicine,* vol. 65, pp. 245-258, 2003.

[18]    R. Den, M. Toda, S. Nagasawa, K. Kitamura, and K. Morimoto, "Circadian rhythm of human salivary chromogranin A," *Biomedical Research,* vol. 28, pp. 57-60, 2007.

[19]    Y. Okada, T. Y. Yoto, T.-a. Suzuki, S. Sakuragawa, H. Sakakibara, K. Shimoi*, et al.*, "Wearable ECG recorder with acceleration sensors for monitoring daily stress," *Journal of Medical and Biological Engineering,* vol. 33, pp. 420-426, 2013.

[20]    J. A. Healey and R. W. Picard, "Detecting stress during real-world driving tasks using physiological sensors," *IEEE Transactions on intelligent transportation systems,* vol. 6, pp. 156-166, 2005.

[21] G. K. Verma and U. S. Tiwary, "Multimodal fusion framework: A multiresolution approach for emotion classification and recognition from physiological signals," *NeuroImage,* vol. 102, pp. 162-172, 2014.

[22] J. F. Thayer, F. Åhs, M. Fredrikson, J. J. Sollers, and T. D. Wager, "A meta-analysis of heart rate variability and neuroimaging studies: implications for heart rate variability as a marker of stress and health," *Neuroscience & Biobehavioral Reviews,* vol. 36, pp. 747-756, 2012.

[23] B. H. Eijckelhof, M. A. Huysmans, B. M. Blatter, P. C. Leider, P. W. Johnson, J. H. van Dieën*, et al.*, "Office workers' computer use patterns are associated with workplace stressors," *Applied ergonomics,* vol. 45, pp. 1660-1667, 2014.

[24] S. Bishop, J. Duncan, M. Brett, and A. D. Lawrence, "Prefrontal cortical function and anxiety: controlling attention to threat-related stimuli," *Nature neuroscience,* vol. 7, pp. 184-188, 2004.

[25] A. F. Arnsten, "Prefrontal cortical network connections: key site of vulnerability in stress and schizophrenia," *International Journal of Developmental Neuroscience,* vol. 29, pp. 215-223, 2011.

[26] C. Liston, B. McEwen, and B. Casey, "Psychosocial stress reversibly disrupts prefrontal processing and attentional control," *Proceedings of the National Academy of Sciences,* vol. 106, pp. 912-917, 2009.

[27] A. F. Arnsten, "Stress weakens prefrontal networks: molecular insults to higher cognition," *Nature neuroscience,* vol. 18, pp. 1376-1385, 2015.

[28] K. Starcke and M. Brand, "Decision making under stress: a selective review," *Neuroscience & Biobehavioral Reviews,* vol. 36, pp. 1228-1248, 2012.

[29] A. F. Arnsten, "Stress signalling pathways that impair prefrontal cortex structure and function," *Nature Reviews Neuroscience,* vol. 10, pp. 410-422, 2009.

[30] K. Dedovic, C. D'Aguiar, and J. C. Pruessner, "What stress does to your brain: a review of neuroimaging studies," *The Canadian Journal of Psychiatry,* vol. 54, pp. 6-15, 2009.

[31] B. Leuner and T. Shors, "Stress, anxiety, and dendritic spines: what are the connections?," *Neuroscience,* vol. 251, pp. 108-119, 2013.

[32]    S. Qin, E. J. Hermans, H. J. van Marle, J. Luo, and G. Fernández, "Acute psychological stress reduces working memory-related activity in the dorsolateral prefrontal cortex," *Biological psychiatry,* vol. 66, pp. 25-32, 2009.

[33]    J. Wang, H. Rao, G. S. Wetmore, P. M. Furlan, M. Korczykowski, D. F. Dinges*, et al.*, "Perfusion functional MRI reveals cerebral blood flow pattern under psychological stress," *Proceedings of the National Academy of Sciences of the United States of America,* vol. 102, pp. 17804-17809, 2005.

[34]    J. C. Pruessner, K. Dedovic, N. Khalili-Mahani, V. Engert, M. Pruessner, C. Buss*, et al.*, "Deactivation of the limbic system during acute psychosocial stress: evidence from positron emission tomography and functional magnetic resonance imaging studies," *Biological psychiatry,* vol. 63, pp. 234-240, 2008.

[35]    J. A. Levine, I. T. Pavlidis, L. MacBride, Z. Zhu, and P. Tsiamyrtzis, "Description and clinical studies of a device for the instantaneous detection of office-place stress," *Work,* vol. 34, pp. 359-364, 2009.

[36]    L. Vézard, P. Legrand, M. Chavent, F. Faïta-Aïnseba, and L. Trujillo, "EEG classification for the detection of mental states," *Applied Soft Computing,* vol. 32, pp. 113-131, 2015.

[37]    C. M. Michel and M. M. Murray, "Towards the utilization of EEG as a brain imaging tool," *Neuroimage,* vol. 61, pp. 371-385, 2012.

[38]    M. Huiku, K. Uutela, M. Van Gils, I. Korhonen, M. Kymäläinen, P. Meriläinen*, et al.*, "Assessment of surgical stress during general anaesthesia," *British Journal of Anaesthesia,* vol. 98, pp. 447-455, 2007.

[39]    G. Chanel, J. Kronegg, D. Grandjean, and T. Pun, "Emotion assessment: Arousal evaluation using EEG's and peripheral physiological signals," *Multimedia content representation, classification and security,* pp. 530-537, 2006.

[40]    T. Takahashi, T. Murata, T. Hamada, M. Omori, H. Kosaka, M. Kikuchi*, et al.*, "Changes in EEG and autonomic nervous activity during meditation and their association with personality traits," *International Journal of Psychophysiology,* vol. 55, pp. 199-207, 2005.

[41]  R. E. Wheeler, R. J. Davidson, and A. J. Tomarken, "Frontal brain asymmetry and emotional reactivity: A biological substrate of affective style," *Psychophysiology,* vol. 30, pp. 82-89, 1993.

[42]  R. S. Lewis, N. Y. Weekes, and T. H. Wang, "The effect of a naturalistic stressor on frontal EEG asymmetry, stress, and health," *Biological psychology,* vol. 75, pp. 239-247, 2007.

[43]  R. W. Hill and E. Castro, *Healing young brains: The neurofeedback solution*: Hampton Roads Publishing, 2009.

[44]  S.-H. Seo and J.-T. Lee, *Stress and EEG*: INTECH Open Access Publisher, 2010.

[45]  Y. Choi, M. Kim, and C. Chun, "Measurement of occupants' stress based on electroencephalograms (EEG) in twelve combined environments," *Building and Environment,* vol. 88, pp. 65-72, 2015.

[46]  M. Thompson and L. Thompson, "Neurofeedback for stress management," *Principles and practice of stress management,* pp. 249-287, 2007.

[47]  J. Alonso, S. Romero, M. Ballester, R. Antonijoan, and M. Mañanas, "Stress assessment based on EEG univariate features and functional connectivity measures," *Physiological measurement,* vol. 36, p. 1351, 2015.

[48]  M. Gärtner, S. Grimm, and M. Bajbouj, "Frontal midline theta oscillations during mental arithmetic: effects of stress," *Frontiers in behavioral neuroscience,* vol. 9, 2015.

[49]  P. Missonnier, F. R. Herrmann, C. Rodriguez, M.-P. Deiber, P. Millet, L. Fazio-Costa*, et al.*, "Age-related differences on event-related potentials and brain rhythm oscillations during working memory activation," *Journal of Neural Transmission,* vol. 118, pp. 945-955, 2011.

[50]  T. Harmony, T. Fernández, J. Silva, J. Bernal, L. Díaz-Comas, A. Reyes*, et al.*, "EEG delta activity: an indicator of attention to internal processing during performance of mental tasks," *International journal of psychophysiology,* vol. 24, pp. 161-171, 1996.

[51]  A. C. Marshall, N. R. Cooper, R. Segrave, and N. Geeraert, "The effects of long-term stress exposure on aging cognition: a behavioral and EEG investigation," *Neurobiology of aging,* vol. 36, pp. 2136-2144, 2015.

[52]    N. L. Lopez-Duran, R. Nusslock, C. George, and M. Kovacs, "Frontal EEG asymmetry moderates the effects of stressful life events on internalizing symptoms in children at familial risk for depression," *Psychophysiology,* vol. 49, pp. 510-521, 2012.

[53]    F. Lotte, M. Congedo, A. Lécuyer, F. Lamarche, and B. Arnaldi, "A review of classification algorithms for EEG-based brain–computer interfaces," *Journal of neural engineering,* vol. 4, p. R1, 2007.

[54]    N. Sharma and T. Gedeon, "Objective measures, sensors and computational techniques for stress recognition and classification: A survey," *Computer methods and programs in biomedicine,* vol. 108, pp. 1287-1301, 2012.

[55]    L. Xin, C. Zetao, Z. Yunpeng, X. Jiali, W. Shuicai, and Z. Yanjun, "Stress State Evaluation by Improved Support Vector Machine," *Journal of Medical Imaging and Health Informatics,* vol. 5, pp. 742-747, 2015.

[56]    G. Chanel, J. J. Kierkels, M. Soleymani, and T. Pun, "Short-term emotion assessment in a recall paradigm," *International Journal of Human-Computer Studies,* vol. 67, pp. 607-627, 2009.

[57]    C. Babiloni, V. Pizzella, C. Del Gratta, A. Ferretti, and G. L. Romani, "Fundamentals of electroencefalography, magnetoencefalography, and functional magnetic resonance imaging," *International review of neurobiology,* vol. 86, pp. 67-80, 2009.

[58]    A. Villringer and B. Chance, "Non-invasive optical spectroscopy and imaging of human brain function," *Trends in neurosciences,* vol. 20, pp. 435-442, 1997.

[59]    A. Sassaroli and S. Fantini, "Comment on the modified Beer? Lambert law for scattering media," *Physics in Medicine and Biology,* vol. 49, p. N255, 2004.

[60]    M. Strait and M. Scheutz, "What we can and cannot (yet) do with functional near infrared spectroscopy," *Frontiers in neuroscience,* vol. 8, 2014.

[61]    A. Villringer, J. Planck, C. Hock, L. Schleinkofer, and U. Dirnagl, "Near infrared spectroscopy (NIRS): a new tool to study hemodynamic changes during activation of brain function in human adults," *Neuroscience letters,* vol. 154, pp. 101-104, 1993.

[62]    Y. Hoshi, "Towards the next generation of near-infrared spectroscopy," *Philosophical Transactions of the Royal Society of London A: Mathematical, Physical and Engineering Sciences,* vol. 369, pp. 4425-4439, 2011.

[63]    T. J. Huppert, R. D. Hoge, A. M. Dale, M. A. Franceschini, and D. A. Boas, "Quantitative spatial comparison of diffuse optical imaging with blood oxygen level-dependent and arterial spin labeling-based functional magnetic resonance imaging," *Journal of biomedical optics,* vol. 11, pp. 064018-064018-16, 2006.

[64]    M. J. Herrmann, M. M. Plichta, A.-C. Ehlis, and A. J. Fallgatter, "Optical topography during a Go–NoGo task assessed with multi-channel near-infrared spectroscopy," *Behavioural brain research,* vol. 160, pp. 135-140, 2005.

[65]    M. Boecker, M. M. Buecheler, M. L. Schroeter, and S. Gauggel, "Prefrontal brain activation during stop-signal response inhibition: an event-related functional near-infrared spectroscopy study," *Behavioural brain research,* vol. 176, pp. 259-266, 2007.

[66]    W. C. Vogt, E. Romero, S. M. LaConte, and C. G. Rylander, "Mechanical indentation improves cerebral blood oxygenation signal quality of functional near-infrared spectroscopy (fNIRS) during breath holding," in *SPIE BiOS*, 2013, pp. 85782K-85782K-7.

[67]    H. Sato, R. Aoki, T. Katura, R. Matsuda, and H. Koizumi, "Correlation of within-individual fluctuation of depressed mood with prefrontal cortex activity during verbal working memory task: optical topography study," *Journal of biomedical optics,* vol. 16, pp. 126007-1260077, 2011.

[68]    R. Takizawa, M. Fukuda, S. Kawasaki, K. Kasai, M. Mimura, S. Pu*, et al.*, "Neuroimaging-aided differential diagnosis of the depressive state," *Neuroimage,* vol. 85, pp. 498-507, 2014.

[69]    M. L. Schroeter, T. Kupka, T. Mildner, K. Uludağ, and D. Y. von Cramon, "Investigating the post-stimulus undershoot of the BOLD signal—a simultaneous fMRI and fNIRS study," *Neuroimage,* vol. 30, pp. 349-358, 2006.

[70]    L. Li, P. Du, T. Li, Q. Luo, and H. Gong, "Design and evaluation of a simultaneous fNIRS/ERP instrument," in *Biomedical Optics (BiOS) 2007*, 2007, pp. 643429-643429-6.

[71]    D. McDuff, A. Karlson, A. Kapoor, A. Roseway, and M. Czerwinski, "AffectAura: an intelligent system for emotional memory," in *Proceedings of the SIGCHI Conference on Human Factors in Computing Systems*, 2012, pp. 849-858.

[72]    F. Custodis, J.-C. Reil, U. Laufs, and M. Böhm, "Heart rate: a global target for cardiovascular disease and therapy along the cardiovascular disease continuum," *Journal of cardiology,* vol. 62, pp. 183-187, 2013.

[73]    N. Hjortskov, D. Rissén, A. K. Blangsted, N. Fallentin, U. Lundberg, and K. Søgaard, "The effect of mental stress on heart rate variability and blood pressure during computer work," *European journal of applied physiology,* vol. 92, pp. 84-89, 2004.

[74]    M. Quazi, S. Mukhopadhyay, N. Suryadevara, and Y. Huang, "Towards the smart sensors based human emotion recognition," in *Instrumentation and Measurement Technology Conference (I2MTC), 2012 IEEE International*, 2012, pp. 2365-2370.

[75]    J. Hernandez, P. Paredes, A. Roseway, and M. Czerwinski, "Under pressure: sensing stress of computer users," in *Proceedings of the SIGCHI conference on Human factors in computing systems*, 2014, pp. 51-60.

[76]    L. M. Vizer, L. Zhou, and A. Sears, "Automated stress detection using keystroke and linguistic features: An exploratory study," *International Journal of Human-Computer Studies,* vol. 67, pp. 870-886, 2009.

[77]    F. Putze, S. Hesslinger, C. Y. Tse, Y. Huang, C. Herff, C. Guan*, et al.*, "Hybrid fNIRS-EEG based classification of auditory and visual perception processes," *Frontiers in Neuroscience,* vol. 8, 2014.

[78]    Y. Blokland, L. Spyrou, D. Thijssen, T. Eijsvogels, W. Colier, M. Floor-Westerdijk*, et al.*, "Combined EEG-fNIRS decoding of motor attempt and imagery for brain switch control: an offline study in patients with tetraplegia," *Neural Systems and Rehabilitation Engineering, IEEE Transactions on,* vol. 22, pp. 222-229, 2014.

[79]     S. Fazli, J. Mehnert, J. Steinbrink, G. Curio, A. Villringer, K.-R. Müller, *et al.*, "Enhanced performance by a hybrid NIRS–EEG brain computer interface," *Neuroimage,* vol. 59, pp. 519-529, 2012.

[80]     D. N. Lenkov, A. B. Volnova, A. R. Pope, and V. Tsytsarev, "Advantages and limitations of brain imaging methods in the research of absence epilepsy in humans and animal models," *Journal of neuroscience methods,* vol. 212, pp. 195-202, 2013.

[81]     A. Machado, J.-M. Lina, J. Tremblay, M. Lassonde, D. K. Nguyen, F. Lesage, *et al.*, "Detection of hemodynamic responses to epileptic activity using simultaneous Electro-EncephaloGraphy (EEG)/Near Infra Red Spectroscopy (NIRS) acquisitions," *Neuroimage,* vol. 56, pp. 114-125, 2011.

[82]     R. M. Sapolsky, L. C. Krey, and B. S. McEwen, "The neuroendocrinology of stress and aging: the glucocorticoid cascade hypothesis*," *Endocrine reviews,* vol. 7, pp. 284-301, 1986.

[83]     K. Palanisamy, M. Murugappan, and S. Yaacob, "Multiple physiological signal-based human stress identification using non-linear classifiers," *Elektronika ir elektrotechnika,* vol. 19, pp. 80-85, 2013.

[84]     N. Sharma, A. Dhall, T. Gedeon, and R. Goecke, "Thermal spatio-temporal data for stress recognition," *EURASIP Journal on Image and Video Processing,* vol. 2014, p. 1, 2014.

[85]     L. Ossewaarde, E. J. Hermans, G. A. van Wingen, S. C. Kooijman, I.-M. Johansson, T. Bäckström, *et al.*, "Neural mechanisms underlying changes in stress-sensitivity across the menstrual cycle," *Psychoneuroendocrinology,* vol. 35, pp. 47-55, 2010.

[86]     M. J. Henckens, E. J. Hermans, Z. Pu, M. Joëls, and G. Fernández, "Stressed memories: how acute stress affects memory formation in humans," *The Journal of Neuroscience,* vol. 29, pp. 10111-10119, 2009.

[87]     H. J. van Marle, E. J. Hermans, S. Qin, and G. Fernández, "From specificity to sensitivity: how acute stress affects amygdala processing of biologically salient stimuli," *Biological psychiatry,* vol. 66, pp. 649-655, 2009.

[88] E. J. Hermans, H. J. van Marle, L. Ossewaarde, M. J. Henckens, S. Qin, M. T. van Kesteren*, et al.*, "Stress-related noradrenergic activity prompts large-scale neural network reconfiguration," *Science,* vol. 334, pp. 1151-1153, 2011.

[89] C. Z. Wei, "Stress emotion recognition based on RSP and EMG signals," in *Advanced Materials Research*, 2013, pp. 827-831.

[90] A. C. Aguiar, M. Kaiseler, H. Meinedo, T. E. Abrudan, and P. R. Almeida, "Speech stress assessment using physiological and psychological measures," in *Proceedings of the 2013 ACM conference on Pervasive and ubiquitous computing adjunct publication*, 2013, pp. 921-930.

[91] L. Schwabe and O. T. Wolf, "Stress prompts habit behavior in humans," *The Journal of Neuroscience,* vol. 29, pp. 7191-7198, 2009.

[92] K. Dedovic, R. Renwick, N. K. Mahani, and V. Engert, "The Montreal Imaging Stress Task: using functional imaging to investigate the effects of perceiving and processing psychosocial stress in the human brain," *Journal of psychiatry & neuroscience: JPN,* vol. 30, p. 319, 2005.

[93] M. Milczarek, E. Schneider, and R. Eusebio, "Stress at work-facts and figures," *European Risk Observatory Report. Luxembourg: European Communities,* 2009.

[94] K. Peternel, M. Pogačnik, R. Tavčar, and A. Kos, "A presence-based context-aware chronic stress recognition system," *Sensors,* vol. 12, pp. 15888-15906, 2012.

[95] J. Wijsman, B. Grundlehner, H. Liu, J. Penders, and H. Hermens, "Wearable physiological sensors reflect mental stress state in office-like situations," in *Affective Computing and Intelligent Interaction (ACII), 2013 Humaine Association Conference on*, 2013, pp. 600-605.

[96] M. Bickford, "Stress in the Workplace: A General Overview of the Causes, the Effects, and the Solutions," *Canadian Mental Health Association Newfoundland and Labrador Division,* pp. 1-3, 2005.

[97] D. Carneiro, J. C. Castillo, P. Novais, A. Fernández-Caballero, and J. Neves, "Multimodal behavioral analysis for non-invasive stress detection," *Expert Systems with Applications,* vol. 39, pp. 13376-13389, 2012.

[98]    R. Cousins*, C. J. Mackay, S. D. Clarke, C. Kelly, P. J. Kelly, and R. H. McCaig, "'Management standards' work-related stress in the UK: Practical development," *Work & Stress,* vol. 18, pp. 113-136, 2004.

[99]    T. Hayashi, Y. Mizuno-Matsumoto, E. Okamoto, M. Kato, and T. Murata, "An fMRI study of brain processing related to stress states," in *World Automation Congress (WAC), 2012*, 2012, pp. 1-6.

[100]   I. Ulstein, T. B. Wyller, and K. Engedal, "High score on the Relative Stress Scale, a marker of possible psychiatric disorder in family carers of patients with dementia," *International journal of geriatric psychiatry,* vol. 22, pp. 195-202, 2007.

[101]   L. Lemyre and R. Tessier, "Measuring psychological stress. Concept, model, and measurement instrument in primary care research," *Canadian Family Physician,* vol. 49, p. 1159, 2003.

[102]   J. E. Wartella, S. M. Auerbach, and K. R. Ward, "Emotional distress, coping and adjustment in family members of neuroscience intensive care unit patients," *Journal of psychosomatic research,* vol. 66, pp. 503-509, 2009.

[103]   S. J. Lupien and F. Seguin, "How to Measure Stress in Humans," Citeseer2007.

[104]   M. Yamaguchi, T. Kanemori, M. Kanemaru, N. Takai, Y. Mizuno, and H. Yoshida, "Performance evaluation of salivary amylase activity monitor," *Biosensors and Bioelectronics,* vol. 20, pp. 491-497, 2004.

[105]   A. S. DeSantis, E. K. Adam, L. C. Hawkley, B. M. Kudielka, and J. T. Cacioppo, "Racial and ethnic differences in diurnal cortisol rhythms: are they consistent over time?," *Psychosomatic medicine,* vol. 77, pp. 6-15, 2015.

[106]   D. A. Granger, K. T. Kivlighan, M. El-SHEIKH, E. B. Gordis, and L. R. Stroud, "Salivary α-amylase in biobehavioral research," *Annals of the New York Academy of Sciences,* vol. 1098, pp. 122-144, 2007.

[107]   V. Engert, S. Vogel, S. I. Efanov, A. Duchesne, V. Corbo, N. Ali*, et al.*, "Investigation into the cross-correlation of salivary cortisol and alpha-amylase responses to psychological stress," *Psychoneuroendocrinology,* vol. 36, pp. 1294-1302, 2011.

[108] N. Rohleder, U. M. Nater, J. M. Wolf, U. Ehlert, and C. Kirschbaum, "Psychosocial stress-induced activation of salivary alpha-amylase: an indicator of sympathetic activity?," *Annals of the New York Academy of Sciences,* vol. 1032, pp. 258-263, 2004.

[109] N. Rohleder and U. M. Nater, "Determinants of salivary α-amylase in humans and methodological considerations," *Psychoneuroendocrinology,* vol. 34, pp. 469-485, 2009.

[110] U. R. Acharya, K. P. Joseph, N. Kannathal, C. M. Lim, and J. S. Suri, "Heart rate variability: a review," *Medical and biological engineering and computing,* vol. 44, pp. 1031-1051, 2006.

[111] B. Cinaz, B. Arnrich, R. La Marca, and G. Tröster, "Monitoring of mental workload levels during an everyday life office-work scenario," *Personal and ubiquitous computing,* vol. 17, pp. 229-239, 2013.

[112] P. Melillo, M. Bracale, and L. Pecchia, "Nonlinear Heart Rate Variability features for real-life stress detection. Case study: students under stress due to university examination," *Biomedical engineering online,* vol. 10, p. 1, 2011.

[113] A. de Santos Sierra, C. S. Ávila, G. B. del Pozo, and J. G. Casanova, "Stress detection by means of stress physiological template," in *Nature and Biologically Inspired Computing (NaBIC), 2011 Third World Congress on*, 2011, pp. 131-136.

[114] F. Seoane, I. Mohino-Herranz, J. Ferreira, L. Alvarez, R. Buendia, D. Ayllón*, et al.*, "Wearable biomedical measurement systems for assessment of mental stress of combatants in real time," *Sensors,* vol. 14, pp. 7120-7141, 2014.

[115] G. Ogedegbe and T. Pickering, "Principles and techniques of blood pressure measurement," *Cardiology clinics,* vol. 28, pp. 571-586, 2010.

[116] C. Maaoui, A. Pruski, and F. Abdat, "Emotion recognition for human-machine communication," in *2008 IEEE/RSJ International Conference on Intelligent Robots and Systems*, 2008, pp. 1210-1215.

[117] M. N. H. Mohd, M. Kashima, K. Sato, and M. Watanabe, "Facial visual-infrared stereo vision fusion measurement as an alternative for physiological measurement," *Journal of Biomedical Image Processing (JBIP),* vol. 1, pp. 34-44, 2014.

[118] J. Wijsman, B. Grundlehner, J. Penders, and H. Hermens, "Trapezius muscle EMG as predictor of mental stress," *ACM Transactions on Embedded Computing Systems (TECS),* vol. 12, p. 99, 2013.

[119] J. Taelman, T. Adriaensen, C. van der Horst, T. Linz, and A. Spaepen, "Textile integrated contactless EMG sensing for stress analysis," in *2007 29th Annual International Conference of the IEEE Engineering in Medicine and Biology Society*, 2007, pp. 3966-3969.

[120] Y. Shi, M. H. Nguyen, P. Blitz, B. French, S. Fisk, F. De la Torre*, et al.*, "Personalized stress detection from physiological measurements," in *International symposium on quality of life technology*, 2010, pp. 28-29.

[121] J. Lázaro, A. Alcaine, E. Gil, P. Laguna, and R. Bailón, "Electrocardiogram derived respiration from QRS slopes," in *2013 35th Annual International Conference of the IEEE Engineering in Medicine and Biology Society (EMBC)*, 2013, pp. 3913-3916.

[122] J. Zhai and A. Barreto, "Stress detection in computer users based on digital signal processing of noninvasive physiological variables," in *Engineering in Medicine and Biology Society, 2006. EMBS'06. 28th Annual International Conference of the IEEE*, 2006, pp. 1355-1358.

[123] H. Chigira, M. Kobayashi, and A. Maeda, "Mouse with photo-plethysmographic surfaces for unobtrusive stress monitoring," in *2012 IEEE Second International Conference on Consumer Electronics-Berlin (ICCE-Berlin)*, 2012.

[124] J. Ramos, J.-H. Hong, and A. K. Dey, "Stress Recognition-A Step Outside the Lab," in *PhyCS*, 2014, pp. 107-118.

[125] W. Liao, W. Zhang, Z. Zhu, and Q. Ji, "A real-time human stress monitoring system using dynamic Bayesian network," in *2005 IEEE Computer Society Conference on Computer Vision and Pattern Recognition (CVPR'05)-Workshops*, 2005, pp. 70-70.

[126] A. Barreto, J. Zhai, N. Rishe, and Y. Gao, "Measurement of pupil diameter variations as a physiological indicator of the affective state in a computer user," *Biomedical sciences instrumentation,* vol. 43, pp. 146-151, 2006.

[127] A. Barreto, J. Zhai, N. Rishe, and Y. Gao, "Significance of pupil diameter measurements for the assessment of affective state in computer users," in *Advances and innovations in systems, computing sciences and software engineering*, ed: Springer, 2007, pp. 59-64.

[128] K. Sakamoto, S. Aoyama, S. Asahara, H. Mizushina, and H. Kaneko, "Relationship between emotional state and pupil diameter variability under various types of workload stress," in *International Conference on Ergonomics and Health Aspects of Work with Computers*, 2009, pp. 177-185.

[129] P. Ren, A. Barreto, J. Huang, Y. Gao, F. R. Ortega, and M. Adjouadi, "Off-line and on-line stress detection through processing of the pupil diameter signal," *Annals of biomedical engineering,* vol. 42, pp. 162-176, 2014.

[130] D. Shastri, M. Papadakis, P. Tsiamyrtzis, B. Bass, and I. Pavlidis, "Perinasal imaging of physiological stress and its affective potential," *IEEE Transactions on Affective Computing,* vol. 3, pp. 366-378, 2012.

[131] T. Chen, P. Yuen, M. Richardson, G. Liu, and Z. She, "Detection of psychological stress using a hyperspectral imaging technique," *IEEE Transactions on Affective Computing,* vol. 5, pp. 391-405, 2014.

[132] A. Kaklauskas, E. K. Zavadskas, M. Seniut, G. Dzemyda, V. Stankevic, C. Simkevičius*, et al.*, "Web-based biometric computer mouse advisory system to analyze a user's emotions and work productivity," *Engineering Applications of Artificial Intelligence,* vol. 24, pp. 928-945, 2011.

[133] C. Epp, M. Lippold, and R. L. Mandryk, "Identifying emotional states using keystroke dynamics," in *Proceedings of the SIGCHI Conference on Human Factors in Computing Systems*, 2011, pp. 715-724.

[134] A. Kołakowska, "A review of emotion recognition methods based on keystroke dynamics and mouse movements," in *2013 6th International Conference on Human System Interactions (HSI)*, 2013, pp. 548-555.

[135] P. Zimmermann, S. Guttormsen, B. Danuser, and P. Gomez, "Affective computing—a rationale for measuring mood with mouse and keyboard," *International journal of occupational safety and ergonomics,* vol. 9, pp. 539-551, 2003.

[136] A. Alhothali, "Modeling user affect using interaction events," 2011.

[137]  M. Gomes, T. Oliveira, F. Silva, D. Carneiro, and P. Novais, "Establishing the relationship between personality traits and stress in an intelligent environment," in *International Conference on Industrial, Engineering and Other Applications of Applied Intelligent Systems*, 2014, pp. 378-387.

[138]  A. Kapoor and R. W. Picard, "Multimodal affect recognition in learning environments," in *Proceedings of the 13th annual ACM international conference on Multimedia*, 2005, pp. 677-682.

[139]  B. Arnrich, C. Setz, R. La Marca, G. Tröster, and U. Ehlert, "What does your chair know about your stress level?," *IEEE Transactions on Information Technology in Biomedicine,* vol. 14, pp. 207-214, 2010.

[140]  N. Ahmad, A. Szymkowiak, and P. Campbell, "Keystroke dynamics in the pre-touchscreen era," *Frontiers in human neuroscience,* vol. 7, p. 835, 2013.

[141]  S. Das and K. Yamada, "Evaluating Instantaneous Psychological Stress from Emotional Composition of a Facial Expression," *JACIII,* vol. 17, pp. 480-492, 2013.

[142]  E. N. No, "EVALUATION OF THE HUMAN VOICE FOR INDICATIONS OF WORKLOAD INDUCED STRESS IN THE AVIATION ENVIRONMENT," 2006.

[143]  H. Lu, D. Frauendorfer, M. Rabbi, M. S. Mast, G. T. Chittaranjan, A. T. Campbell*, et al.*, "Stresssense: Detecting stress in unconstrained acoustic environments using smartphones," in *Proceedings of the 2012 ACM Conference on Ubiquitous Computing*, 2012, pp. 351-360.

[144]  P. Adams, M. Rabbi, T. Rahman, M. Matthews, A. Voida, G. Gay*, et al.*, "Towards personal stress informatics: Comparing minimally invasive techniques for measuring daily stress in the wild," in *Proceedings of the 8th International Conference on Pervasive Computing Technologies for Healthcare*, 2014, pp. 72-79.

[145]  G. Demenko and M. Jastrzebska, "Analysis of voice stress in call centers conversations," in *Proceedings of Speech Prosody, 6th International Conference, Shanghai, China*, 2012.

[146]  H. Kurniawan, A. V. Maslov, and M. Pechenizkiy, "Stress detection from speech and galvanic skin response signals," in *Proceedings of the 26th IEEE*

*International Symposium on Computer-Based Medical Systems*, 2013, pp. 209-214.

[147] S. Lloyd-Fox, A. Blasi, and C. Elwell, "Illuminating the developing brain: the past, present and future of functional near infrared spectroscopy," *Neuroscience & Biobehavioral Reviews,* vol. 34, pp. 269-284, 2010.

[148] R. N. Aslin, M. Shukla, and L. L. Emberson, "Hemodynamic correlates of cognition in human infants," *Annual review of psychology,* vol. 66, pp. 349-379, 2015.

[149] R. B. Buxton, E. C. Wong, and L. R. Frank, "Dynamics of blood flow and oxygenation changes during brain activation: the balloon model," *Magnetic resonance in medicine,* vol. 39, pp. 855-864, 1998.

[150] J. D. Watson, "Images of the working brain: understanding human brain function with positron emission tomography," *Journal of neuroscience methods,* vol. 74, pp. 245-256, 1997.

[151] N. K. Logothetis, J. Pauls, M. Augath, T. Trinath, and A. Oeltermann, "Neurophysiological investigation of the basis of the fMRI signal," *Nature,* vol. 412, pp. 150-157, 2001.

[152] F. Scholkmann, S. Kleiser, A. J. Metz, R. Zimmermann, J. M. Pavia, U. Wolf, *et al.*, "A review on continuous wave functional near-infrared spectroscopy and imaging instrumentation and methodology," *Neuroimage,* vol. 85, pp. 6-27, 2014.

[153] E. Puterman, A. O'Donovan, N. E. Adler, A. J. Tomiyama, M. Kemeny, O. M. Wolkowitz, *et al.*, "Physical activity moderates stressor-induced rumination on cortisol reactivity," *Psychosomatic medicine,* vol. 73, p. 604, 2011.

[154] Y. Tran, R. Thuraisingham, N. Wijesuriya, H. Nguyen, and A. Craig, "Detecting neural changes during stress and fatigue effectively: a comparison of spectral analysis and sample entropy," in *Neural Engineering, 2007. CNE'07. 3rd International IEEE/EMBS Conference on*, 2007, pp. 350-353.

[155] K. S. Rahnuma, A. Wahab, N. Kamaruddin, and H. Majid, "EEG analysis for understanding stress based on affective model basis function," in *Consumer Electronics (ISCE), 2011 IEEE 15th International Symposium on*, 2011, pp. 592-597.

[156] J. Russel, "Acircumplexmodelofafect," *Journal ofPersonalityand,* 1980.

[157] H. Zhang, Y. Zhu, J. Maniyeri, and C. Guan, "Detection of variations in cognitive workload using multi-modality physiological sensors and a large margin unbiased regression machine," in *2014 36th Annual International Conference of the IEEE Engineering in Medicine and Biology Society*, 2014, pp. 2985-2988.

[158] K. Ishino and M. Hagiwara, "A feeling estimation system using a simple electroencephalograph," in *Systems, Man and Cybernetics, 2003. IEEE International Conference on*, 2003, pp. 4204-4209.

[159] K. Ryu and R. Myung, "Evaluation of mental workload with a combined measure based on physiological indices during a dual task of tracking and mental arithmetic," *International Journal of Industrial Ergonomics,* vol. 35, pp. 991-1009, 2005.

[160] G. Chanel, K. Ansari-Asl, and T. Pun, "Valence-arousal evaluation using physiological signals in an emotion recall paradigm," in *Systems, Man and Cybernetics, 2007. ISIC. IEEE International Conference on*, 2007, pp. 2662-2667.

[161] C.-T. Lin, K.-L. Lin, L.-W. Ko, S.-F. Liang, B.-C. Kuo, and I.-F. Chung, "Nonparametric single-trial EEG feature extraction and classification of driver's cognitive responses," *EURASIP Journal on Advances in Signal Processing,* vol. 2008, pp. 1-10, 2008.

[162] S. A. Hosseini and M. A. Khalilzadeh, "Emotional stress recognition system using EEG and psychophysiological signals: Using new labelling process of EEG signals in emotional stress state," in *Biomedical Engineering and Computer Science (ICBECS), 2010 International Conference on*, 2010, pp. 1-6.

[163] A. Saidatul, M. P. Paulraj, S. Yaacob, and M. A. Yusnita, "Analysis of EEG signals during relaxation and mental stress condition using AR modeling techniques," in *Control System, Computing and Engineering (ICCSCE), 2011 IEEE International Conference on*, 2011, pp. 477-481.

[164] R. Khosrowabadi, C. Quek, K. K. Ang, S. W. Tung, and M. Heijnen, "A Brain-Computer Interface for classifying EEG correlates of chronic mental

stress," in *Neural Networks (IJCNN), The 2011 International Joint Conference on*, 2011, pp. 757-762.

[165] N. Sharma and T. Gedeon, "Modeling stress recognition in typical virtual environments," in *Pervasive Computing Technologies for Healthcare (PervasiveHealth), 2013 7th International Conference on*, 2013, pp. 17-24.

[166] A. Subasi, M. K. Kiymik, A. Alkan, and E. Koklukaya, "Neural network classification of EEG signals by using AR with MLE preprocessing for epileptic seizure detection," *Mathematical and Computational Applications,* vol. 10, pp. 57-70, 2005.

[167] D. Cvetkovic, E. D. Übeyli, and I. Cosic, "Wavelet transform feature extraction from human PPG, ECG, and EEG signal responses to ELF PEMF exposures: A pilot study," *Digital signal processing,* vol. 18, pp. 861-874, 2008.

[168] S. Soltani, "On the use of the wavelet decomposition for time series prediction," *Neurocomputing,* vol. 48, pp. 267-277, 2002.

[169] M. Unser and A. Aldroubi, "A review of wavelets in biomedical applications," *Proceedings of the IEEE,* vol. 84, pp. 626-638, 1996.

[170] N. Hazarika, J. Z. Chen, A. C. Tsoi, and A. Sergejew, "Classification of EEG signals using the wavelet transform," in *Digital Signal Processing Proceedings, 1997. DSP 97., 1997 13th International Conference on*, 1997, pp. 89-92.

[171] A. Sassaroli and S. Fantini, "Comment on the modified Beer-Lambert law for scattering media," *Physics in Medicine and Biology,* vol. 49, pp. N255-N257, 2004.

[172] R. C. Bendall and C. Thompson, "Functional near-infrared spectroscopy: An emerging neuroimaging technique successful in studying the interaction between emotion and cognition," *PsyPAG Quarterly,* vol. 96, pp. 14-17.

[173] D. A. Boas, A. M. Dale, and M. A. Franceschini, "Diffuse optical imaging of brain activation: approaches to optimizing image sensitivity, resolution, and accuracy," *Neuroimage,* vol. 23, pp. S275-S288, 2004.

[174] N. Naseer, M. J. Hong, and K.-S. Hong, "Online binary decision decoding using functional near-infrared spectroscopy for the development of brain–

computer interface," *Experimental brain research,* vol. 232, pp. 555-564, 2014.

[175] D. A. Boas, T. Gaudette, G. Strangman, X. Cheng, J. J. Marota, and J. B. Mandeville, "The accuracy of near infrared spectroscopy and imaging during focal changes in cerebral hemodynamics," *Neuroimage,* vol. 13, pp. 76-90, 2001.

[176] R. B. Buxton, *Introduction to functional magnetic resonance imaging: principles and techniques*: Cambridge university press, 2009.

[177] M. Stangl, G. Bauernfeind, J. Kurzmann, R. Scherer, and C. Neuper, "A haemodynamic brain-computer interface based on real-time classification of near infrared spectroscopy signals during motor imagery and mental arithmetic," *J. Near Infrared Spectrosc,* vol. 21, pp. 157-171, 2013.

[178] H. Kawaguchi and E. Okada, "Evaluation of image reconstruction algorithm for near infrared topography by virtual head phantom," in *European Conference on Biomedical Optics*, 2007, pp. 662906-662906-10.

[179] H. Sato, M. Kiguchi, A. Maki, Y. Fuchino, A. Obata, T. Yoro*, et al.*, "Within-subject reproducibility of near-infrared spectroscopy signals in sensorimotor activation after 6months," *Journal of biomedical optics,* vol. 11, pp. 014021-014021-8, 2006.

[180] M. Schweiger, I. Nissilä, D. A. Boas, and S. R. Arridge, "Image reconstruction in optical tomography in the presence of coupling errors," *Applied optics,* vol. 46, pp. 2743-2756, 2007.

[181] A. Harrivel, T. McKay, A. Hylton, J. King, K. Latorella, S. Peltier*, et al.*, "Toward Improved Headgear for Monitoring with Functional Near Infrared Spectroscopy," *NeuroImage,* vol. 47, p. S141, 2009.

[182] L. M. Hirshfield, R. Gulotta, S. Hirshfield, S. Hincks, M. Russell, R. Ward*, et al.*, "This is your brain on interfaces: enhancing usability testing with functional near-infrared spectroscopy," in *Proceedings of the SIGCHI Conference on Human Factors in Computing Systems*, 2011, pp. 373-382.

[183] S. D. Power, A. Kushki, and T. Chau, "Automatic single-trial discrimination of mental arithmetic, mental singing and the no-control state from prefrontal

activity: toward a three-state NIRS-BCI," *BMC research notes,* vol. 5, p. 141, 2012.

[184] S. Cutini, S. B. Moroa, and S. Biscontib, "Functional near infrared optical imaging in cognitive neuroscience: an introductory," 2012.

[185] R. Sitaram, H. Zhang, C. Guan, M. Thulasidas, Y. Hoshi, A. Ishikawa*, et al.*, "Temporal classification of multichannel near-infrared spectroscopy signals of motor imagery for developing a brain–computer interface," *NeuroImage,* vol. 34, pp. 1416-1427, 2007.

[186] H.-J. Hwang, J.-H. Lim, D.-W. Kim, and C.-H. Im, "Evaluation of various mental task combinations for near-infrared spectroscopy-based brain-computer interfaces," *Journal of biomedical optics,* vol. 19, pp. 077005-077005, 2014.

[187] G. Derosière, S. Dalhoumi, S. Perrey, G. Dray, T. Ward, and Y. Hoshi, "Towards a near infrared spectroscopy-based estimation of operator attentional state," *PloS one,* vol. 9, p. e92045, 2014.

[188] B. Horwitz, "The elusive concept of brain connectivity," *Neuroimage,* vol. 19, pp. 466-470, 2003.

[189] K. J. Friston, "Functional and effective connectivity: a review," *Brain connectivity,* vol. 1, pp. 13-36, 2011.

[190] G. Pfurtscheller and C. Andrew, "Event-related changes of band power and coherence: methodology and interpretation," *Journal of clinical neurophysiology,* vol. 16, p. 512, 1999.

[191] M. Kaminski, A. Brzezicka, J. Kaminski, and K. J. Blinowska, "Measures of Coupling between Neural Populations Based on Granger Causality Principle," *Frontiers in computational neuroscience,* vol. 10, 2016.

[192] A. F. Arnsten, C. D. Paspalas, N. J. Gamo, Y. Yang, and M. Wang, "Dynamic Network Connectivity: A new form of neuroplasticity," *Trends in cognitive sciences,* vol. 14, pp. 365-375, 2010.

[193] M. Naito, Y. Michioka, K. Ozawa, M. KIGUCHI, and T. KANAZAWA, "A communication means for totally locked-in ALS patients based on changes in cerebral blood volume measured with near-infrared light," *IEICE transactions on information and systems,* vol. 90, pp. 1028-1037, 2007.

[194] S. D. Power, T. H. Falk, and T. Chau, "Classification of prefrontal activity due to mental arithmetic and music imagery using hidden Markov models and frequency domain near-infrared spectroscopy," *Journal of neural engineering,* vol. 7, p. 026002, 2010.

[195] S. D. Power, A. Kushki, and T. Chau, "Towards a system-paced near-infrared spectroscopy brain? computer interface: differentiating prefrontal activity due to mental arithmetic and mental singing from the no-control state," *Journal of neural engineering,* vol. 8, p. 066004, 2011.

[196] G. Bauernfeind, R. Scherer, G. Pfurtscheller, and C. Neuper, "Single-trial classification of antagonistic oxyhemoglobin responses during mental arithmetic," *Medical & biological engineering & computing,* vol. 49, pp. 979-984, 2011.

[197] B. Abibullaev, J. An, and J.-I. Moon, "Neural network classification of brain hemodynamic responses from four mental tasks," *International Journal of Optomechatronics,* vol. 5, pp. 340-359, 2011.

[198] L. Holper and M. Wolf, "Single-trial classification of motor imagery differing in task complexity: a functional near-infrared spectroscopy study," *Journal of neuroengineering and rehabilitation,* vol. 8, p. 1, 2011.

[199] S. D. Power, A. Kushki, and T. Chau, "Intersession consistency of single-trial classification of the prefrontal response to mental arithmetic and the no-control state by NIRS," *PloS one,* vol. 7, p. e37791, 2012.

[200] S. D. Power and T. Chau, "Automatic single-trial classification of prefrontal hemodynamic activity in an individual with Duchenne muscular dystrophy," *Developmental neurorehabilitation,* vol. 16, pp. 67-72, 2013.

[201] L. C. Schudlo and T. Chau, "Dynamic topographical pattern classification of multichannel prefrontal NIRS signals: II. Online differentiation of mental arithmetic and rest," *Journal of neural engineering,* vol. 11, p. 016003, 2013.

[202] M. J. Khan, M. J. Hong, and K.-S. Hong, "Decoding of four movement directions using hybrid NIRS-EEG brain-computer interface," *Frontiers in human neuroscience,* vol. 8, p. 244, 2014.

[203] K.-S. Hong, N. Naseer, and Y.-H. Kim, "Classification of prefrontal and motor cortex signals for three-class fNIRS–BCI," *Neuroscience letters,* vol. 587, pp. 87-92, 2015.

[204] N. Naseer, N. K. Qureshi, F. M. Noori, and K.-S. Hong, "Analysis of Different Classification Techniques for Two-Class Functional Near-Infrared Spectroscopy-Based Brain-Computer Interface," *Computational Intelligence and Neuroscience,* vol. 2016, 2016.

[205] J. Shin, K.-R. Müller, and H.-J. Hwang, "Near-infrared spectroscopy (NIRS)-based eyes-closed brain-computer interface (BCI) using prefrontal cortex activation due to mental arithmetic," *Scientific Reports,* vol. 6, 2016.

[206] M. Ding, S. L. Bressler, W. Yang, and H. Liang, "Short-window spectral analysis of cortical event-related potentials by adaptive multivariate autoregressive modeling: data preprocessing, model validation, and variability assessment," *Biological cybernetics,* vol. 83, pp. 35-45, 2000.

[207] L. Zhang, J. Sun, B. Sun, C. Gao, and H. Gong, "detecting bilateral functional connectivity in the prefrontal cortex during a Stroop task by near-infrared spectroscopy," *Journal of Innovative Optical Health Sciences,* vol. 6, p. 1350031, 2013.

[208] M. D. Greicius, K. Supekar, V. Menon, and R. F. Dougherty, "Resting-state functional connectivity reflects structural connectivity in the default mode network," *Cerebral cortex,* vol. 19, pp. 72-78, 2009.

[209] B. Biswal, F. Zerrin Yetkin, V. M. Haughton, and J. S. Hyde, "Functional connectivity in the motor cortex of resting human brain using echo-planar mri," *Magnetic resonance in medicine,* vol. 34, pp. 537-541, 1995.

[210] R. Salvador, A. Martinez, E. Pomarol-Clotet, J. Gomar, F. Vila, S. Sarró*, et al.*, "A simple view of the brain through a frequency-specific functional connectivity measure," *Neuroimage,* vol. 39, pp. 279-289, 2008.

[211] L. Sugiura, S. Ojima, H. Matsuba-Kurita, I. Dan, D. Tsuzuki, T. Katura*, et al.*, "Sound to language: different cortical processing for first and second languages in elementary school children as revealed by a large-scale study using fNIRS," *Cerebral Cortex,* p. bhr023, 2011.

[212] S. Sasai, F. Homae, H. Watanabe, and G. Taga, "Frequency-specific functional connectivity in the brain during resting state revealed by NIRS," *Neuroimage,* vol. 56, pp. 252-257, 2011.

[213] B. R. White, A. Z. Snyder, A. L. Cohen, S. E. Petersen, M. E. Raichle, B. L. Schlaggar*, et al.*, "Resting-state functional connectivity in the human brain revealed with diffuse optical tomography," *Neuroimage,* vol. 47, pp. 148-156, 2009.

[214] H. Zhang, Y.-J. Zhang, C.-M. Lu, S.-Y. Ma, Y.-F. Zang, and C.-Z. Zhu, "Functional connectivity as revealed by independent component analysis of resting-state fNIRS measurements," *Neuroimage,* vol. 51, pp. 1150-1161, 2010.

[215] H. Niu and Y. He, "Resting-state functional brain connectivity: lessons from functional near-infrared spectroscopy," *The Neuroscientist,* vol. 20, pp. 173-188, 2014.

[216] C.-J. Huang, P.-H. Chou, H.-L. Wei, and C.-W. Sun, "Functional connectivity during phonemic and semantic verbal fluency test: a multichannel near infrared spectroscopy study," *IEEE Journal of Selected Topics in Quantum Electronics,* vol. 22, pp. 43-48, 2016.

[217] L. Zhang, J. Sun, B. Sun, Q. Luo, and H. Gong, "Studying hemispheric lateralization during a Stroop task through near-infrared spectroscopy-based connectivity," *Journal of biomedical optics,* vol. 19, pp. 057012-057012, 2014.

[218] R. C. Mesquita, M. A. Franceschini, and D. A. Boas, "Resting state functional connectivity of the whole head with near-infrared spectroscopy," *Biomedical optics express,* vol. 1, pp. 324-336, 2010.

[219] H. Zhu, Y. Fan, H. Guo, D. Huang, and S. He, "Reduced interhemispheric functional connectivity of children with autism spectrum disorder: evidence from functional near infrared spectroscopy studies," *Biomedical optics express,* vol. 5, pp. 1262-1274, 2014.

[220] S. Koike, Y. Nishimura, R. Takizawa, N. Yahata, and K. Kasai, "Near-infrared spectroscopy in schizophrenia: a possible biomarker for predicting clinical outcome and treatment," 2013.

[221] K. Ikezawa, M. Iwase, R. Ishii, M. Azechi, L. Canuet, K. Ohi*, et al.*, "Impaired regional hemodynamic response in schizophrenia during multiple prefrontal activation tasks: a two-channel near-infrared spectroscopy study," *Schizophrenia research,* vol. 108, pp. 93-103, 2009.

[222] L. Xu, B. Wang, G. Xu, W. Wang, Z. Liu, and Z. Li, "Functional connectivity analysis using fNIRS in healthy subjects during prolonged simulated driving," *Neuroscience Letters,* 2017.

[223] H. Tsunashima and K. Yanagisawa, "Measurement of brain function of car driver using functional near-infrared spectroscopy (fNIRS)," *Computational intelligence and neuroscience,* vol. 2009, 2009.

[224] L. Duan, Y.-J. Zhang, and C.-Z. Zhu, "Quantitative comparison of resting-state functional connectivity derived from fNIRS and fMRI: a simultaneous recording study," *Neuroimage,* vol. 60, pp. 2008-2018, 2012.

[225] C.-M. Lu, Y.-J. Zhang, B. B. Biswal, Y.-F. Zang, D.-L. Peng, and C.-Z. Zhu, "Use of fNIRS to assess resting state functional connectivity," *Journal of neuroscience methods,* vol. 186, pp. 242-249, 2010.

[226] M. L. Schroeter, O. Schmiedel, and D. Y. von Cramon, "Spontaneous low-frequency oscillations decline in the aging brain," *Journal of Cerebral Blood Flow & Metabolism,* vol. 24, pp. 1183-1191, 2004.

[227] A. V. Medvedev, "Does the resting state connectivity have hemispheric asymmetry? A near-infrared spectroscopy study," *Neuroimage,* vol. 85, pp. 400-407, 2014.

[228] T. Katura, N. Tanaka, A. Obata, H. Sato, and A. Maki, "Quantitative evaluation of interrelations between spontaneous low-frequency oscillations in cerebral hemodynamics and systemic cardiovascular dynamics," *Neuroimage,* vol. 31, pp. 1592-1600, 2006.

[229] C. W. Wu, H. Gu, H. Lu, E. A. Stein, J.-H. Chen, and Y. Yang, "Frequency specificity of functional connectivity in brain networks," *Neuroimage,* vol. 42, pp. 1047-1055, 2008.

[230] Q. Tan, M. Zhang, Y. Wang, M. Zhang, Y. Wang, Q. Xin*, et al.*, "Frequency-specific functional connectivity revealed by wavelet-based coherence analysis

in elderly subjects with cerebral infarction using NIRS method," *Medical physics,* vol. 42, pp. 5391-5403, 2015.

[231]   S. Geng, X. Liu, B. B. Biswal, and H. Niu, "Effect of Resting-State fNIRS Scanning Duration on Functional Brain Connectivity and Graph Theory Metrics of Brain Network," *Frontiers in neuroscience,* vol. 11, 2017.

[232]   B. Koo, H.-G. Lee, Y. Nam, H. Kang, C. S. Koh, H.-C. Shin*, et al.*, "A hybrid NIRS-EEG system for self-paced brain computer interface with online motor imagery," *Journal of neuroscience methods,* vol. 244, pp. 26-32, 2015.

[233]   X. Yin, B. Xu, C. Jiang, Y. Fu, Z. Wang, H. Li*, et al.*, "A hybrid BCI based on EEG and fNIRS signals improves the performance of decoding motor imagery of both force and speed of hand clenching," *Journal of Neural Engineering,* vol. 12, 2015.

[234]   F. Wallois, M. Mahmoudzadeh, A. Patil, and R. Grebe, "Usefulness of simultaneous EEG–NIRS recording in language studies," *Brain and language,* vol. 121, pp. 110-123, 2012.

[235]   H. Morioka, A. Kanemura, S. Morimoto, T. Yoshioka, S. Oba, M. Kawanabe*, et al.*, "Decoding spatial attention by using cortical currents estimated from electroencephalography with near-infrared spectroscopy prior information," *Neuroimage,* vol. 90, pp. 128-139, 2014.

[236]   S. Ahn, T. Nguyen, H. Jang, J. G. Kim, and S. C. Jun, "Exploring neuro-physiological correlates of drivers' mental fatigue caused by sleep deprivation using simultaneous EEG, ECG, and fNIRS data," *Frontiers in human neuroscience,* vol. 10, 2016.

[237]   V. D. Calhoun, T. Adali, N. Giuliani, J. Pekar, K. Kiehl, and G. Pearlson, "Method for multimodal analysis of independent source differences in schizophrenia: combining gray matter structural and auditory oddball functional data," *Human brain mapping,* vol. 27, pp. 47-62, 2006.

[238]   P. A. Valdes-Sosa, J. M. Sanchez-Bornot, R. C. Sotero, Y. Iturria-Medina, Y. Aleman-Gomez, J. Bosch-Bayard*, et al.*, "Model driven EEG/fMRI fusion of brain oscillations," *Human brain mapping,* vol. 30, pp. 2701-2721, 2009.

[239]  S. M. Plis, V. D. Calhoun, M. P. Weisend, T. Eichele, and T. Lane, "MEG and fMRI fusion for non-linear estimation of neural and BOLD signal changes," *Frontiers in neuroinformatics,* vol. 4, 2010.

[240]  H. Yang, J. Liu, J. Sui, G. Pearlson, and V. D. Calhoun, "A hybrid machine learning method for fusing fMRI and genetic data: combining both improves classification of schizophrenia," *Frontiers in human neuroscience,* vol. 4, 2010.

[241]  P. Skudlarski, K. Jagannathan, K. Anderson, M. C. Stevens, V. D. Calhoun, B. A. Skudlarska*, et al.*, "Brain connectivity is not only lower but different in schizophrenia: a combined anatomical and functional approach," *Biological psychiatry,* vol. 68, pp. 61-69, 2010.

[242]  B. Horwitz and D. Poeppel, "How can EEG/MEG and fMRI/PET data be combined?," *Human brain mapping,* vol. 17, pp. 1-3, 2002.

[243]  V. D. Calhoun and T. Adali, "Feature-based fusion of medical imaging data," *IEEE Transactions on Information Technology in Biomedicine,* vol. 13, pp. 711-720, 2009.

[244]  V. D. Calhoun, T. Adali, G. Pearlson, and K. Kiehl, "Neuronal chronometry of target detection: fusion of hemodynamic and event-related potential data," *Neuroimage,* vol. 30, pp. 544-553, 2006.

[245]  Q.-S. Sun, S.-G. Zeng, P.-A. Heng, and D.-S. Xia, "Feature fusion method based on canonical correlation analysis and handwritten character recognition," in *Control, Automation, Robotics and Vision Conference, 2004. ICARCV 2004 8th*, 2004, pp. 1547-1552.

[246]  S. Tak and J. C. Ye, "Statistical analysis of fNIRS data: a comprehensive review," *Neuroimage,* vol. 85, pp. 72-91, 2014.

[247]  V. N. Vapnik and V. Vapnik, *Statistical learning theory* vol. 1: Wiley New York, 1998.

[248]  C.-C. Chang and C.-J. Lin, "LIBSVM: a library for support vector machines," *ACM Transactions on Intelligent Systems and Technology (TIST),* vol. 2, p. 27, 2011.

[249]  T. Fawcett, "An introduction to ROC analysis," *Pattern recognition letters,* vol. 27, pp. 861-874, 2006.

[250] S. G. Hart and L. E. Staveland, "Development of NASA-TLX (Task Load Index): Results of empirical and theoretical research," *Advances in psychology,* vol. 52, pp. 139-183, 1988.

[251] A. Delorme and S. Makeig, "EEGLAB: an open source toolbox for analysis of single-trial EEG dynamics including independent component analysis," *Journal of neuroscience methods,* vol. 134, pp. 9-21, 2004.

[252] R. N. Khushaba, S. Kodagoda, S. Lal, and G. Dissanayake, "Driver drowsiness classification using fuzzy wavelet-packet-based feature-extraction algorithm," *Biomedical Engineering, IEEE Transactions on,* vol. 58, pp. 121-131, 2011.

[253] S. Sutoko, H. Sato, A. Maki, M. Kiguchi, Y. Hirabayashi, H. Atsumori*, et al.*, "Tutorial on platform for optical topography analysis tools," *Neurophotonics,* vol. 3, pp. 010801-010801, 2016.

[254] V. D. Calhoun, J. Liu, and T. Adalı, "A review of group ICA for fMRI data and ICA for joint inference of imaging, genetic, and ERP data," *Neuroimage,* vol. 45, pp. S163-S172, 2009.

[255] N. M. Correa, Y.-O. Li, T. Adali, and V. D. Calhoun, "Canonical correlation analysis for feature-based fusion of biomedical imaging modalities and its application to detection of associative networks in schizophrenia," *IEEE journal of selected topics in signal processing,* vol. 2, pp. 998-1007, 2008.

[256] T. W. Lee, M. Girolami, and T. J. Sejnowski, "Independent component analysis using an extended infomax algorithm for mixed subgaussian and supergaussian sources," *Neural Computation,* vol. 11, pp. 417-441, 1999.

[257] D. Weenink, "Canonical correlation analysis," in *Proceedings of the Institute of Phonetic Sciences of the University of Amsterdam*, 2003, pp. 81-99.

[258] W. Krzanowski, *Principles of multivariate analysis*: OUP Oxford, 2000.

[259] E. Glerean, J. Salmi, J. M. Lahnakoski, I. P. Jääskeläinen, and M. Sams, "Functional magnetic resonance imaging phase synchronization as a measure of dynamic functional connectivity," *Brain connectivity,* vol. 2, pp. 91-101, 2012.

[260] K. Dedovic, R. Renwick, N. K. Mahani, V. Engert, S. J. Lupien, and J. C. Pruessner, "The Montreal Imaging Stress Task: using functional imaging to

investigate the effects of perceiving and processing psychosocial stress in the human brain," *Journal of Psychiatry and Neuroscience,* vol. 30, p. 319, 2005.

[261] R. Thibodeau, R. S. Jorgensen, and S. Kim, "Depression, anxiety, and resting frontal EEG asymmetry: a meta-analytic review," *Journal of abnormal psychology,* vol. 115, p. 715, 2006.

[262] M. Tops, J. M. van Peer, A. E. Wester, A. A. Wijers, and J. Korf, "State-dependent regulation of cortical activity by cortisol: an EEG study," *Neuroscience letters,* vol. 404, pp. 39-43, 2006.

[263] C. Quaedflieg, T. Meyer, F. Smulders, and T. Smeets, "The functional role of individual-alpha based frontal asymmetry in stress responding," *Biological psychology,* vol. 104, pp. 75-81, 2015.

[264] L. Ossewaarde, S. Qin, H. J. Van Marle, G. A. van Wingen, G. Fernández, and E. J. Hermans, "Stress-induced reduction in reward-related prefrontal cortex function," *Neuroimage,* vol. 55, pp. 345-352, 2011.

[265] A. F. Arnsten and P. S. Goldman-Rakic, "Noise stress impairs prefrontal cortical cognitive function in monkeys: evidence for a hyperdopaminergic mechanism," *Archives of general psychiatry,* vol. 55, pp. 362-368, 1998.

[266] F. Tian, A. Yennu, A. Smith-Osborne, F. Gonzalez-Lima, C. S. North, and H. Liu, "Prefrontal responses to digit span memory phases in patients with post-traumatic stress disorder (PTSD): a functional near infrared spectroscopy study," *NeuroImage: Clinical,* vol. 4, pp. 808-819, 2014.

[267] A. F. Arnsten, M. A. Raskind, F. B. Taylor, and D. F. Connor, "The effects of stress exposure on prefrontal cortex: translating basic research into successful treatments for post-traumatic stress disorder," *Neurobiology of stress,* vol. 1, pp. 89-99, 2015.

[268] T. Jovanovic, T. Ely, N. Fani, E. M. Glover, D. Gutman, E. B. Tone*, et al.*, "Reduced neural activation during an inhibition task is associated with impaired fear inhibition in a traumatized civilian sample," *Cortex,* vol. 49, pp. 1884-1891, 2013.